\documentclass[twocolumn,english,pra]{revtex4-1}
\usepackage[T1]{fontenc}
\usepackage[latin9]{inputenc}
\usepackage{xcolor}
\usepackage{array}
\usepackage{float}
\usepackage{multirow}
\usepackage{amsmath}
\usepackage{amssymb}
\usepackage{graphicx}
\usepackage{wasysym}

\makeatletter

\newcommand{\noun}[1]{\textsc{#1}}
\providecommand{\tabularnewline}{\\}

\makeatother

\usepackage{babel}
\begin{document}

\title{Full multipartite steering inseparability, genuine multipartite steering
and monogamy for continuous variable systems}

\author{Run Yan Teh$^{1}$, Manuel Gessner$^{2,3}$, Margaret D. Reid$^{1}$
and Matteo Fadel$^{4}$}

\affiliation{$^{1}$Centre for Quantum Science and Technology Theory, Swinburne
University of Technology, Melbourne 3122, Australia}

\affiliation{$^{2}$Laboratoire Kastler Brossel, ENS-Universit\'{e} PSL, CNRS,
Sorbonne Universit\'{e}, College de France, 24 Rue Lhomond, 75005,
Paris, France}

\affiliation{$^{3}$ICFO-Institut de Ci\`{e}ncies Fot\`{o}niques, The Barcelona
Institute of Science and Technology, Av. Carl Friedrich Gauss 3, 08860,
Castelldefels (Barcelona), Spain}

\affiliation{$^{4}$Department of Physics, University of Basel, Klingelbergstrasse
82, 4056 Basel, Switzerland}
\begin{abstract}
We derive inequalities sufficient to detect the genuine $N$-partite
steering of $N$ distinct systems. Here, we are careful to distinguish
between the concepts of full $N$-partite steering inseparability
(where steering is confirmed individually for all bipartitions of
the $N$ systems, thus negating the bilocal hidden state model for
each bipartition) and genuine $N$-partite steering (which excludes
all convex combinations of the bilocal hidden state models). Other
definitions of multipartite steering are possible and we also derive
inequalities to detect a stricter genuine $N$-partite steering where
only one site needs to be trusted. The inequalities are expressed
as variances of quadrature phase amplitudes and thus apply to continuous
variable systems. We show how genuine $N$-partite steerable
states can be created and detected for the nodes of a network formed
from a single-mode squeezed state passed through a sequence of $N-1$
beam splitters. A stronger genuine $N$-partite steering is created,
if one uses two squeezed inputs, or $N$ squeezed inputs. We  are
able to confirm that genuine tripartite steering (by the above definition
and by the stricter definition) has been realised experimentally.
Finally, we analyze how bipartite steering and entanglement are distributed
among the systems in the tripartite case, illustrating with monogamy
inequalities. While we use Gaussian states to benchmark the criteria,
the inequalities derived in this paper are not based on the assumption
of Gaussian states, which gives advantage for quantum communication
protocols.
\end{abstract}
\maketitle

\section{Introduction}

From the perspective of both  fundamental and applied physics,
understanding whether $N$ distinct systems can be genuinely entangled
is considered important \citep{vanLoock_application_PRL2000,Horodecki_RMP2009,Jing_application_PRL2003,Jungnitsch_biseparable_PRL2011,Bancal_application_PRL2011,Hillery_application_PRA1999,Bose_application_PRA1998,Briegel_application_Nature2009,Dur_PRA2000,vanLoock_PRA2003}.
Svetlichny first showed that a genuine tripartite nonlocality can
be shared among three systems \citep{Svetlichny_PRD1987}. Greenberger,
Horne and Zeilinger (GHZ) \citep{Greenberger1989} and Mermin \citep{Mermin_PRL1990}
investigated the nonlocal properties of $N$ genuinely entangled systems
created in an extreme quantum superposition state.  Multipartite
entangled systems have applications in the field of quantum information.
The tripartite-entangled GHZ states were proposed for quantum secret
sharing, where two parties must collaborate to uncover a cryptographic
key \citep{Hillery_application_PRA1999}. Usually, the certification
of entanglement required to ensure security against eavesdropping
involves assumptions about the instruments at the given sites. When
entanglement is confirmed by way of violation of a Bell inequality,
however, fewer assumptions are needed, resulting in device-independent
security \citep{mayers1998quantum,Mayers_selftesting2004,Barrett_QKD_PRL2005,acin_PRL2006,acin_PRL2007,Bancal_application_PRL2011}.
A quantum network consisting of mutually entangled systems may form
the basis for a quantum communication network. Recent papers motivate
the use of multipartite Bell nonlocality for device-independent security
on a network \citep{Rosset_PRL2016,Moreno_Chaves_PRA2020,Holz_Bell_PRR2020}.

Einstein-Podolsky-Rosen (EPR) steering is a type of entanglement associated
with the nonlocality of the EPR paradox \citep{Einstein_EPR_PhysRev1935}.
The concept of EPR-steering was motivated by the arguments put forward
by Schr\"{o}dinger in response to the EPR-paradox paper \citep{Jones_PRA2007,Wiseman_PRL2007,Cavalcanti_PRA2009,Uola_RMP2020}.
Based on the negation of asymmetric local hidden state models, steering
is useful for one-sided device-independent quantum security, where
one has control over some, but not all, of the devices on the nodes
of the network \citep{Branciard_PRA2012,He_Reid_steering_PRL2013,Huang_Nori_PRA2019}.
Implementing steering, as opposed to entanglement, thus increases
the potential for ultra-secure communication \citep{Branciard_PRA2012,Opanchuk_PRA2014,Walk_optica2016,Kocsis_Nature2015}.
Steering  has also been proposed as a resource for other applications,
including secure quantum teleportation \citep{He_PRL2015}, quantum
metrology \citep{Yadin_NatureCom2021} and secret sharing \citep{wilkinson2021quantum}.
However, the set of steerable states is a strict subset of the set
of entangled states \citep{Jones_PRA2007}. The detection of $N$-partite
steering is therefore more challenging and standard witnesses for
entanglement will not suffice. An early criterion for steering was
an inequality applied to the EPR paradox \citep{Reid_PRA1989,Cavalcanti_PRA2009,Wiseman_PRL2007}.
Numerous steering criteria have since been derived, but most refer
to the bipartite case \citep{Uola_RMP2020}. This highlights the need
for criteria to confirm steering shared among $N$ systems, derived
with minimal assumptions about the states being measured.

The concept of multipartite steering was introduced by He and Reid
\citep{He_Reid_steering_PRL2013} and developed for Gaussian states
by Kogias \emph{et al}. \citep{Kogias_PRL2015}. Experiments followed
\citep{Armstrong_Nature2015,Cavalcanti_steering_Nature2015,Fadel409,Kunkel413,Opanchuk_steering_PRA2019,Deng_monogamy_PRL2017},
 which  motivated studies of the monogamy of steering \citep{Reid_monogamy_PRA2013,Kogias_PRL2015,Cheng_PRA2016,Deng_monogamy_PRL2017}.
Marian and Marian have recently derived criteria involving EPR variances
that are sufficient to certify multimode steering, and have linked
these results to Gaussian states \citep{Marian_PRA2021}. However,
as for multipartite entanglement and nonlocality \citep{vanLoock_PRA2003,Guhne:2009aa,Shalm_Nature2012,PhysRevA.90.062337,Gallego_Navascues_PRL2012,Bancal_multipartite_PRA2013,Szalay_quantum2019,Ren_PRL2021},
different definitions of multipartite steering are possible. This
is particularly true for steering, where subtleties enter into the
definitions, because different nodes on a network can have different
levels of trust in devices.

In this paper, we further address gaps in the literature, by deriving
criteria sufficient to detect $N$-partite steering. We follow \citep{gessner_steering,Shalm_Nature2012,PhysRevA.90.062337}
to provide a treatment where classes of multipartite steering are
clearly distinguished. In particular, we consider \emph{full $N$-partite
steering inseparability }(where there is steering across each of
the bipartitions of the $N$ systems), and distinguish this from the
stricter definition, \emph{genuine $N$-partite steering} (which excludes
convex mixtures of hidden states allowing for bilocality along different
bipartitions). Similar to the earlier results of van Loock and
Furusawa for the $N$-partite entanglement of continuous variable
(CV) systems \citep{vanLoock_PRA2003,PhysRevA.90.062337,Armstrong_Nature2015,Shalm_Nature2012},
we envisage that local measurements can be performed on each system
and derive inequalities that are only violated if the systems are
mutually steerable. The inequalities are in terms of the variances
of the local field quadrature phase amplitudes. We follow the bilocal
approach of Svetlichny \citep{Svetlichny_PRD1987} and Collins \emph{et
al}. \citep{Collins_PRL2002}, noting that other approaches might
also be considered \citep{Gallego_Navascues_PRL2012,Bancal_multipartite_PRA2013,Dutta_PRA2020}.

As we show in this paper, the variance criteria are useful to detect
genuine multipartite steering for $N$-partite Gaussian systems, where
one creates Gaussian states for $N$ field modes from squeezed states
and beam splitters. This is relevant given recent CV multipartite
entanglement experiments \citep{Aoki_PRL2003,Coelho_Science2009,Pfister_PRL2011,Armstrong_Nature2012,pfister_PRL2014,Gerke_Fabre_PRL2015,Armstrong_Nature2015}
and Gaussian boson sampling experiments \citep{Zhong_science2020},
some of which realized networks for very large $N$. Three types of
$N$-partite Gaussian states are considered in this paper: the CV
GHZ (and cluster, for $N=3$), CV EPR and CV split-squeezed (SS) states.
Although we use the $N$-partite Gaussian states to benchmark our
criteria, the inequalities are valid in principle to detect steering
in non-Gaussian systems. This is recognized in the bipartite case
where it has been shown that variance inequalities can detect steering
for NOON states \citep{Teh_PRA2016} and for cat states \citep{Cavalcanti_PRA2008,Rosales-Zarate_JOSA2015}.
Our work, different to some previous work \citep{Kogias_PRL2015},
is not based on the underlying assumption that the systems are prepared
in Gaussian states. This is an important and necessary requirement
for one-sided device-independent protocols \citep{Branciard_PRA2012,Opanchuk_PRA2014,Walk_optica2016,Kocsis_Nature2015}.
Using values reported in the literature, we apply the inequalities
to confirm that genuine tripartite steering has been realized experimentally
for optical fields. We also derive inequalities which allow for $N-1$
untrusted sites, and deduce that the genuine tripartite steering has
been confirmed for $N=3$, with the requirement that only one site
needs to be trusted.

We also examine monogamy relations for tripartite systems, where $N=3$.
These relations give constraints on the amount of bipartite steering
and entanglement between any two of the nodes of the tri-party network.
We derive a new relation that constrains the amount of bipartite entanglement,
showing that for the CV GHZ, CV EPR and CV SS states, although a limited
amount of bipartite entanglement is possible, no bipartite steering
can be observed as measured by the two-mode Gaussian steering parameter.

The continuous-variable criteria of this paper will apply to systems
of large collective spin, in certain limits \citep{Julsgaard:2001aa,Korolkova_polarization_entanglement_PRA2002,PhysRevA.67.052104,Teh_PRA2019,Teh_erratum_PRA2020}.
Furthermore, since the inequalities of this paper are derived from
uncertainty relations, the methods presented may be generalized for
multipartite spin systems, as done for Bohm's EPR paradox \citep{Cavalcanti_optexp2009}.
Although multiparticle and multi-mode entanglement and steering can
be inferred from spin squeezing \citep{Sorensen_Nature2001,Sorensen_spin_squeezing_RPL2001,Gross_Nature2010,Zarate_steering_PRA2018,Fadel_splitBEC_PRA2020,Schmied_Science2016},
Fisher information \citep{Frowis_RMP2018,Pezze_RMP2018,Qin:2019aa,Yadin_NatureCom2021},
or (using superselection rules) interference \citep{Dalton_PhysicaScripta2017_2,Opanchuk_steering_PRA2019,Dalton_PRA2020},
to demonstrate nonlocality in a strict way requires spatial separation
and local measurements \citep{Fadel_PRA2020,Opanchuk_double_wellBEC_PRA2012,He_NJP2012,Gross_Oberthaler_Nature2011,He_EPRsteering_double_wellBEC_PRA2012,Peise_Nature2015}.
 Two-particle entanglement and Bell correlations have been studied
and reported for separated atoms \citep{Rosenfeld_PRL2017,Shin_bell_nature2019,Bonneau_PRA2018,Bergschneider_Nature2019},
and bipartite EPR steering and genuine $N$-partite entanglement has
been observed for separated atomic clouds containing several hundreds
of atoms \citep{Fadel409,Kunkel413}. Yet, it remains a challenge
to demonstrate $N$-partite steering ($N>2$) for spatially separated
atomic systems. The results of this paper may be useful for this
purpose.

\textbf{\emph{Summary of paper:}} In Section II, we summarize the
definitions of $N$-partite steering, distinguishing between full
$N$-partite steering inseparability and the more strict definition
given by genuine $N$-partite steering. In this section, we derive
steering inequalities that if violated will reveal the presence of
genuine $N$-partite steering. These inequalities involve a single
set of gain parameters optimized to detect the EPR steering of a single
preferred party, and are similar to those considered recently by Marian
and Marian in independent work \citep{Marian_PRA2021}.

Continuing, in Section III, we focus on tripartite systems. We expand
the set of criteria, also deriving inequalities closely related to
those of van Loock and Furusawa that have been widely used to study
multipartite entanglement \citep{Armstrong_Nature2012,Villar_PRL2006,Coelho_Science2009}.
These inequalities include those with a broader set of gain parameters,
which allow certification of genuine tripartite steering once the
EPR steering of each party is sufficiently strong (as measured by
vanishing conditional inference variances). The work presented in
Sections II and III extends the work of He and Reid \citep{He_Reid_steering_PRL2013},
who derived inequalities based on a weaker form of genuine $N$-partite
steering. We confirm that those earlier inequalities (derived in \citep{He_Reid_steering_PRL2013,PhysRevA.90.062337})
will also certify the stricter form of genuine tripartite steering
defined in the present paper. In addition, we derive new inequalities
that if violated will certify an even stricter form of genuine tripartite
steering, giving a method to detect genuine tripartite entanglement
and steering where only one node needs to be trusted.

In Section IV, we show how the criteria derived in Sections II and
III with suitably optimized gains are useful to detect the genuine
$N$-partite steering of four types of CV Gaussian states. These are
the CV split-squeezed states, the CV EPR states, and the CV GHZ (and
cluster) states, created using one, two and $N$ squeezed-vacuum states
incident on beam splitters. We examine details for $N=3$ in Section
V. Using the criteria, we infer full tripartite steering inseparability
and strict genuine tripartite steering from experimental results reported
in the literature. Full tripartite steering inseparability has been
detected in at least two experiments \citep{Armstrong_Nature2012,Armstrong_Nature2015},
for the correlations generated from CV EPR and cluster states. 
For CV GHZ and cluster states, the stricter form of genuine tripartite
steering (including where two sites can be untrusted) is also detectable
using the inequalities, and we are able to confirm the experimental
realization for tripartite CV cluster states in one experiment \citep{Armstrong_Nature2012}
where values for the van Loock-Furusawa variances were measured. Finally,
in Section VI, we give an analysis of the monogamy properties of the
CV tripartite steerable states. A conclusion is given in Section VII.

\section{Definitions and criteria for $N$-partite steering}

\subsection{Preliminaries}

Consider an $N$-partite system consisting of $N$ modes, where each
mode labelled $i=1,..,N$ is described by bosonic annihilation operators
$a_{i}$ satisfying canonical commutation relations. We introduce
the generalized quadrature  operators with phase $\varphi$ as 
\begin{equation}
q_{i}(\varphi)=e^{i\varphi}a_{i}+e^{-i\varphi}a_{i}^{\dagger}\thinspace,\label{eq:q}
\end{equation}
noting we will later define $x_{i}=a_{i}+a_{i}^{\dagger}$ and $p_{i}=(a_{i}-a_{i}^{\dagger})/i$
which implies $\Delta x_{i}\Delta p_{i}\geq1$.

We use $\Lambda=A-B$ to denote a bipartition of the $N$ subsystems
into two groups, $A$ and $B$. Proving entanglement for such a bipartition
would require to falsify any separable model of the form \citep{Werner_PRA1989}
\begin{equation}
\rho_{A-B}=\sum_{R}P_{R}\rho_{A}^{(R)}\rho_{B}^{(R)}\label{eq:insep-1-23}
\end{equation}
where $\sum_{R}P_{R}=1$ ($P_{R}>0$), $\rho_{A-B}$ represents a
density operator for the combined systems, $\rho_{A}^{(R)}$ is an
arbitrary density operator for the subsystem $A$, and $\rho_{B}^{R}$
is an arbitrary density operator for the subsystem $B$. Here, no
assumption of separability is assumed between any subsystems of either
$A$ or $B$. However, to confirm \emph{steering} of system $A$ (by
$B$), the falsification is to be achieved without the explicit assumption
that $\rho_{B}^{(R)}$ would necessarily correspond to a quantum state
described by a quantum density operator \citep{Wiseman_PRL2007}.
To make this distinction symbolically, we denote the density operator
$\rho_{A}^{(R)}$ by $\rho_{AQ}^{(R)}$, but omit the subscript for
$\rho_{B}^{(R)}$. Thus, we demonstrate the steering of $A$ by system
$B$ if we falsify the biseparable local hidden state model \emph{symbolized}
as
\begin{equation}
\rho_{B\rightarrow A}=\sum_{R}P_{R}\rho_{AQ}^{(R)}\rho_{B}^{(R)}\thinspace.\label{eq:lhs-1-23}
\end{equation}
Where more convenient, we will also symbolize as $\rho_{AQ-B}$.

As rigorously formalized in \citep{Wiseman_PRL2007,Jones_PRA2007,Cavalcanti_PRA2009},
the definition of steering concerns certain probabilistic models,
called local hidden state (LHS) models, rather than density operators
defined within quantum theory. Such LHS models are local hidden variable
models, where extra constraints are given for the local hidden variable
states describing the local system at some of the sites. Thus, to
determine steering of $A$ by $B$, we negate all models giving the
probability distribution for joint measurements at sites $A$ and
$B$ as
\begin{equation}
P(x_{A},x_{B}|\theta,\phi)=\int\rho(\lambda)d\lambda P_{Q}(x_{A}|\lambda,\theta)P(x_{B}|\lambda,\phi)\thinspace,\label{eq:lhs-1-23-int}
\end{equation}
where $x_{A}$ and $x_{B}$ are the results of measurements on each
system. Here, $\lambda$ are hidden variables, $\rho(\lambda)$ is
the corresponding probability density, and $\theta$ and $\phi$ are
local measurement settings at the sites of systems $A$ and $B$ respectively.
$P(x_{A}|\lambda,\theta)$ is the probability of an outcome $x_{A}$,
given the parameters $\lambda$ and the settings $\theta$. For steering
of $A$, it is required to negate the model where $P_{Q}(x_{A}|\lambda,\theta)$
is consistent with a local quantum density operator $\rho_{A}$ for
system $A$, as denoted by the subscript $Q$. There is thus a constraint
on the distributions for $A$, so that quantum uncertainty relations
are satisfied. If the criteria to negate the LHS models are to be
valid, one must be sure to use valid quantum devices and measurements
at that site. Hence the quantum sites are referred to as\emph{ trusted}
sites. Otherwise, the sites are said to be\emph{ untrusted}.

In this paper, for notational convenience we use Eq. (\ref{eq:lhs-1-23}),
the meaning of Eqs. (\ref{eq:lhs-1-23}) and (\ref{eq:lhs-1-23-int})
being equivalent. We will denote the bipartition $A-B$ where the
system $A$ is to be trusted as $AQ-B$. Similarly, the bipartition
$A-BQ$ denotes that $B$ is trusted. If the systems $A$ and $B$
comprise more than one subsystem, or mode, then more options of trust
are available. Suppose system $A$ is just one mode labelled $k$
and system $B$ is two modes labelled $l$ and $m$. If both subsystems
$l$ and $m$ are trusted, but system $k$ is not, then the bipartition
is denoted $k-(lmQ)$. The bipartition $k-lm$ where only the site
$l$ is trusted will be denoted $k-(lQ)m$.

\subsection{Steering in one partition}

We follow van Loock and Furusawa \citep{vanLoock_PRA2003} and introduce
the following linear combinations of the operators for the $N$ systems
\begin{align}
u=\sum_{i=1}^{N}g_{i}q_{i}(\varphi_{i})\thinspace,\qquad v=\sum_{i=1}^{N}h_{i}q_{i}(\chi_{i})\thinspace.\label{eq:quad-gh}
\end{align}
Here, $g_{i}$ and $h_{i}$ are real numbers that will be optimized
to reduce the variances in $u$ and $v$, as we will later see.

Let us first focus on the factorized description $\rho_{AQ}^{(R)}\rho_{B}^{(R)}$.
In this case the variance for the state denoted by $R$ in the expansion
Eq. (\ref{eq:lhs-1-23}) reads 
\begin{align}
(\Delta u)_{R}^{2} & =(\Delta u_{A})_{AQ}^{2}+(\Delta u_{B})_{B}^{2}\nonumber \\
 & \geq\left(\Delta u_{A}\right)_{AQ}^{2}\thinspace,
\end{align}
where 
\begin{align}
u_{I}=\sum_{i\in I}g_{i}q_{i}(\varphi_{i})\thinspace.
\end{align}
Here we use the notation that $(\Delta u_{A})_{AQ}^{2}$ is the variance
with respect to the operators of system $A$, evaluated assuming the
system is described as a quantum state $\rho_{A}^{(R)}$. The variance
associated with a system $B$ in a hidden variable state is constrained
only by the condition $(\Delta u_{B})_{B}^{2}\geq0$. Generally, we
use the notation $(\Delta x)^{2}$ to denote the variance of $x$,
and $\Delta x$ to denote the standard deviation.

Moreover, the LHS model symbolized by $\rho_{AQ}\rho_{B}$ as defined
by Eq. (\ref{eq:lhs-1-23}) must satisfy the uncertainty relation
\begin{align}
(\Delta u_{A})_{AQ}(\Delta v_{A})_{AQ}\geq C_{A}\thinspace,
\end{align}
where we define 
\begin{align}
C_{I}=\left|\sum_{i\in I}g_{i}h_{i}\sin(\varphi_{i}-\chi_{i})\right|
\end{align}
and we use 
\begin{align}
[u_{I},v_{I}]%
 & =2i\sum_{i\in I}g_{i}h_{i}\sin(\varphi_{i}-\chi_{i})\thinspace.
\end{align}
This arises from the Heisenberg uncertainty relation $\left[x_{j},p_{j}\right]=2i$.

Now we assume a LHS description of the form of Eq. (\ref{eq:lhs-1-23}).
For a system in a mixture, the overall variance of $u$ is \citep{Hofmann_PRA2003}
\begin{align}
(\Delta u)^{2} & \geq\sum_{R}P_{R}(\Delta u)_{R}^{2}\thinspace.\label{eq:var_u-2-1-1}
\end{align}
The Cauchy-Schwarz inequality implies
\begin{eqnarray}
(\Delta u)^{2}(\Delta v)^{2} & \geq & (\sum_{R}P_{R}(\Delta u)_{R}^{2})(\sum_{R}P_{R}(\Delta v)_{R}^{2})\nonumber \\
 & \geq & \{\sum_{R}P_{R}(\Delta u)_{R}(\Delta v)_{R}\}^{2}\thinspace.\label{eq:23-1-3}
\end{eqnarray}
Thus, using the concavity of the variance, the Cauchy-Schwarz inequality,
and the above results, we find for the system described by the LHS
model $\rho_{AQ}\rho_{B}$, it is always true that
\begin{align}
(\Delta u)(\Delta v) & \geq\sum_{R}P_{R}(\Delta u)_{R}(\Delta v)_{R}\nonumber \\
 & \geq\sum_{R}P_{R}(\Delta u)_{AQ}(\Delta v)_{AQ}\nonumber \\
 & \geq C_{A}\thinspace.
\end{align}
If this condition is violated, then one observes falsification of
the LHS model $\rho_{AQ}\rho_{B}$, and therefore steering of $A$
by $B$.

Analogously, the model $\rho_{A}\rho_{BQ}$ always implies 
\begin{align}
(\Delta u)(\Delta v)\geq C_{B}\thinspace.\label{eq:condAtoB}
\end{align}
If this condition is violated, then one observes falsification of
the LHS model $\rho_{A}\rho_{BQ}$, and therefore steering of $B$
by $A$.

To summarize our results so far, we detect \emph{one-way steering
in a particular direction} of a partition $\Lambda$ by one of the
criteria above. If either one of the two conditions, i.e., if 
\begin{align}
(\Delta u)(\Delta v)\geq\max\{C_{A},C_{B}\}\label{eq:oneway}
\end{align}
is not satisfied, we have observed \emph{one-way steering} in the
partition $\Lambda$. If, however, we see that both conditions do
not hold, i.e., if 
\begin{align}
(\Delta u)(\Delta v)\geq\min\{C_{A},C_{B}\}\label{eq:twoway}
\end{align}
is false, we have detected \emph{two-way steering} in the partition
$\Lambda$.

\subsection{Full steering inseparability}

To demonstrate the full $N$-partite inseparability of $N$ systems,
it is necessary to prove the failure of each separable model $\rho_{A-B}$,
for \emph{all} the bipartitions $\Lambda$ of the $N$ systems \citep{vanLoock_PRA2003}.
Full $N$-partite steering inseparability is to be defined in a similar
way. However, we see that because of the asymmetry in the definition
of steering, there is the possibility that steering across a bipartition
is in one direction only (``one-way steering'') \citep{Midgley_Olsen_PRA2010,Handchen_Nature2012}.
This would imply that one of the LHS models $\rho_{A\rightarrow B}$
or $\rho_{B\rightarrow A}$ is valid, while the other can be negated.
A consequence is that different definitions of multipartite steering
are possible, as formalized in the following.

\textbf{\textit{Definition: }}We conclude that a system displays \emph{full
$N$-partite steering inseparability} if one may demonstrate steering
for \emph{all} bipartitions $\Lambda=A-B$ of the $N$ systems. Specifically,
this means demonstrating, for each bipartition, that there is steering
at least in one direction i.e. for each $\Lambda$, either all models
denoted by $\rho_{A\rightarrow B}$, or all models denoted by $\rho_{B\rightarrow A}$,
or both, can be negated.

We say that we demonstrate \emph{full $N$-partite steering two-way
inseparability} if, for each bipartition, \emph{all }LHS models $\rho_{A_{\Lambda}\rightarrow B_{\Lambda}}$
\emph{and} $\rho_{B_{\Lambda}\rightarrow A_{\Lambda}}$ are negated.

\emph{Criteria}: A criterion sufficient to confirm full $N$-partite
steering inseparability reads as the violation of
\begin{align}
(\Delta u)(\Delta v)\geq\min_{\Lambda}\max\{C_{A},C_{B}\}\thinspace,\label{eq:fulloneway}
\end{align}
where the minimum includes all bipartitions $\Lambda=A-B$ of the
system (we make sure that $A-B$ and $B-A$ are considered as the
same partition). Similarly, a violation of the condition 
\begin{align}
(\Delta u)(\Delta v)\geq\min_{\Lambda}\min\{C_{A},C_{B}\}\thinspace,\label{eq:fulltwoway}
\end{align}
implies full two-way steering inseparability. The proof follows straightforwardly
from the definitions. $\square$

\subsection{Genuine multipartite steering}

\subsubsection{Definitions}

An even stronger condition is given if instead of excluding each bipartition
separately, we can exclude also more general LHS models that are constructed
from convex combinations of LHS models across all the \emph{different}
bipartitions. Formally, this is described in the notation of Section
II.A by a distribution 
\begin{align}
\rho=\sum_{\rho_{\Lambda}}P_{\Lambda}\rho_{\Lambda}\thinspace.\label{eq:modelgensteer}
\end{align}
Here, $\sum_{\Lambda}P_{\Lambda}=1$, $P_{\Lambda}>0$, and each $\rho_{\Lambda}$
describes an LHS model for at least one (both) direction(s) of the
partition $\Lambda$ in order to exclude genuine multipartite (two-way)
steering. We conclude that we demonstrate genuine $N$-partite steering
if the above model is negated.

Here, we already have two definitions, and as we discuss in Section
II.E and below, other definitions are also possible. In this paper,
we will adopt the stricter of the above definitions, and refer to
this as Definition 1 for genuine tripartite steering. This definition
is based on the concept to ensure steering across all bipartitions,
in both directions.

\textbf{\emph{Definition 1:}}\emph{ }We say that a system displays
\emph{genuine $N$-partite steering }if the above model, Eq. (\ref{eq:modelgensteer}),
can be negated, where $\rho_{\Lambda}$ denotes the LHS model in \emph{both}
directions ($A\rightarrow B$ and $B\rightarrow A$). Specifically,
we negate that the correlations can be modeled by a theory that expresses
the joint probability for a set of outcomes $\mathbf{x_{I}}=(x_{1},..x_{N})$
as 
\begin{equation}
P(\mathbf{x_{I}}|\mathbf{a})=\sum_{\lambda}P_{\lambda}P(\mathbf{x_{I}}|\lambda,\mathbf{a})\thinspace.\label{eq:problhsgenuine-1}
\end{equation}
Here, $x_{i}$ is an outcome of a measurement performed on the subsystem
$i$, and $\mathbf{a}$ is a set of numbers that denotes the measurement
settings for each subsystem. The $\lambda$ indexes all the LHS models
symbolized by $AQ-B$ and $BQ-A$ for each bipartition $\Lambda=A-B$.
Here,  $P(\mathbf{x_{I}}|\lambda,\mathbf{a})$ is the probability
for outcomes $\mathbf{x_{I}}=(x_{1},..,x_{N})$, given the system
is in the LHS state denoted by $\lambda$ and with measurement settings
\textbf{$\mathbf{a}$}. $P_{\lambda}\geq0$ is the probability for
the LHS state $\lambda$, and $\sum_{\lambda}P_{\lambda}=1$.

\textbf{\emph{Definition 2:}} Genuine $N$-partite one-way steering
is confirmed if the above model is negated, where for each bipartition
$\Lambda$, we consider only one of the directions for steering i.e.
only one of $AQ-B$ and $BQ-A$ is included in the convex combination.

\textbf{\emph{Definition 3}}\textbf{:} \textbf{\emph{(One trusted
site)}} We may propose as a very strict yet convincing definition
of genuine $N$-partite steering that one negates all LHS models (and
their convex combinations) where we consider that only one site can
be trusted.  In the tripartite case, we would seek to negate $kQ-lm$,
and also the models $k-(lQ)m$ and $k-l(mQ)$ (and all convex combinations).
This definition is stricter than the Definition 1, which only requires
negation of $kQ-lm$, $k-(lmQ)$. The negation of $kQ-lm$, $k-(lQ)m$
and $k-l(mQ)$ implies negation of $kQ-lm$, $k-(lmQ)$, and hence
genuine $N$-partite steering by Definition 3 implies genuine $N$-partite
steering according to Definition 1. The Definition 1 that requires
negation of $k-(lmQ)$ does not exclude that there is steering between
the system $l$ and $m$. This is because in this case the LHS model
has two trusted sites $l$ and $m$. By contrast, the Definition 3
requires negation of $k-l(mQ)$ and $k-(lQ)m$ which excludes steering
for the combined systems $lm$.

\subsubsection{Criteria}

Let us first consider genuine multipartite steering (Definition 2)
and use methods as above, together with Eq. (\ref{eq:oneway}), to
obtain the condition 
\begin{align}
(\Delta u)(\Delta v) & \geq\sum_{\rho_{\Lambda}}P_{\Lambda}\max\{C_{A}^{(\Lambda)},C_{B}^{(\Lambda)}\}\nonumber \\
 & \geq\min_{\Lambda}\max\{C_{A}^{(\Lambda)},C_{B}^{(\Lambda)}\}\label{eq:crit-gen-steer-1-generalN}
\end{align}
where $C_{A}^{(\Lambda)}$ and $C_{B}^{(\Lambda)}$ are the values
of $C_{A}$ and $C_{B}$ for the bipartition $\Lambda$. Analogously,
we obtain the condition for genuine multipartite two-way steering
(Definition 1) from Eq. (\ref{eq:twoway}) as 
\begin{align}
(\Delta u)(\Delta v)\geq\min_{\Lambda}\min\{C_{A}^{(\Lambda)},C_{B}^{(\Lambda)}\}\thinspace.\label{eq:crit-ge-steer-2-generalN}
\end{align}

The definition used in Eq. (\ref{eq:lhs-1-23-int}) with respect to
the untrusted sites $l$ and $m$ is along the lines introduced by
Svetlichny \citep{Svetlichny_PRD1987} and Collins \emph{et al}. \citep{Collins_PRL2002},
in relation to genuine tripartite nonlocality. There is no assumption
of locality made between the two systems, $l$ and $m$, and indeed
the combined bipartite system denoted $lm$ may be nonlocal. The Svetlichny
model corresponds to Eq. (\ref{eq:lhs-1-23-int}) but without the
assumption of a trusted quantum state for $k$, and is referred to
as a \emph{bilocal} model.

\subsection{Discussion of definitions}

Recent analyses indicate that for consistency with operational definitions
of genuine multipartite nonlocality, a weaker definition of genuine
tripartite nonlocality corresponding to a stricter subset of models
is preferable \citep{Gallego_Navascues_PRL2012,Bancal_multipartite_PRA2013,Dutta_PRA2020}.
This definition takes into account no-signaling and time ordering
between measurements made by the untrusted parties. While this is
an important issue that may lead to more sensitive criteria, we do
not address this in this paper. We derive criteria sufficient to negate
the Svetlichny-type models (Eq. (\ref{eq:lhs-1-23-int})), noting
that such criteria will also be sufficient to rule out the stricter
subset.

Other definitions may also become applicable, where we anticipate
that for future applications, there is a strategy for trust that means
not all LHS models will be relevant. A question relevant to secret
sharing applications is whether the steering of at least one of the
single systems requires \emph{all} of the remaining $N-1$ parties.
Following \citep{gessner_steering}, this value $N-1$ is called the
\emph{depth of steering parties}. It can be shown that genuine $N$-partite
steering as given by Definition 1 is a necessary condition for $N$
systems to have a depth of steering parties of $N-1$.

If we anticipate application of the steerable states to scenarios
with $N=3$ where there will always be at least \emph{two trusted
sites among three}, then the relevant LHS models that describe separability
for the system are $(kQ)-(lQ)m$, $(kQ)-l(mQ)$ and $k-(lmQ)$ ($k=1,2,3$).
Since negation of $(kQ)-lm$ implies negation of $(kQ)-(mQ)l$ and
$(kQ)-m(lQ)$, we see that the negation of Eq. (\ref{eq:problhsgenuine-1})
will imply negation of all these LHS models, based on two trusted
sites. Hence, the criteria derived in this paper according to Definition
1 will negate this type of genuine tripartite steering. This is
useful, for example, if there are two fixed trusted sites on a network.
Definitions based on sites with fixed trust have been given in \citep{Cavalcanti_steering_Nature2015}.

More generally, the Definition 1 for genuine $N$-partite steering
will negate all LHS models (and convex combinations), where $N-1$
of the subsystems are trusted. However, if we restrict to just \emph{one}
trusted site on a network, we would wish to negate the LHS models
$kQ-lm$, $k-(lQ)m$ and $k-l(mQ)$. This then requires Definition
3. The criteria derived in this paper according to Definition 1 do
\emph{not} detect this type of genuine tripartite steering.

It is also possible to consider negation of the LHS models (and convex
mixtures of them) relevant to secret sharing: $kQ-lm$. This is where
one does not trust the two collaborating (steering) parties $l$ and
$m$, which gives the alternative definition (Definition 2) of genuine
tripartite steering, referred to in \citep{He_Reid_steering_PRL2013}.
The criteria derived in this paper will negate all such models, and
therefore confirm genuine tripartite steering according to this earlier
definition, which we call Definition 2.

Definition 3 gives a strict definition of genuine $N$-partite steering.
In the tripartite case, this implies to negate the model $k-(lQ)m$,
since the bipartite system composed of systems $l$ and $m$ (where
only $l$ is trusted) can show steering. The Definition 1 of genuine
$N$-partite steering does \emph{not} negate this model, and hence
may not exclude that there is steering among $N-1$ subsystems. Criteria
for genuine tripartite steering with the Definition 3 will also be
presented in this paper, in Section III.D.

\subsection{Steering, security against an eavesdropper, and the EPR paradox}

One may demonstrate the Einstein-Podolsky-Rosen (EPR) paradox \citep{Einstein_EPR_PhysRev1935}
if we are able to select two observables $\hat{O}_{B}^{X}$ and $\hat{O}_{B}^{P}$
of $B$, such that \citep{Reid_PRA1989}
\begin{equation}
\mathcal{S}_{A|B}<1\thinspace.\label{eq:epr-genS-1}
\end{equation}
Here $\mathcal{S}_{A|B}=\Delta(x_{A}-O_{B}^{X})\Delta(p_{A}-O_{B}^{P})$
and $x_{A}$ and $p_{A}$ are observables for system $A$, such that
the uncertainty relation gives $\Delta x_{A}\Delta p_{A}\geq1$. The
observables $O_{B}^{X}$ and $O_{B}^{P}$ are general, although in
this paper, we will use linear combinations of quadrature phase amplitudes.
The condition is also sufficient to demonstrate steering of system
$A$, by the steering parties of a distinct system $B$ \citep{Wiseman_PRL2007,Cavalcanti_PRA2009}.

Consider $u$ and $v$ as defined in Eq. (\ref{eq:quad-gh}), where
we have three systems identified as $k$, $l$ and $m$, so that $i=1,2,3$
corresponds to $i=k,l,m$. Let $h_{k}=g_{k}=1$. Denoting the values
of $u$ and $v$ in this case by $u_{k}$ and $v_{k}$, so that
\begin{eqnarray}
u_{k} & = & x_{k}+h_{l}x_{l}+h_{m}x_{m}\nonumber \\
v_{k} & = & p_{k}+g_{l}p_{l}+g_{m}p_{m}\thinspace,\label{eq:uv-1-1-3-3}
\end{eqnarray}
then an inference variance product is defined
\begin{equation}
S_{k|lm}=\Delta u_{k}\Delta v_{k}\thinspace.\label{eq:eprS-3}
\end{equation}
Here we use the notation $S$ instead of $\mathcal{S}$, to make clear
we have specified the observables $O_{B}^{X}$ and $O_{B}^{P}$ to
be used by the steering parties. The violation of the inequality
\begin{equation}
S_{k|lm}\geq1\label{eq:epr-cond-1}
\end{equation}
then implies an Einstein-Podolsky-Rosen (EPR) paradox, where the ``elements
of reality'' referred to in the EPR argument relate to the system
$k$. This constitutes a steering of system $k$ by the systems $l$
and $m$, because the local hidden state model of Eq. (\ref{eq:lhs-1-23-int})
is falsified. This condition is readily generalized to $N$ systems.

In the set-up to measure $S_{k|lm}<1$, the observers at the combined
systems $l$ and $m$ are in some way (either alone or together) making
measurements at their sites, in order to predict or infer the results
of measurements made by an observer of system $k$. An EPR steering
paradox is obtained when the errors in the inference, given by $\Delta u_{k}$
and $\Delta v_{k}$, have a product that goes below the value of the
Heisenberg uncertainty bound for system $k$, because a local state
with these variances in $x_{k}$ and $p_{k}$ is not possible according
to quantum mechanics. 

Full tripartite steering inseparability is a necessary condition for
genuine tripartite steering and is the key to some important applications.
We consider a quantum secret sharing scenario, where two observers
$2$ and $3$ collaborate to infer the result of a measurement of
the spin of a third party, $1$ for the purpose of sharing a secret
key \citep{Hillery_application_PRA1999,Tittel_PRA2001,Lance:2003aa,Lance_PRL2004,wilkinson2021quantum}.
The observers later collaborate via public channels, to confirm for
entanglement between the groups, thus determining if there has been
intervention by an eavesdropper. Extra security is possible, if steering
of the third system by the two parties $2$ and $3$ is confirmed
\citep{He_Reid_steering_PRL2013}. This is because then there are
minimal assumptions made about the steering parties, while the station
of the third party $1$ is kept secure \citep{Branciard_PRA2012,Opanchuk_PRA2014}.

Thus, if one is able to demonstrate that for certain states there
is steering across all bipartitions, we have a system that can be
used for this purpose in a symmetrical fashion among the different
observers and sites. Tripartite steering inseparability does not
necessarily ensure however that the steering cannot be generated by
mixing states where there is steering between just two parties (refer
to \citep{Shalm_Nature2012,PhysRevA.90.062337} for similar discussions
in relation to entanglement). The desirable feature to ensure that
two observers are required to observe the steering, as opposed to
one, can be confirmed for a particular bipartition, in principle,
by negating steering of the third party $1$ by $2$ (or $3$) \citep{Mondal_PRA2019}.

\subsection{Sum inequalities}

Previous papers on multipartite entanglement have considered inequalities
involving the sum of the variances, ${\color{black}(\Delta u)^{2}+(\Delta v)^{2}}$,
where $u$ and $v$ are defined by Eq. (\ref{eq:quad-gh}) \citep{vanLoock_PRA2003}.
Such sum criteria can be derived from the above Criteria 1 and 2,
using that $x^{2}+y^{2}\geq2xy$ for all real $x$ and $y$. Specifically,
if the Criterion involving the products is of the form $\Delta u\Delta v\geq I$
where $I$ is an expression involving $g_{i}$ and $h_{i}$, then
we obtain the criterion
\begin{equation}
\frac{{\color{black}(\Delta u)^{2}+(\Delta v)^{2}}}{2}\geq{\color{black}\Delta u\Delta v\geq I}\label{eq:prod-sum}
\end{equation}
for the sum of the variances. We see however from this relation that
the violation of sum criterion Eq. (\ref{eq:prod-sum}) will always
imply violation of the product inequality $\Delta u\Delta v\geq I$.

\section{Criteria for continuous variable tripartite steering \label{sec:tripartite-1}}

\subsection{Single inference inequalities}

For concreteness, we consider three systems and use the results
of the previous section to obtain product inequalities for the certification
of genuine tripartite steering. One may define the linear combination
\citep{vanLoock_PRA2003}
\begin{eqnarray}
u & = & h_{1}x_{1}+h_{2}x_{2}+h_{3}x_{3}\nonumber \\
v & = & g_{1}p_{1}+g_{2}p_{2}+g_{3}p_{3}\label{eq:uv-1-1}
\end{eqnarray}
where $h_{j}$ and $g_{j}$ ($j=1,2,3$) are real numbers. We note
that the phase of the quadratures $x_{j}$ and $p_{j}$ can be adjusted
independently at each site and we may also define, for example:
\begin{eqnarray}
u' & = & h_{1}x_{1}-h_{2}p_{2}+h_{3}x_{3}\nonumber \\
v' & = & g_{1}p_{1}+g_{2}x_{2}+g_{3}p_{3}\thinspace.\label{eq:uv-1-1-2-2}
\end{eqnarray}
 Assuming the Heisenberg uncertainty relation $\Delta x_{j}\Delta p_{j}\geq1$
for each $j=1,2,3$, the separability assumption of Eq. (\ref{eq:insep-1-23})
written as ($k,l,m\in\{1,2,3\}$ and $k\neq l\neq m$)
\begin{equation}
\rho_{lm,k}=\sum_{R}P_{R}\rho_{lm,Q}^{R}\rho_{k,Q}^{R}\label{eq:sep-1}
\end{equation}
 implies \citep{PhysRevA.90.062337}
\begin{equation}
\Delta u\Delta v\geq|h_{k}g_{k}|+|h_{l}g_{l}+h_{m}g_{m}|\thinspace.\label{eq:uvineq-1-3}
\end{equation}
Similarly, based on the uncertainty relation $\Delta\left(-h_{2}p_{2}+h_{m}x_{m}\right)\Delta\left(g_{2}x_{2}+g_{m}p_{m}\right)\geq\left|g_{2}h_{2}+g_{m}h_{m}\right|$,
 the separability assumption implies \citep{PhysRevA.90.062337}
\begin{equation}
\Delta u'\Delta v'\geq|h_{k}g_{k}|+|h_{l}g_{l}+h_{m}g_{m}|\thinspace.\label{eq:uvineq-1-1-2}
\end{equation}
This brings us to inequalities for steering.

\textbf{\emph{Lemma 1:}} Violation of
\begin{equation}
\Delta u\Delta v\geq\left|g_{k}h_{k}\right|\label{eq:prod}
\end{equation}
implies EPR steering of system $k$ by the systems $l$ and $m$.
The violation of
\begin{equation}
\Delta u\Delta v\geq\left|g_{l}h_{l}+g_{m}h_{m}\right|\label{eq:prod2}
\end{equation}
implies EPR steering of systems $l$ and $m$, by system $k$. The
violation of the inequality
\begin{align}
\Delta u\Delta v & \geq\text{min}\left\{ \left|g_{k}h_{k}\right|,\left|g_{l}h_{l}+g_{m}h_{m}\right|\right\} \,\label{eq:two-way_arbitrary_bipartition-1-1-1}
\end{align}
thus implies two-way EPR steering across the bipartition $A|B$ where
$A=\{k\}$ and $B=\{l,m\}$. Here, $l\neq k\neq m$ and $l,k,m\in\{1,2,3\}$.
In the steering of system $k$, the observers of the combined systems
$l$ and $m$ infer values for the $x_{k}$ and $p_{k}$ of system
$k$, in order to violate the inequality Eq. (\ref{eq:prod}). In
the steering of combined systems $l$ and $m$, the observer at \emph{k
}infers values for the $h_{l}x_{l}+h_{m}x_{m}$ and $g_{l}p_{l}+g_{m}p_{m}$
of the combined systems $l$ and $m$, in order to violate the inequality
Eq. (\ref{eq:prod2}). The two-way steering can be established by
selecting different values of $g_{i}$ and $h_{i}$ for \emph{each}
direction of steering. However, here, we are interested to rule out
all bipartitions with a \emph{single} inequality. In this way, one
can immediately also rule out the mixtures associated with different
bipartitions and LHS models, and hence deduce genuine tripartite steering.
We note the same result will apply to $u'$ and $v'$ defined by Eq.
(\ref{eq:uv-1-1-2-2}).

\emph{Proof:} The proof of the Lemma follows from Eq. (\ref{eq:fulltwoway})
taking $C_{A}=|g_{k}h_{k}|$ and $C_{B}=|g_{l}h_{l}+g_{m}h_{m}|$.
$\square$

Now we arrive at conditions sufficient for genuine tripartite steering,
using Definition 1 of Section II.D.

\textbf{\textit{Criterion 1.}}\textbf{ }Violation of the inequality
\begin{align}
S_{3}\equiv{\color{black}\Delta u\Delta v} & {\color{black}\geq\text{min}\Bigl\{\left|g_{1}h_{1}\right|,{\color{red}{\color{black}\left|g_{2}h_{2}+g_{3}h_{3}\right|}},}\nonumber \\
{\color{black}} & \left|g_{2}h_{2}\right|,\left|g_{1}h_{1}+g_{3}h_{3}\right|,\nonumber \\
{\color{black}} & \left|g_{3}h_{3}\right|,{\color{black}{\color{black}{\color{red}{\color{black}\left|g_{1}h_{1}+g_{2}h_{2}\right|}}}}\Bigl\}\label{eq:criterion-1-4-2}
\end{align}
is sufficient to confirm full two-way tripartite steering inseparability
and also to confirm genuine tripartite steering. An identical result
holds for ${\color{black}\Delta u'\Delta v'}$. The proof follows
from Lemma 1 and Eq. (\ref{eq:crit-ge-steer-2-generalN}). $\square$

Letting $g_{1}=h_{1}=1$ and $g_{2}=-h_{2}=1/\sqrt{2}$, $g_{3}=-h_{3}=1/\sqrt{2}$,
Criterion 1 implies that observation of
\begin{align}
S_{3}=\Delta\left[x_{1}-\frac{\left(x_{2}+x_{3}\right)}{\sqrt{2}}\right]\Delta\left[p_{1}+\frac{\left(p_{2}+p_{3}\right)}{\sqrt{2}}\right] & <0.5\,\label{eq:old-crit}
\end{align}
is sufficient to confirm genuine tripartite steering. This inequality
has been used previously to certify genuine tripartite steering, in
the work of \citep{Armstrong_Nature2015,He_Reid_steering_PRL2013,PhysRevA.90.062337}.
The definition of genuine tripartite steering in those works however
was based on Definition 2 (refer Section II.D). The results of this
paper show that the inequality Eq. (\ref{eq:old-crit}) (and the associated
criterion for the sum of the variances, see Eq. (\ref{eq:prod-sum}))
also implies the stronger genuine tripartite steering, defined according
to Definition 1 (Section II.D). Generally, where ${\color{black}\left|g_{l}h_{l}+g_{m}h_{m}\right|}\geq\left|g_{k}h_{k}\right|$,
Criterion 1 reduces to $\Delta u\Delta v\geq\text{min}\left\{ \left|g_{1}h_{1}\right|,\left|g_{2}h_{2}\right|,\left|g_{3}h_{3}\right|\right\} $,
which equates to that derived using the Definition 2.

\textbf{\textit{Criterion 2.}}\textbf{ }Violation of the inequality
\begin{align}
S_{3}\equiv{\color{black}\Delta u\Delta v} & {\color{black}\geq\text{min}\Bigl\{\max\{\left|g_{1}h_{1}\right|,{\color{red}{\color{black}\left|g_{2}h_{2}+g_{3}h_{3}\right|\}}},}\nonumber \\
{\color{black}} & \max\{\left|g_{2}h_{2}\right|,\left|g_{1}h_{1}+g_{3}h_{3}\right|\},\nonumber \\
{\color{black}} & \max\{\left|g_{3}h_{3}\right|,{\color{black}{\color{black}{\color{red}{\color{black}\left|g_{1}h_{1}+g_{2}h_{2}\right|\}}}}}\Bigl\}\label{eq:criterion-1-4-1-1}
\end{align}
is sufficient to confirm full tripartite steering inseparability.
The proof is given as for Eq. (\ref{eq:crit-gen-steer-1-generalN}).
$\square$

\subsection{Average-variance inequalities}

The above approach uses a \emph{single} inequality to confirm $N$-partite
steering, and hence there is no optimization of the gains $g_{i}$,
$h_{i}$ for each bipartition. Another approach introduced in \citep{He_Reid_steering_PRL2013}
is to take averages over the variances associated with each bipartition,
thus allowing one to individually optimize gains for each bipartition.
The approach also allows one to see that if it is possible to show
sufficient steering of \emph{each} system $k=1,2,3$, so that each
of the EPR variances $S_{k|lm}$ becomes sufficiently small, then
genuine tripartite steering must follow.

This brings us to the following criterion to certify genuine tripartite
steering.

\textbf{\emph{Criterion 3:}} We define 
\begin{eqnarray}
S_{k|lm} & \equiv & \Delta(x_{k}+h_{k,l}x_{l}+h_{k,m}x_{m})\Delta(p_{k}+g_{k,l}p_{l}+g_{k,m}p_{m})\nonumber \\
\label{eq:defn}
\end{eqnarray}
where the $h_{k,l}$, $h_{k,m}$, $g_{k,l}$ and $g_{k,m}$ are real
constants. Here, $l\neq k\neq m$ and $l,k,m\in\{1,2,3\}$. Violation
of the following inequality will certify genuine tripartite steering
among the three systems:
\begin{eqnarray}
S_{1|23}+S_{2|13}+S_{3|12} & \geq & \min(1,\left|g_{2,1}h_{2,1}+g_{2,3}h_{2,3}\right|,\nonumber \\
 &  & \left|g_{3,2}h_{3,2}+g_{3,1}h_{3,1}\right|,\nonumber \\
 &  & \left|g_{1,2}h_{1,2}+g_{1,3}h_{1,3}\right|)\thinspace.\label{eq:stineqgen}
\end{eqnarray}
Provided $\left|g_{k,l}h_{k,l}+g_{k,m}h_{k,m}\right|\geq1$, for each
$k$ where $k\neq l\neq m$, this simplifies to the inequality
\begin{equation}
S_{1|23}+S_{2|13}+S_{3|12}\geq1\thinspace.\label{eq:stinequ1}
\end{equation}
\emph{Proof:} We consider that the system is described by mixtures
of the type 
\begin{eqnarray}
\rho_{mix} & = & P'_{1}\rho_{1Q-23}+P''_{1}\rho_{1-(23Q)}\nonumber \\
 &  & +P'_{2}\rho_{2Q-13}+P''_{2}\rho_{2-(13Q)}\nonumber \\
 &  & +P'_{3}\rho_{3Q-12}+P''_{3}\rho_{3-(12Q)}\thinspace,\label{eq:mix-gen-lhs-3}
\end{eqnarray}
where we use the abbreviated notation for LHS mixtures, given in Section
II. Here, $P'_{k}$, $P''_{k}$ and $P_{k}$ are probabilities, such
that $P_{k}=P'_{k}+P_{k}''$ and $P_{1}+P_{2}+P_{3}=1$. Using Definition
1 of Section II.D for genuine tripartite steering, we wish to negate
all such models $\rho_{mix}$. First, we note that if $S_{k|lm}<1$,
we certify steering of system $k$ (Lemma 1). If $S_{k|lm}<\left|g_{k,l}h_{k,l}+g_{k,m}h_{k,m}\right|$,
then we certify steering of the combined systems $lm$. This is because
the LHS models (denoted $\rho_{k-(lmQ)}$ and $\rho_{kQ-lm}$) associated
with that bipartition $k-lm$ imply $S_{k|lm}\geq1$ and $S_{lm|k}\geq\left|g_{l}h_{l}+g_{m}h_{m}\right|$,
respectively. More generally, each LHS model associated with the bipartition
$k-lm$ implies
\begin{equation}
S_{k|lm}\geq\min\left\{ 1,\left|g_{k,l}h_{k,l}+g_{k,m}h_{k,m}\right|\right\} \thinspace.\label{eq:sstep}
\end{equation}
This is also true if for any particular $k$ we define
\begin{eqnarray}
S_{k|lm} & \equiv & \Delta(x_{k}-h_{k,l}p_{l}+h_{k,m}x_{m})\Delta(p_{k}+g_{k,l}x_{l}+g_{k,m}p_{m})\nonumber \\
\label{eq:defn-4}
\end{eqnarray}
or as
\begin{eqnarray}
S_{k|lm} & \equiv & \Delta(x_{k}-g_{k,l}p_{l}+g_{k,m}p_{m})\Delta(-p_{k}+h_{k,l}x_{l}+h_{k,m}x_{m})\nonumber \\
\label{eq:defn-4-1-1-1}
\end{eqnarray}
as we see from the results of Eqs. (\ref{eq:uv-1-1-2-2}) and (\ref{eq:uvineq-1-1-2}).
Abbreviating the notation to define $S_{k}\equiv S_{k|lm}$, we assume
the system is described by (\ref{eq:mix-gen-lhs-3}) and we use Eq.
(\ref{eq:23-1-3}). We see that the value of $S_{1}$ would be taken
as the average over the three different bipartitions $k-lm$, and
similarly for $S_{2}$ and $S_{3}$, which leads to the inequality
\begin{eqnarray}
S_{1}+S_{2}+S_{3} & \geq & P_{1}S_{1,1}+P_{2}S_{1,2}+P_{3}S_{1,3}\nonumber \\
 &  & +P_{1}S_{2,1}+P_{2}S_{2,2}+P_{3}S_{2,3}\nonumber \\
 &  & +P_{1}S_{3,1}+P_{2}S_{3,2}+P_{3}S_{3,3}\thinspace.\label{eq:step3}
\end{eqnarray}
Here, we denote $S_{k',k}$ as the value of $S_{k'}$ for the bipartition
$k-lm$. The system will be in a given bipartition with a fixed probability,
$P_{i}$. Recognizing each $S_{k',k}$ to be positive, we can therefore
write 
\begin{eqnarray}
S_{1}+S_{2}+S_{3} & \geq & P_{1}S_{1,1}+P_{2}S_{2,2}+P_{3}S_{3,3}\thinspace.\label{eq:step}
\end{eqnarray}
If the system is in the bipartition labelled $1$, i.e. the bipartition
$1-23$, then $S_{1,1}\geq\min(1,\left|g_{1,2}h_{1,2}+g_{1,3}h_{1,3}\right|)$.
Applying the conditions on the $S_{k,k}$ from Lemma 1, this becomes
\begin{eqnarray}
S_{1}+S_{2}+S_{3} & \geq & P_{1}\min(1,\left|g_{1,2}h_{1,2}+g_{1,3}h_{1,3}\right|)\nonumber \\
 &  & +P_{2}\min(1,\left|g_{2,1}h_{2,1}+g_{2,3}h_{2,3}\right|)\nonumber \\
 &  & +P_{3}\min(1,\left|g_{3,2}h_{3,2}+g_{3,1}h_{3,1}\right|)\thinspace,\nonumber \\
\label{eq:stepbetween}
\end{eqnarray}
which leads to
\begin{eqnarray}
S_{1}+S_{2}+S_{3} & \geq & \min(1,\left|g_{2,1}h_{2,1}+g_{2,3}h_{2,3}\right|,\nonumber \\
 &  & \left|g_{3,2}h_{3,2}+g_{3,1}h_{3,1}\right|,\nonumber \\
 &  & \left|g_{1,2}h_{1,2}+g_{1,3}h_{1,3}\right|)\thinspace.\label{eq:step4-1-1-2}
\end{eqnarray}
We note that provided $\left|g_{k,l}h_{k,l}+g_{k,m}h_{k,m}\right|\geq1$,
the LHS bipartition $k-lm$ implies $S_{k}\geq1$. Here we use that
for the bipartition $k-lm$, for both LHS models $S_{k,k}\geq1$,
provided $\left|g_{k,l}h_{k,l}+g_{k,m}h_{k,m}\right|\geq1$. $\square$

The above Criterion 3 justifies the use of 
\begin{equation}
S_{1}+S_{2}+S_{3}\geq1\label{eq:sum}
\end{equation}
where we take $\left|g_{k,l}h_{k,l}+g_{k,m}h_{k,m}\right|\geq1$
for each $k$. The condition $S_{1}+S_{2}+S_{3}<1$ was stated as
sufficient to detect genuine tripartite steering in \citep{He_Reid_steering_PRL2013},
based on the Definition 2 of genuine tripartite steering. We now see
that this condition also holds as a criterion, according to the stricter
Definition 1 used in this paper.

We note that the phases of $u$ and $v$ can be adjusted according
to Eq. (\ref{eq:uv-1-1-2-2}) and Eqs. (\ref{eq:defn-4}) or (\ref{eq:defn-4-1-1-1}).
To be explicit, in Section V.A.4 we will define $S_{1|23}$ and $S_{3|12}$
according to Eq. (\ref{eq:defn-4}) and $S_{2|13}$ according to Eq.
(\ref{eq:defn-4-1-1-1}). Criterion 3 also applies for this choice
of phase. It is also possible to extend Criterion 3, by deriving
an inequality that keeps more terms at the step Eq. (\ref{eq:step3})
in the derivation. However, this gave no advantage for the states
examined in this paper (refer Supplemental Materials).

\subsection{Van Loock-Furusawa-type steering criteria}

The work of van Loock and Furusawa motivated experimental measurements
of sums of variances, which enabled detection of genuine multipartite
entanglement. The application to steering of these inequalities was
considered by Teh and Reid \citep{PhysRevA.90.062337}, using Definition
2 for genuine multipartite steering. We are thus motivated to examine
these criteria using the stricter Definition 1.

In their paper \citep{vanLoock_PRA2003}, van Loock and Furusawa consider
quantities
\begin{eqnarray}
B_{I}\equiv[\Delta(x_{1}-x_{2})]^{2}+[\Delta(p_{1}+p_{2}+g_{3}p_{3})]^{2}\nonumber \\
B_{II}\equiv[\Delta(x_{2}-x_{3})]^{2}+[\Delta(g_{1}p_{1}+p_{2}+p_{3})]^{2}\nonumber \\
B_{III}\equiv[\Delta(x_{1}-x_{3})]^{2}+[\Delta(p_{1}+g_{2}p_{2}+p_{3})]^{2}\nonumber \\
\label{eq:threeineq}
\end{eqnarray}
defined for arbitrary real parameters $g_{1}$, $g_{2}$ and $g_{3}$.
They consider the biseparable bipartitions denoted $\rho_{kQ-(lmQ)}$
in our notation, for which the following uncertainty relation holds
\begin{equation}
\Delta u\Delta v\geq|h_{k}g_{k}|+|h_{l}g_{l}+h_{m}g_{m}|\thinspace.\label{eq:uvineq-1-2}
\end{equation}
They then use Eq. (\ref{eq:uvineq-1-2}), to show that inequality
$B_{I}\geq4$ is implied by both the biseparable states $ $$\rho_{(13)Q,2Q}$
and $\rho_{(23Q),1Q}$, which assume separability between systems
$1$ and $2$. Similarly, a second inequality $B_{II}\geq4$ is implied
by the biseparable states $ $$\rho_{(13Q),2Q}$ and $\rho_{(12Q),3Q}$,
while a third inequality $B_{III}\geq4$ follows from biseparable
states $ $$\rho_{(12Q),3Q}$ and $\rho_{(23Q),1Q}$. Van Loock and
Furusawa derived a condition for full tripartite inseparability, based
on the violation of two of the above inequalities. 

Here, we will prove a similar result for steering. We follow \citep{PhysRevA.90.062337}
and define the product inequalities that were used for multipartite
entanglement
\begin{eqnarray}
S_{I}\equiv\Delta(x_{1}-x_{2})\Delta(p_{1}+p_{2}+g_{3}p_{3})\nonumber \\
S_{II}\equiv\Delta(x_{2}-x_{3})\Delta(g_{1}p_{1}+p_{2}+p_{3})\nonumber \\
S_{III}\equiv\Delta(x_{1}-x_{3})\Delta(p_{1}+g_{2}p_{2}+p_{3})\thinspace.\label{eq:threeineq-1-1}
\end{eqnarray}
First, following the proofs of \citep{vanLoock_PRA2003} and \citep{PhysRevA.90.062337},
we note that the inequality $S_{I}\geq2$ is implied by the biseparable
states associated with bipartitions $13-2$ and $23-1$. Similarly,
the second inequality $S_{II}\geq2$ is implied by the biseparable
states associated with bipartitions $13-2$ and $12-3$; and the third
inequality $S_{III}\geq2$ by separable states over bipartitions $12-3$
and $23-1$. This means that the violation of any two of the inequalities
$S_{I}\geq2$, $S_{II}\geq2$ and $S_{III}\geq2$ is sufficient to
prove full tripartite inseparability. 

We now extend the result for tripartite steering.

\textbf{\emph{Criterion 4\label{Criterion-4}:}} The violation of
any two of the inequalities
\begin{equation}
S_{I}\geq1\thinspace,S_{II}\geq1\thinspace,S_{III}\geq1\thinspace,\label{eq:bineq}
\end{equation}
implies full tripartite two-way steering inseparability.

\emph{Proof:} Here, we use the results with $h_{3}=0$ , $h_{1}=g_{1}=g_{2}=1$
and $h_{2}=-1$ to show that the LHS model $\rho_{1Q}\rho_{23}$ and
$\rho_{1}\rho_{23Q}$ implies $S_{I}\geq1$, and similarly the LHS
models $\rho_{2Q}\rho_{13}$ and $\rho_{2}\rho_{13,Q}$ imply
\begin{equation}
S_{I}\geq1\thinspace.\label{step5}
\end{equation}
This follows from Lemma 1: Considering bipartition $\{k-lm\}$ where
$k=1$, $l=2$ and $m=3$, we find that violation of the inequality
$\Delta u\Delta v\geq1\,$ implies two-way steering across the bipartition
$\{1-23\}$, and similarly across bipartition $\{2-13\}$. Similarly,
we see that LHS models $\rho_{3Q}\rho_{12}$ and $\rho_{3}\rho_{12,Q}$
(and the LHS models $\rho_{2Q}\rho_{13}$ and $\rho_{2}\rho_{13,Q}$)
imply
\begin{equation}
S_{II}\geq1\thinspace.\label{eq:step6}
\end{equation}
The LHS models $\rho_{3Q}\rho_{12}$ and $\rho_{3}\rho_{12,Q}$ (and
the LHS models $\rho_{1Q}\rho_{23}$ and $\rho_{1}\rho_{23,Q}$) imply
\begin{equation}
S_{III}\geq1\thinspace.\label{eq:step7}
\end{equation}
Hence, if two inequalities are violated, all three bipartitions show
two-way steering, and the result follows. $\square$

\textbf{\emph{Criterion 5:}}\textbf{ }We confirm genuine tripartite
steering, if the inequality 
\begin{equation}
S_{I}+S_{II}+S_{III}\geq2\thinspace\label{eq:threesum-3}
\end{equation}
is violated.

\emph{Proof:}\textbf{ }For $N=3$ parties, there are three bipartitions,
and three corresponding biseparable states $\rho_{1,23}$, $\rho_{2,13}$,
and $\rho_{3,12}$ that we index by $k=1,2,3$ respectively. Consider
any mixture of the form Eq. (\ref{eq:mix-gen-lhs-3}). We consider
the three types of bipartitions, grouping the two associated LHS models
together. We use that for mixtures $\rho_{mix}=\sum P_{R}\rho^{(R)}$,
the result Eq. (\ref{eq:23-1-3}) follows, where here the subscript
denotes that the averages are over the state $\rho^{(R)}$. We can
then write
\begin{eqnarray}
S_{I} & \geq & P_{1}S_{I,1}+P_{2}S_{I,2}+P_{3}S_{I,3}\nonumber \\
 & \geq & P_{1}S_{I,1}+P_{2}S_{I,2}\geq P_{1}+P_{2}\label{eq:pr45}
\end{eqnarray}
where here we define $S_{I,k}$ as the expected value of $S_{I}$
for the bipartition $\rho_{k,lm}$ with probability $P_{k}$. This
uses that we know from the proof of Criterion 4 above that both of
the LHS models $\rho_{1Q}\rho_{23}$ and $\rho_{1}\rho_{23Q}$ with
bipartition $k=1$ will satisfy $S_{I}\geq1$, and also $S_{III}\geq1$.
Similarly, both LHS models $\rho_{2Q}\rho_{13}$ and $\rho_{2}\rho_{13,Q}$
satisfy $S_{I}\geq1$ and $S_{II}\geq1$, and both LHS models $\rho_{3Q}\rho_{12}$
and $\rho_{3}\rho_{12,Q}$ satisfy $S_{II}\geq1$ and $S_{III}\geq1$.
Hence, for any mixture $S_{I}\geq P_{1}+P_{2}$. Similarly, $S_{II}\geq P_{2}+P_{3}$
and $S_{III}\geq P_{1}+P_{3}$. Then we see that since $\sum_{k=1}^{3}P_{k}=1$,
for any mixture it must be true that $S_{I}+S_{II}+S_{III}\geq2$.
$\square$

Immediately, from the result Eq. (\ref{eq:prod-sum}), one arrives
at the corresponding Criteria for the original van Loock-Furusawa
inequalities, involving summations. We choose to write these explicitly,
in the event this may be useful, because the quantities have been
experimentally reported.

\textbf{\emph{Criterion 4b\label{Criterion-4b}:}} The violation of
any two of the inequalities
\begin{equation}
B_{I}\geq2,B_{II}\geq2,B_{III}\geq2\thinspace,\label{eq:bineq-1-1}
\end{equation}
implies full tripartite two-way steering. The proof follows from Criterion
$4$ and the result Eq. (\ref{eq:prod-sum}).

\textbf{\emph{Criterion 5b\label{Criterion-5b}:}}\textbf{ }We confirm
genuine tripartite steering, if the inequality 
\begin{equation}
B_{I}+B_{II}+B_{III}\geq4\label{eq:threesum-3-1-1}
\end{equation}
is violated, where $B_{I}\geq2$, $B_{II}\geq2$ and $B_{III}\geq2$
are the van Loock-Furusawa steering inequalities, Eq. (\ref{eq:bineq-1-1}).
The proof follows from Criterion 5 and the result Eq. (\ref{eq:prod-sum}).

This result confirms the validity of the criteria given as (35) and
Result (4) stated in \citep{PhysRevA.90.062337} and \citep{He_Reid_steering_PRL2013}
respectively (based on Definition 2), for the Definition 1, used in
this paper.

We also note criteria involving just two of the van Loock-Furusawa
inequalities. These inequalities are an adaption of the similar criterion
derived for entanglement in \citep{Shalm_Nature2012,PhysRevA.90.062337}.

\textbf{\textit{Criterion 5c: }}We can confirm genuine tripartite
steering, if with $g_{1}=g_{2}=g_{3}=1$ the inequality 
\begin{equation}
S_{I}+S_{II}<1\label{eq:critasym-1-1-1}
\end{equation}
 is satisfied (or $S_{I}+S_{III}<1$, or $S_{II}+S_{III}<1$).

\emph{Proof:} This follows from the proof of Criterion 5.

\textbf{\textit{Criterion 6c: }}We can confirm genuine tripartite
steering, if with $g_{1}=g_{2}=g_{3}=1$ the inequality
\begin{equation}
B_{I}+B_{II}<2\label{eq:critasym-1-1-1-1}
\end{equation}
 is satisfied (or $B_{I}+B_{III}<2$, or $B_{II}+B_{III}<2$).

\emph{Proof:} This follows from Criterion 5c.

\subsection{Strict genuine tripartite steering: Definition 3}

We now give criteria for the stricter definition of genuine tripartite
steering, discussed in Section II.E. This definition allows the inference
of genuine tripartite steering with only one trusted site, and negates
all LHS models that allow steering among two parties.

\textbf{\emph{Criterion 7:}}\textbf{ }We confirm genuine tripartite
steering by Definition 3, if either of the following inequalities
is violated with $g_{i}=1$ ($i=1,2,3)$:
\begin{equation}
S_{I}+S_{II}+S_{III}\geq2\label{eq:threesum-3-2}
\end{equation}
or $B_{I}+B_{II}+B_{III}\geq4$. Similarly, we conclude genuine tripartite
steering by Definition 3 if any one of $S_{I}+S_{II}\geq1$, $S_{II}+S_{III}\geq1$,
$S_{I}+S_{III}\geq1$ (or $B_{I}+B_{II}\geq2$ or $B_{II}+B_{III}\geq2$,
$B_{I}+B_{III}\geq2$) is violated, with $g_{i}=1$ ($i=1,2,3)$.

\emph{Proof:}\textbf{ }For $N=3$ parties, there are three bipartitions,
and three corresponding biseparable states $\rho_{1,23}$, $\rho_{2,13}$,
and $\rho_{3,12}$ that we index by $k=1,2,3$ respectively. For Definition
3, the relevant LHS models for $\rho_{1,23}$ ($k=1$) are $\rho_{1Q}\rho_{23}$,
$\rho_{1}\rho_{2(3Q)}$ and $\rho_{1}\rho_{(2Q)3}$; and similarly
for each $k=2$ and $3$. As for the proof of Criterion 5, we consider
any convex mixture of these LHS states $\rho_{mix}=\sum_{k}P_{k}\rho^{(k)}$
where $P_{k}\rho^{(k)}=P_{k,k}\rho_{kQ}\rho_{lm}+P_{k,l}\rho_{k}\rho_{(lQ)m}+P_{k,m}\rho_{k}\rho_{l(mQ)}$
and $P_{k,k}+P_{k,l}+P_{k,m}=P_{k}$, $\sum_{k}P_{k}=1$, with each
$P_{k,i}\geq0$. From Eq. (\ref{eq:uvineq-1-3}), the uncertainty
relation for $\rho_{kQ}\rho_{(lm)Q}$ is 
\begin{equation}
\Delta u\Delta v\geq|h_{k}g_{k}|+|h_{l}g_{l}+h_{m}g_{m}|\thinspace.\label{eq:uvineq-1-3-1-1}
\end{equation}
For $\rho_{1Q}\rho_{23}$, using $h_{1}=1$, $h_{2}=-1$, $h_{3}=0$
and $g_{1}=g_{2}=g_{3}=1$, the relation becomes
\begin{equation}
\Delta(x_{1}-x_{2})\Delta(p_{1}+p_{2}+p_{3})\geq|h_{1}g_{1}|=1\thinspace,
\end{equation}
implying $S_{I}\geq1$. For $\rho_{1}\rho_{(2Q)3}$, using $h_{1}=1$,
$h_{2}=-1$, $h_{3}=0$ and $g_{1}=g_{2}=g_{3}=1$, we see that 
\begin{equation}
\Delta(x_{1}-x_{2})\Delta(p_{1}+p_{2}+p_{3})\geq|h_{2}g_{2}|=1
\end{equation}
i.e. $S_{I}\geq1$. However, for $\rho_{1}\rho_{2(3Q)}$, using $h_{1}=1$,
$h_{2}=1$, $h_{3}=0$ and $g_{1}=g_{2}=g_{3}=1$, 
\begin{equation}
\Delta(x_{1}-x_{2})\Delta(p_{1}+p_{2}+p_{3})\geq|h_{3}g_{3}|=0
\end{equation}
i.e. $S_{I}\geq0$. Proceeding similarly, for $\rho_{1Q}\rho_{23}$,
using $h_{1}=1$, $h_{3}=-1$, $h_{2}=0$ and $g_{1}=g_{2}=g_{3}=1$,
\begin{equation}
\Delta(x_{1}-x_{3})\Delta(p_{1}+p_{2}+p_{3})\geq|h_{1}g_{1}|
\end{equation}
i.e. $S_{III}\geq1$. Continuing, we find for $\rho_{1}\rho_{(2Q)3}$
that $S_{III}\geq0$; and for $\rho_{1}\rho_{2(3Q)}$ that $S_{III}\geq1$.
Proceeding similarly, for $\rho_{1Q}\rho_{23}$, using $h_{2}=1$,
$h_{3}=-1$, $h_{1}=0$ and $g_{1}=g_{2}=g_{3}=1$, we see from
\begin{equation}
\Delta(x_{2}-x_{3})\Delta(p_{1}+p_{2}+p_{3})\geq|h_{1}g_{1}|
\end{equation}
that $S_{II}\geq0$. For $\rho_{1}\rho_{(2Q)3}$, we find $S_{II}\geq1$;
and for $\rho_{1}\rho_{2(3Q)}$, we see that $S_{III}\geq1$. In
summary: for $\rho_{1Q}\rho_{23}$ and $\rho_{1}\rho_{(2Q)3}$, $S_{I}\geq1$;
for $\rho_{1Q}\rho_{23}$ and $\rho_{1}\rho_{2(3Q)}$, $S_{III}\geq1$;
for $\rho_{1}\rho_{(2Q)3}$ and $\rho_{1}\rho_{2(3Q)}$, $S_{II}\geq1$.

Similarly, for $\rho_{2Q}\rho_{13}$ and $\rho_{2}\rho_{(1Q)3}$,
$S_{I}\geq1$; for $\rho_{2Q}\rho_{13}$ and $\rho_{2}\rho_{1(3Q)}$,
$S_{II}\geq1$; for $\rho_{2}\rho_{(1Q)3}$ and $\rho_{2}\rho_{1(3Q)}$,
$S_{III}\geq1$. For $\rho_{3Q}\rho_{21}$ and $\rho_{3}\rho_{(2Q)1}$,
$S_{II}\geq1$; for $\rho_{3Q}\rho_{21}$ and $\rho_{3}\rho_{2(1Q)}$,
$S_{III}\geq1$; for $\rho_{3}\rho_{(2Q)1}$ and $\rho_{3}\rho_{2(1Q)}$,
$S_{I}\geq1$. Hence, for $\rho_{mix}$, using that for mixtures the
variance is given according to Eq. (\ref{eq:var_u-2-1-1}), we find
\begin{eqnarray}
S_{I} & \geq & P_{1,1}+P_{1,2}+P_{2,1}+P_{2,2}+P_{3,2}+P_{3,1}\nonumber \\
S_{II} & \geq & P_{1,2}+P_{1,3}+P_{2,3}+P_{2,2}+P_{3,3}+P_{3,2}\nonumber \\
S_{III} & \geq & P_{1,1}+P_{1,3}+P_{2,3}+P_{2,1}+P_{3,3}+P_{3,1}\thinspace.
\end{eqnarray}
Hence,
\begin{eqnarray}
S_{I}+S_{II}+S_{III} & \geq & 2P_{1,1}+2P_{1,2}+2P_{1,3}\nonumber \\
 &  & +2P_{2,2}+2P_{2,1}+2P_{2,3}\nonumber \\
 &  & +2P_{3,3}+2P_{3,2}+2P_{3,1}\geq2\thinspace.\nonumber \\
\label{eq:vanloockproof2}
\end{eqnarray}
The proof of the second inequality follows from Eq. (\ref{eq:prod-sum}).
We also see that 
\begin{eqnarray}
S_{I}+S_{II} & \geq & P_{1,1}+2P_{1,2}+P_{1,3}\nonumber \\
 &  & +2P_{2,2}+P_{2,1}+P_{2,3}\nonumber \\
 &  & +P_{3,3}+2P_{3,2}+P_{3,1}\geq1\thinspace,\label{eq:vanloockproof2-1}
\end{eqnarray}
and similarly, $S_{I}+S_{III}\geq1$ and $S_{II}+S_{III}\geq1$. The
results for the sum inequalities $B_{I}$, $B_{II}$ and $B_{III}$
follow from Eq. (\ref{eq:prod-sum}). $\square$

\section{Genuine $N$-partite steerable states}

We now give an analysis of how to generate and detect genuine $N$-partite
steering. We consider three types of states that we refer to as the
CV EPR, CV split squeezed state and CV GHZ states. These are generated
by a network of $N-1$ beam splitters, using two, one and $N$ squeezed-vacuum
input states, respectively.

\subsection{CV EPR state \label{subsec:CVEPR_steer}}

\begin{figure}[H]
\begin{centering}
\includegraphics[width=0.8\columnwidth]{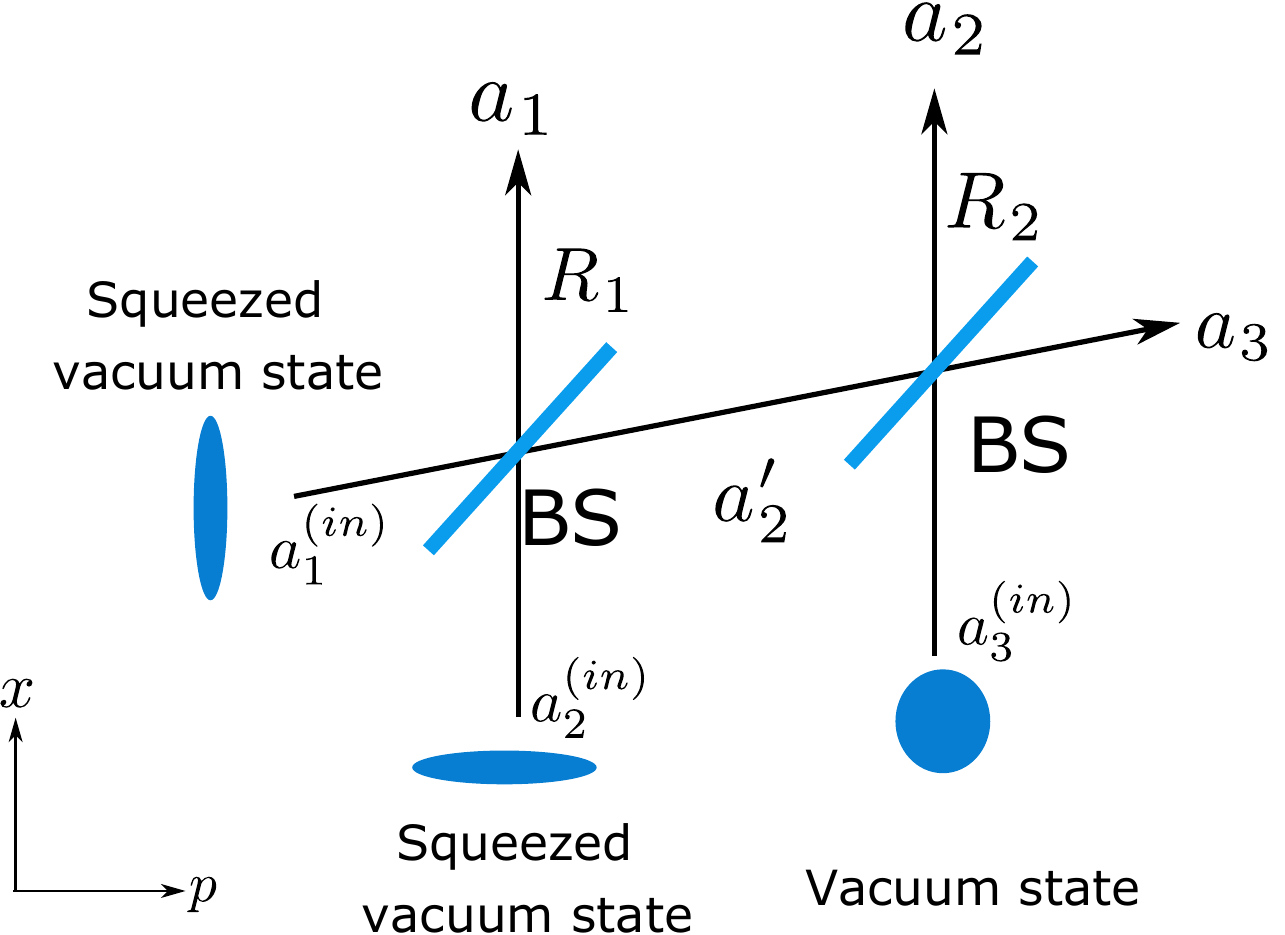}
\par\end{centering}
\caption{Generation of the tripartite-entangled CV EPR state. The configuration
uses two squeezed-vacuum inputs $a_{1}^{(in)}$ and $a_{2}^{(in)}$,
and two beam splitters (BS) with reflectivities $R_{1}=R_{2}=1/2$.
The $x_{i}$ and $p_{i}$ are the two orthogonal quadrature-phase
amplitudes of the spatially separated optical modes, denoted by $a_{i}$
$(i=1,2,3)$. \label{fig:tripartite_ent_EPR-2}}
\end{figure}

We begin with the CV EPR state. This state has been used to generate
genuine tripartite entanglement \citep{Armstrong_Nature2012}, and
can be generated following the schemes suggested in \citep{PhysRevA.90.062337,vanLoock_PRA2003}.
The set-up for $N=3$ is illustrated in Figure \ref{fig:tripartite_ent_EPR-2}.
We will show that the schemes also produce a genuinely $N$-partite
steerable state that can be detected using Criterion 1, derived in
the previous section.

Two orthogonally squeezed inputs are placed through a 50/50 beam splitter
$BS1$, to produce EPR entangled fields with boson operators $a_{1}$
and $a'_{2}$ at the two outputs. We show this by writing the output
fields as
\begin{eqnarray}
a_{1} & = & \sqrt{R_{1}}a_{1}^{(in)}+\sqrt{T_{1}}a_{2}^{(in)}\nonumber \\
a'_{2} & = & \sqrt{T_{1}}a_{1}^{(in)}-\sqrt{R_{1}}a_{2}^{(in)}\label{eq:BS}
\end{eqnarray}
where $R_{1}+T_{1}=1$, $R_{1}$ being the reflectivity of the beam
splitter, and $a_{1}^{(in)}$ and $a_{2}^{(in)}$ are the inputs to
the beam splitter (Figure 1). This gives $a_{1}^{(in)}=\sqrt{R_{1}}a_{1}+\sqrt{T_{1}}a'_{2}$
and $a_{2}^{(in)}=\sqrt{T_{1}}a_{1}+\sqrt{R_{1}}a'_{2}$. Thus, $\sqrt{T_{1}}x_{1}-\sqrt{R_{1}}x'_{2}=x_{2}^{(in)}$
and $\sqrt{R_{1}}p_{1}+\sqrt{T_{1}}p'_{2}=p_{1}^{(in)}$, where $x_{j}$,
$x'_{j}$ and $p_{j}$, $p'_{j}$ are the quadratures of the fields
$a_{j}$, $a_{j}'$ respectively, and $x_{j}^{(in)}$ and $p_{j}^{(in)}$
are the quadratures of the field $a_{j}^{(in)}$. If the input $a_{2}^{(in)}$
is a squeezed input with $\Delta^{2}x_{2}^{(in)}=e^{-2r}$, then
\begin{equation}
\Delta^{2}(\sqrt{T_{1}}x_{1}-\sqrt{R_{1}}x'_{2})=\Delta^{2}x_{2}^{(in)}=e^{-2r}\label{eq:var1}
\end{equation}
where we use the notation $\Delta^{2}x\equiv(\Delta x)^{2}$ to simplify
the use of brackets. If the input $a_{1}^{(in)}$ is squeezed in $p$,
so that $\Delta^{2}p_{1}^{(in)}=e^{-2r}$, then
\begin{equation}
\Delta^{2}(\sqrt{R_{1}}p_{1}+\sqrt{T_{1}}p'_{2})=e^{-2r}\thinspace.\label{eq:varp}
\end{equation}
Choosing $R_{1}=\frac{1}{2}$, these fields satisfy 
\begin{eqnarray}
\Delta^{2}(x_{1}-x'_{2}) & = & 2e^{-2r}\nonumber \\
\Delta^{2}(p_{1}+p'_{2}) & = & 2e^{-2r}\label{eq:epr-2}
\end{eqnarray}
where $x_{i}$ and $p_{i}$ are the quadratures associated with each
mode $a_{i}$. Entanglement is detected when \citep{Giovannetti_PRA2003}
\begin{equation}
\Delta(x_{1}-x_{1}')\Delta(p_{1}+p'_{1})<2\thinspace,\label{eq:ent-1}
\end{equation}
 implying the fields to be entangled for all $r>0$.

\begin{figure}[H]
\begin{centering}
\includegraphics[width=0.93\columnwidth]{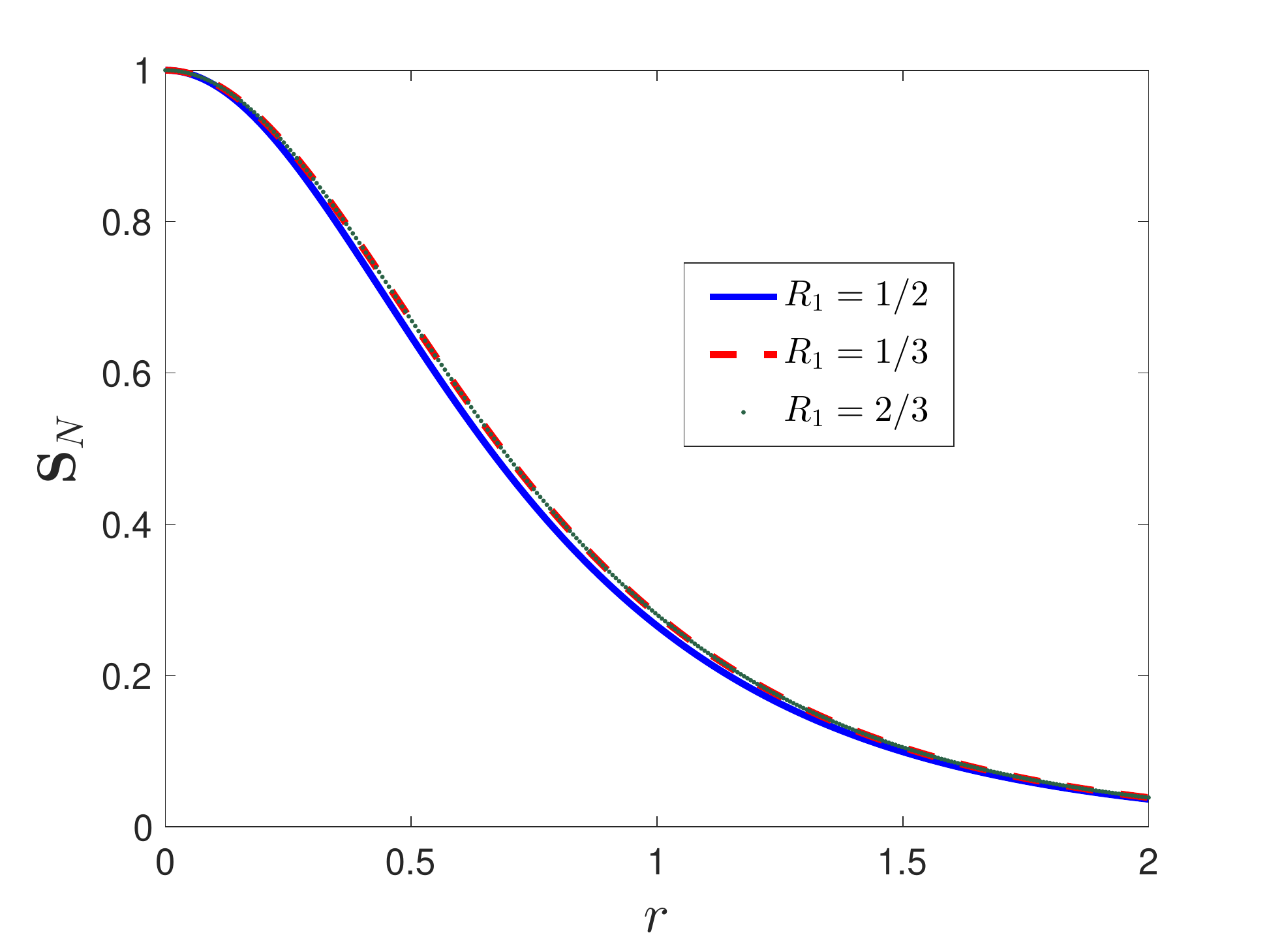}
\par\end{centering}
\caption{Genuine $N$-partite steering for the CV EPR state: The plots shows
the optimal steering parameter $\mathbf{S}_{N}$ defined in Eqs. (\ref{eq:S_npartite})
and (\ref{eq:steer-varN}) as a function of the squeezing parameter
$r$ and for different reflectivities $R_{1}$. Steering of system
$1$ is obtained when $\mathbf{S}_{N}\equiv S_{1|\{2,..,N\}}<1$.
It is evident that genuine $N$-partite steering is obtained for sufficiently
large $r$, implying mutual steering between all subsystems, as given
by Criterion 1b and Eq. (\ref{eq:gen-N}). \label{fig:steering-product}\textcolor{blue}{}}
\end{figure}

\begin{figure}[H]
\begin{centering}
\includegraphics[width=0.93\columnwidth]{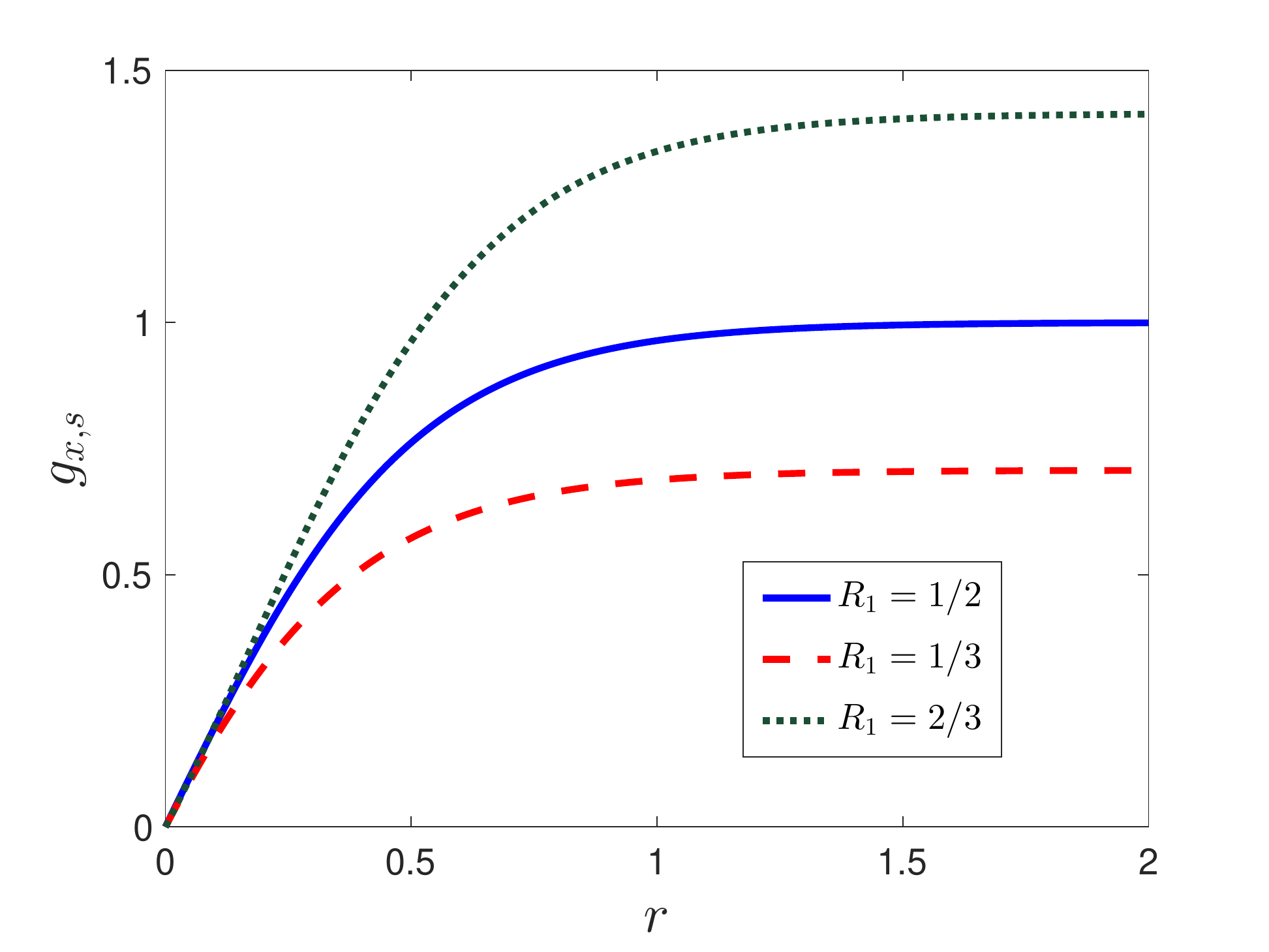}
\par\end{centering}
\caption{Optimal gains: The gains $g_{x,s}$ required for the steering in Figure
\ref{fig:steering-product} as given by Eq. (\ref{eq:g_xs}), as a
function of the squeezing parameter $r$ for different reflectivities
$R_{1}$. \label{fig:gains}\textcolor{red}{}The plots for $g_{p,s}$
as in Eq. (\ref{eq:g_ps}), are obtained on replacing $R_{1}$ with
$T_{1}$.}
\end{figure}

More generally, to investigate the EPR steering correlations as in
\citep{Reid_PRA1989}, we find
\begin{eqnarray*}
\Delta^{2}(x_{1}-g_{x,s}x'_{2}) & = & g_{x,s}^{2}(T_{1}e^{2r}+R_{1}e^{-2r})+R_{1}e^{2r}+T_{1}e^{-2r}\\
 &  & -2\sqrt{R_{1}T_{1}}g_{x,s}(e^{2r}-e^{-2r})
\end{eqnarray*}
which is minimum for 
\begin{equation}
g_{x,s}=\frac{\sqrt{R_{1}T_{1}}(e^{2r}-e^{-2r})}{(T_{1}e^{2r}+R_{1}e^{-2r})}\thinspace.\label{eq:g_xs}
\end{equation}
Similarly, 
\begin{eqnarray*}
\Delta^{2}(p_{1}+g_{p,s}p'_{2}) & = & g_{p,s}^{2}(T_{1}e^{-2r}+R_{1}e^{2r})+R_{1}e^{-2r}+T_{1}e^{2r}\\
 &  & -2\sqrt{R_{1}T_{1}}g_{p,s}(e^{2r}-e^{-2r})
\end{eqnarray*}
which is minimum for 
\begin{equation}
g_{p,s}=\frac{\sqrt{R_{1}T_{1}}(e^{2r}-e^{-2r})}{(R_{1}e^{2r}+T_{1}e^{-2r})}\thinspace.\label{eq:g_ps}
\end{equation}
This gives
\begin{eqnarray}
\Delta^{2}(x_{1}-g_{x,s}x'_{2}) & = & \frac{1}{\left(1-R_{1}\right)e^{2r}+R_{1}e^{-2r}}\nonumber \\
\Delta^{2}(p_{1}+g_{p,s}p'_{2}) & = & \frac{1}{R_{1}e^{2r}+\left(1-R_{1}\right)e^{-2r}}\,.\label{eq:var-steer-x}
\end{eqnarray}
For all values of beam splitter reflectivity $R_{1}$, there is
EPR steering whenever $r>0$, and perfect EPR correlation as the variances
become zero, as $r\rightarrow\infty$. The optimal EPR steering product
as defined by Eq. (\ref{eq:eprS-3}) for the two output modes is 
\begin{equation}
\mathbf{S}_{2}\equiv S_{1|2}=\Delta u_{1}\Delta v_{1}=\Delta(x_{1}-g_{x,s}x'_{2})\Delta(p_{1}+g_{p,s}p'_{2})\label{eq:S_npartite}
\end{equation}
where here for Eq. (\ref{eq:eprS-3}) we identify $h_{2}=-g_{x,s}$
and $g_{2}=g_{p,s}$. We plot $\mathbf{S}_{2}$ in Figure \ref{fig:steering-product}
for the optimal choice of gains $g_{x,s}$ and $g_{p,s}$, for various
values of reflectivity $R_{1}$. One may prove by differentiation
that the optimal choice to minimize $S_{2}$ is $R_{1}=1/2$. The
optimal steering product then becomes 
\begin{equation}
\mathbf{S}_{2}\equiv S_{1|2}=\Delta u_{1}\Delta v_{1}=\frac{1}{\cosh2r}\label{eq:epr-solnS}
\end{equation}
for optimal gains given by
\begin{equation}
g_{x,s}=g_{p,s}=\tanh2r\label{eq:gains-epr-soln}
\end{equation}
In Figure \ref{fig:gains}, we plot the optimal gains versus $r$.

As shown in \citep{PhysRevA.90.062337,vanLoock_PRA2003}, to produce
tripartite entangled fields, the field $2$ is then split using a
beam splitter labelled $BS2$, with a vacuum input $a_{3}^{(in)}$,
to give two new outputs $a_{2}=(\sqrt{R_{2}}a_{2}'+\sqrt{T_{2}}a_{3}^{(in)})$
and $a_{3}=(\sqrt{T_{2}}a'_{2}-\sqrt{R_{2}}a_{3}^{(in)})$, where
$R_{2}+T_{2}=1$. The set-up for $N=3$ is shown in Figure \ref{fig:tripartite_ent_EPR-2}.
We see that $a_{2}'=\sqrt{R_{2}}a_{2}+\sqrt{T_{2}}a_{3}$. Thus 
\begin{eqnarray}
x'_{2} & = & \sqrt{R_{2}}x_{2}+\sqrt{T_{2}}x_{3}\nonumber \\
p'_{2} & = & \sqrt{R_{2}}p_{2}+\sqrt{T_{2}}p_{3}\label{eq:bs2x,p}
\end{eqnarray}
where $x_{j}$ and $p_{j}$ are the quadratures of the field $a_{j}$.
Taking $R_{1}=\frac{1}{2}$, from Eq. (\ref{eq:epr-2}) we see that
$\Delta^{2}(x_{1}-\frac{1}{\sqrt{2}}(x_{2}+x_{3}))=2e^{-2r}$ and
$\Delta^{2}(p_{1}+\frac{1}{\sqrt{2}}(p_{2}+p_{3}))=2e^{-2r}$, implying
that the variances are zero for large $r$. The output fields satisfy
the condition
\begin{equation}
\Delta(x_{1}-\frac{1}{\sqrt{2}}(x_{2}+x_{3}))\Delta(p_{1}+\frac{1}{\sqrt{2}}(p_{2}+p_{3}))<1\label{eq:cond-ent3}
\end{equation}
for genuine tripartite entanglement given by equation (17) of \citep{PhysRevA.90.062337},
with equal gains for the second and third modes. The product becomes
zero, indicating maximum EPR entanglement, in the limit of large $r$.

However, we are interested to examine the tripartite EPR steering.
On taking $R_{2}=\frac{1}{2}$, from Eq. (\ref{eq:var-steer-x}) we
see on substituting Eq. (\ref{eq:bs2x,p}) that 
\begin{eqnarray}
\Delta^{2}(x_{1}-\frac{g_{x,s}}{\sqrt{2}}(x_{2}+x_{3})) & = & \frac{1}{\cosh2r}\nonumber \\
\Delta^{2}(p_{1}+\frac{g_{p,s}}{\sqrt{2}}(p_{2}+p_{3})) & = & \frac{1}{\cosh2r}\thinspace.\label{eq:steer-var}
\end{eqnarray}
Hence the steering product defined by Eq. (\ref{eq:eprS-3}) for
the three output modes becomes
\begin{equation}
\mathbf{S}_{3}\equiv S_{1|23}=\Delta u_{1}\Delta v_{1}=\frac{1}{\cosh2r}\label{eq:steer3}
\end{equation}
where here for Eq. (\ref{eq:eprS-3}) we identify $h_{2}=h_{3}=-g_{x,s}/\sqrt{2}$
and $g_{2}=g_{3}=g_{p,s}/\sqrt{2}$. There is steering of system $1$
for all values of $r$, as we see by examining the steering condition
Eq. (\ref{eq:epr-cond-1}). We will also see that Criterion 1 for
genuine tripartite steering is satisfied for $r>0.76$.

Continuing, it is possible to select the reflectivities of the string
of beam splitters so that $x'_{2}=\frac{1}{\sqrt{N}}(\sum_{i}^{N-1}x_{i})$
and $p'_{2}=\frac{1}{\sqrt{N}}(\sum_{i}^{N-1}p_{i})$. In this case,
on substituting in Eq. (\ref{eq:epr-2}), we obtain 
\begin{eqnarray}
\Delta^{2}(x_{1}-\frac{1}{\sqrt{N-1}}(x_{2}+x_{3}+..x_{N})) & = & 2e^{-2r}\nonumber \\
\Delta^{2}(p_{1}+\frac{1}{\sqrt{N-1}}(p_{2}+p_{3}+..p_{N})) & = & 2e^{-2r}\thinspace.\label{eq:ent}
\end{eqnarray}
The criterion
\begin{equation}
S_{N}=\Delta u\Delta v<\frac{1}{N-1},\label{eq:cond-ent1}
\end{equation}
defined for $u$ and $v$ with $g_{1}=h_{1}=1$ and $g_{i}=g$ and
$h_{i}=h$ for $i>1$,  is sufficient to confirm genuine $N$-partite
entanglement, as shown in \citep{PhysRevA.90.062337,vanLoock_PRA2003}.
This criterion is clearly satisfied for large $r$.

To analyze genuine $N$-partite steering, we use Eq. (\ref{eq:var-steer-x})
to obtain
\begin{align}
\Delta^{2}(x_{1}-\frac{g_{x,s}}{\sqrt{N-1}}(x_{2}+x_{3}+..x_{N})) & =\frac{1}{\cosh2r}\nonumber \\
\Delta^{2}(p_{1}+\frac{g_{p,s}}{\sqrt{N-1}}(p_{2}+p_{3}+..p_{N})) & =\frac{1}{\cosh2r}\thinspace.\label{eq:steer-result}
\end{align}
Hence the steering product as defined by Eq. (\ref{eq:eprS-3}) for
the $N$ output modes becomes
\begin{equation}
\mathbf{S}_{N}\equiv S_{1|\{2,..,N\}}=\Delta u_{1}\Delta v_{1}=\frac{1}{\cosh2r}\label{eq:steer-varN}
\end{equation}
where here for Eq. (\ref{eq:eprS-3}) we identify for $i>1$ that
$h_{i}=-g_{x,s}/\sqrt{N-1}$ and $g_{i}=g_{p,s}/\sqrt{N-1}$. There
is steering of system $1$ for all values of $r$, as we see by examining
the steering condition Eq. (\ref{eq:epr-cond-1}). We also see that
the Criterion 1 as extended for genuine $N$-partite steering below
is satisfied for large $r$. We consider the state created by selecting
for the beam splitters, $R_{N-1}=1/2$, $R_{N-2}=\frac{1}{3}$, $R_{N-r}=\frac{1}{r+1}$
for $r<N-1$, as explained in \citep{PhysRevA.90.062337,vanLoock_PRA2003},
with $R_{1}=1/2$. The state produced shows genuine $N$-partite steering.

\begin{figure}[H]
\begin{centering}
\includegraphics[width=0.93\columnwidth]{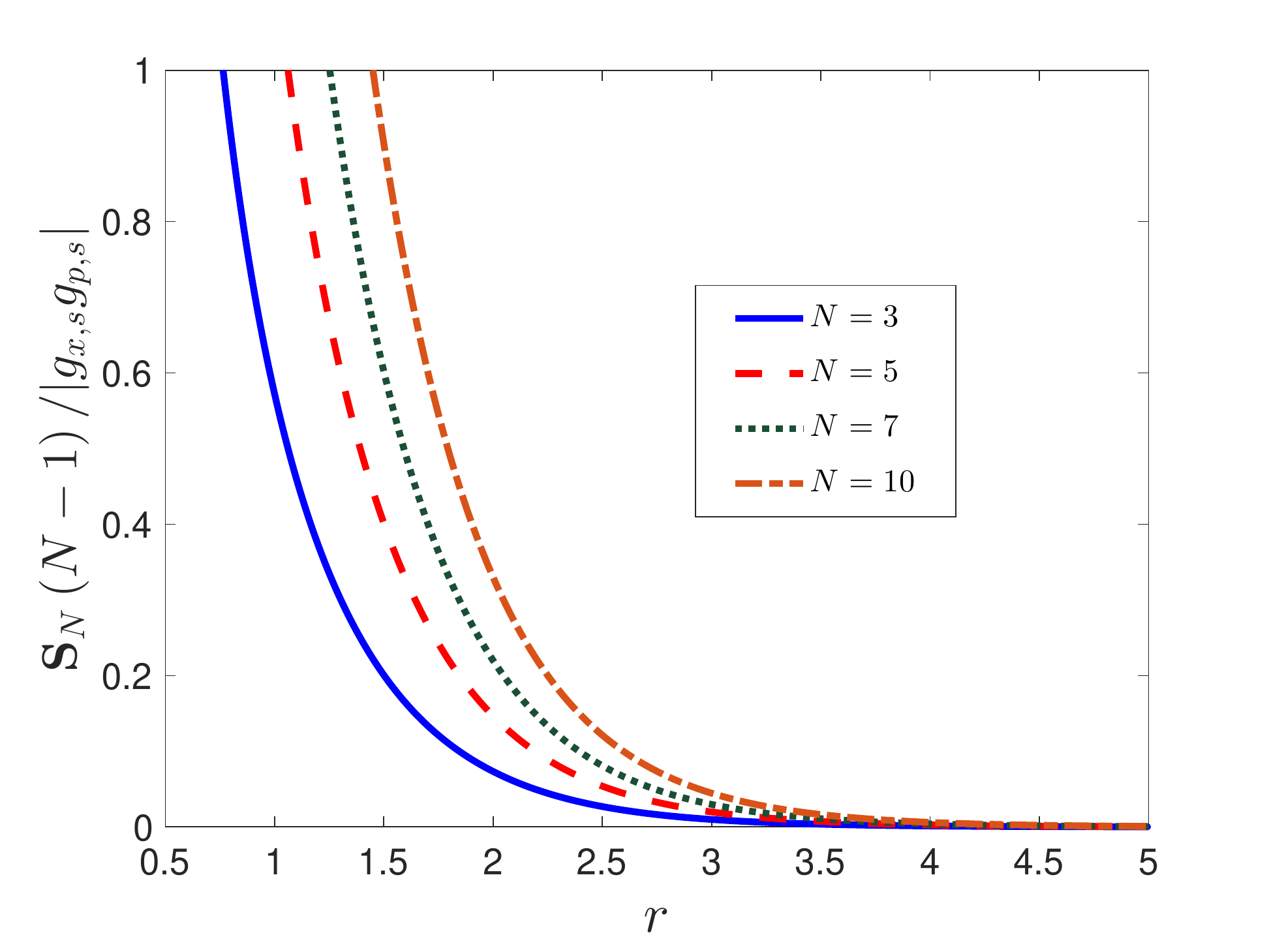}
\par\end{centering}
\caption{Genuine $N$-partite steering for the CV EPR state: The value of $\mathbf{S}_{N}$
divided by the bound $|g_{x,s}g_{p,s}|/\left(N-1\right)$ as provided
in Criterion \textit{\emph{1b}}, for the CV EPR state. The analytical
expression for $\mathbf{S}_{N}$ and the optimal gains $g_{x,s}$
and $g_{p,s}$ are given by Eq. (\ref{eq:gains-epr-soln}). When the
value is less than $1$, there is genuine $N$-partite steering according
to Criterion \textit{\emph{1b}}, given by Eq. (\ref{eq:steering-cond-gen}).
\label{fig:S_N_cvepr-bound}}
\end{figure}

The Criterion 1 can be generalized to $N$ parties, to give the following
criterion. 

\textbf{\emph{Criterion 1b}}\textbf{:} Selecting $u_{1}=x_{1}+h(x_{2}+x_{3}+..x_{N})$
and $v_{1}=p_{1}+g(p_{2}+p_{3}+..p_{N})$, we see that the corresponding
gains for the Criterion 1 are $g_{1}=h_{1}=1$ and $g_{i}=g$ and
$h_{i}=h$ ($i>1)$. Since $0\leq g,-h\leq1$, one simultaneously
confirms two-way steering along all bipartitions if with this choice
of $u$ and $v$, one can confirm
\begin{align}
\mathbf{S}_{N} & \equiv\Delta u_{1}\Delta v_{1}\nonumber \\
 & <\min\Bigl\{1,|(N-1)gh|;|gh|,|1+(N-2)gh|;\nonumber \\
 & |2gh|,|1+(N-3)gh|;|3gh|,|1+(N-4)gh|;...;\nonumber \\
 & |1+gh|,|(N-2)gh|;|1+2gh|,|(N-3)gh|;...\Bigl\}\nonumber \\
 & <\min\Bigl\{|gh|,|1-(N-2)|gh||\Bigl\}\thinspace.
\end{align}
Where the gains are $g_{i}=\frac{g_{p,s}}{\sqrt{N-1}}$ and $h_{i}=-\frac{g_{x,s}}{\sqrt{N-1}}$,
so that $u_{1}=x_{1}-\frac{g_{x,s}}{\sqrt{N-1}}(x_{2}+x_{3}+..x_{N})$
and $v_{1}=p_{1}+\frac{g_{p,s}}{\sqrt{N-1}}(p_{2}+p_{3}+..p_{N})$,
this reduces to
\begin{equation}
\mathbf{S}_{N}\equiv\Delta u_{1}\Delta v_{1}<\min\{\frac{g_{x,s}g_{p,s}}{N-1},1-\frac{N-2}{N-1}g_{x,s}g_{p,s}\}\thinspace.\label{eq:steering-cond-gen}
\end{equation}
This inequality therefore gives a criterion for genuine $N$-partite
steering. With the optimal choice of gains, the inequality becomes
 $\mathbf{S}_{N}<\frac{\tanh^{2}2r}{N-1}$, since the second term
is greater.

For the choice of gains $g_{x,s}$ and $g_{p,s}$ given by Eqs. (\ref{eq:g_xs})
and (\ref{eq:g_ps}) that optimize for steering of system $1$, the
value of $\mathbf{S}_{N}$ is plotted in Figure \ref{fig:steering-product}.
The criterion $\mathbf{S}_{N}<1$ is clearly satisfied for all $r>0$,
and as $r\rightarrow\infty$, $\mathbf{S}_{N}\rightarrow0$, implying
maximal EPR steering of system $1$. Using the expressions in
Eq. (\ref{eq:ent}) and the inequality Eq. (\ref{eq:steering-cond-gen}),
the corresponding condition on $r$ for genuine $N$-partite steering
according to Criterion 1b with this choice of gains is given by 
\begin{eqnarray}
\mathbf{S}_{N} & \equiv & S_{1|\{2,..,N\}}=\Delta u_{1}\Delta v_{1}=\frac{1}{\cosh2r}<\frac{\tanh^{2}2r}{N-1}\thinspace,\nonumber \\
\label{eq:gen-N}
\end{eqnarray}
which is satisfied for $r>0.76$ for $N=3$. The normalized value
given by $\mathbf{S}_{N}$ divided by the bound $|g_{x,s}g_{p,s}|/\left(N-1\right)$
is plotted in Figure \ref{fig:S_N_cvepr-bound}. The minimum squeezing
parameter required to show genuine steering according to Eq. (\ref{eq:gen-N})
satisfies $\cosh2r>\frac{\left(N-1\right)+\sqrt{\left(N-1\right)^{2}+4}}{2}$.
We note that it is likely genuine $N$-partite steering can be detected
for smaller $r$ values if the gains are chosen to optimize the inequality,
rather than to optimize for the steering of system $1$.

\subsection{CV split squeezed state}

\begin{figure}[H]
\begin{centering}
\includegraphics[width=0.8\columnwidth]{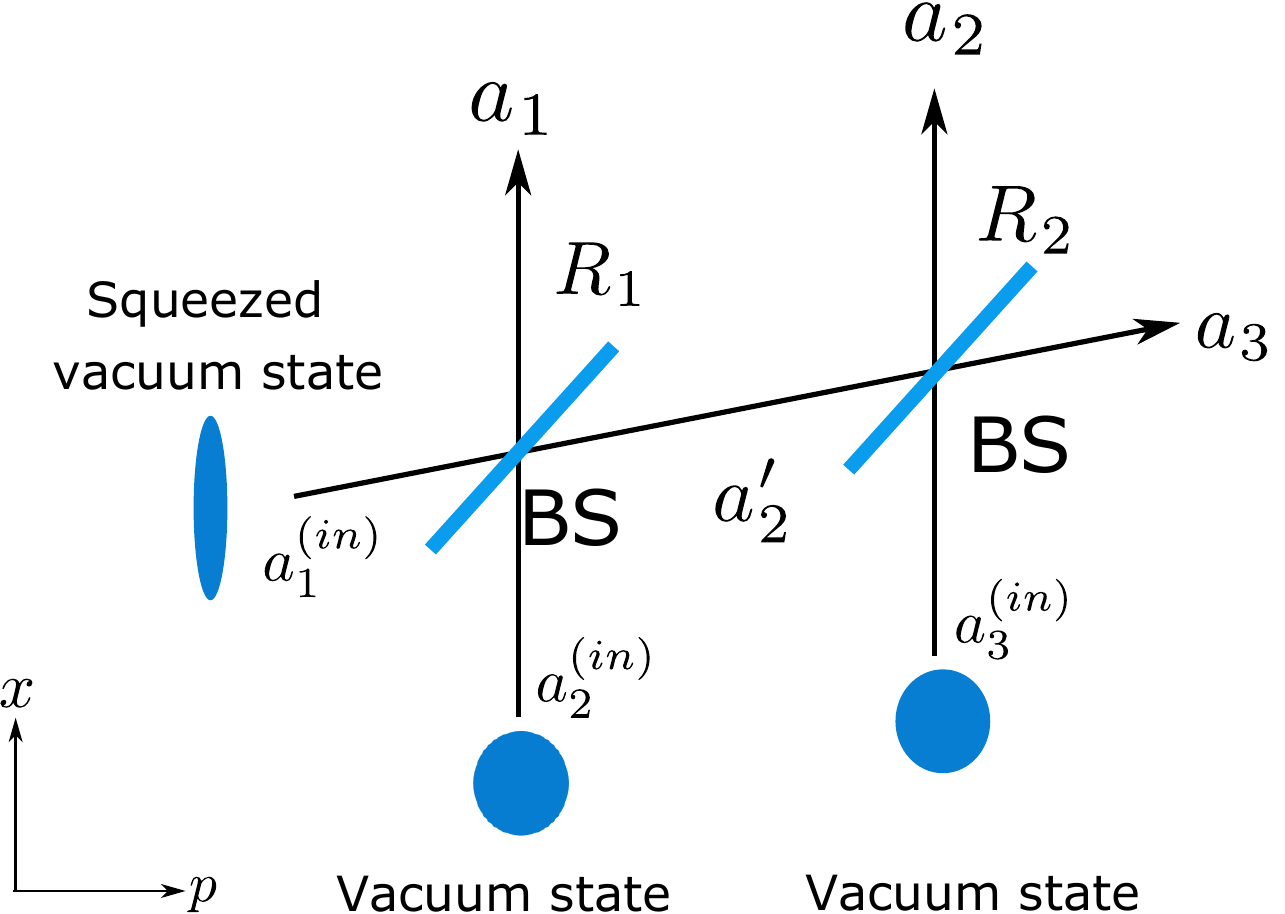}
\par\end{centering}
\caption{Generation of the tripartite-entangled CV split squeezed state. The
configuration uses a single squeezed-vacuum input $a_{1}^{(in)}$
and two beam splitters (BS) with reflectivities $R_{1}=1/2$ and $R_{2}=1/2$.
\textcolor{red}{}The $x_{i}$ and $p_{i}$ are the two orthogonal
quadrature-phase amplitudes of the spatially separated optical modes
$i$ $(i=1,2,3)$. \label{fig:tripartite_ent_EPR-1-1}}
\end{figure}

Genuine $N$-partite entanglement and steering can also be generated
from a network with just one \emph{single} squeezed-state vacuum input
(Figure \ref{fig:tripartite_ent_EPR-1-1}). This is possible because
the two outputs of a beam splitter with a single squeezed vacuum input
are EPR correlated \citep{Reid_PRA1989}. A high degree of squeezing
$r$ is required for the input however, in order to generate a feasible
amount of multipartite entanglement.

We calculate the steering correlations explicitly. We first consider
the bipartite steering created where one squeezed input is placed
through a beam splitter $BS1$ with inputs $a_{1}^{(in)}$ and $a_{2}{}^{(in)}$.
This produces EPR entangled fields with boson operators $a_{1}$ and
$a'_{2}$ at the two outputs. Using the procedure from Section IV.A,
if $a_{2}^{(in)}$ is a vacuum then $\Delta^{2}x_{2}^{(in)}=1$,
implying
\begin{equation}
\Delta^{2}(\sqrt{T_{1}}x_{1}-\sqrt{R_{1}}x'_{2})=\Delta^{2}x_{2}^{(in)}=1\thinspace.\label{eq:123}
\end{equation}
If $a_{1}^{(in)}$ is squeezed in $p$, so that $\Delta^{2}p_{1}^{(in)}=e^{-2r}$,
we have
\begin{equation}
\Delta^{2}(\sqrt{R_{1}}p_{1}+\sqrt{T_{1}}p'_{2})=e^{-2r}\thinspace.\label{eq:1234}
\end{equation}
Choosing $R_{1}=\frac{1}{2}$, these fields satisfy 
\begin{eqnarray}
\Delta^{2}(x_{1}-x'_{2}) & = & 2\nonumber \\
\Delta^{2}(p_{1}+p'_{2}) & = & 2e^{-2r}\label{eq:epr-solns}
\end{eqnarray}
where $x_{i}$ and $p_{i}$ are the quadratures associated with each
mode $a_{i}$. Entanglement is detected when $\Delta(x_{1}-x_{1}')\Delta(p_{1}+p'_{1})<2$
\citep{Giovannetti_PRA2003} so that the fields are entangled for
all $r>0$.

We next examine the EPR steering between the two modes. For arbitrary
$R_{1}$, we obtain
\begin{eqnarray}
\Delta^{2}(x_{1}-g_{x,s}x'_{2}) & = & g_{x,s}^{2}(T_{1}e^{2r}+R_{1})+R_{1}e^{2r}+T_{1}\nonumber \\
 &  & -2\sqrt{R_{1}T_{1}}g_{x,s}(e^{2r}-1)\label{eq:soln1}
\end{eqnarray}
which is minimum for 
\begin{equation}
g_{x,s}=\frac{\sqrt{R_{1}T_{1}}(e^{2r}-1)}{(T_{1}e^{2r}+R_{1})}\thinspace.\label{eq:gsoln}
\end{equation}
Similarly,
\begin{eqnarray}
\Delta^{2}(p_{1}+g_{p,s}p'_{2}) & = & g_{p,s}^{2}(T_{1}e^{-2r}+R_{1})+R_{1}e^{-2r}+T_{1}\nonumber \\
 &  & -2\sqrt{R_{1}T_{1}}g_{p,s}(1-e^{-2r})\label{eq:var-soln-p}
\end{eqnarray}
which is minimum for 
\begin{equation}
g_{p,s}=\frac{\sqrt{R_{1}T_{1}}(1-e^{-2r})}{(R_{1}+T_{1}e^{-2r})}\thinspace.\label{eq:soln-gp}
\end{equation}
This gives
\begin{eqnarray}
\Delta^{2}(x_{1}-g_{x,s}x'_{2}) & = & \frac{e^{2r}}{\left(1-R_{1}\right)e^{2r}+R_{1}}\thinspace,\label{eq:u_CVSS}
\end{eqnarray}
which implies that for $r=0$, $\Delta^{2}(x_{1}-g_{x,s}x'_{2})=1$.
Similarly,
\begin{eqnarray}
\Delta^{2}(p_{1}+g_{p,s}p'_{2}) & = & \frac{e^{-2r}}{R_{1}+\left(1-R_{1}\right)e^{-2r}}\thinspace,\label{eq:v_CVSS-3}
\end{eqnarray}
which is $1$ for $r=0$ and for large $r$ becomes $0$. This is
true for all values of reflectivity $R_{1}$. Hence, the optimal steering
product $\mathbf{S}_{2}=\Delta(x_{1}-g_{x,s}x'_{2})\Delta(p_{1}-g_{p,s}p'_{2})$
defined by Eq. (\ref{eq:eprS-3}) for the two output modes is given
by

\begin{align}
\mathbf{S}_{2}^{2} & \equiv(S_{1|2})^{2}=\frac{1}{1+4R_{1}\left(1-R_{1}\right)\sinh^{2}r}\thinspace.\label{eq:sr}
\end{align}
As long as $R_{1}\neq0$, $\mathbf{S}_{2}\rightarrow0$ for large
$r$.

\begin{figure}[H]
\begin{centering}
\includegraphics[width=0.93\columnwidth]{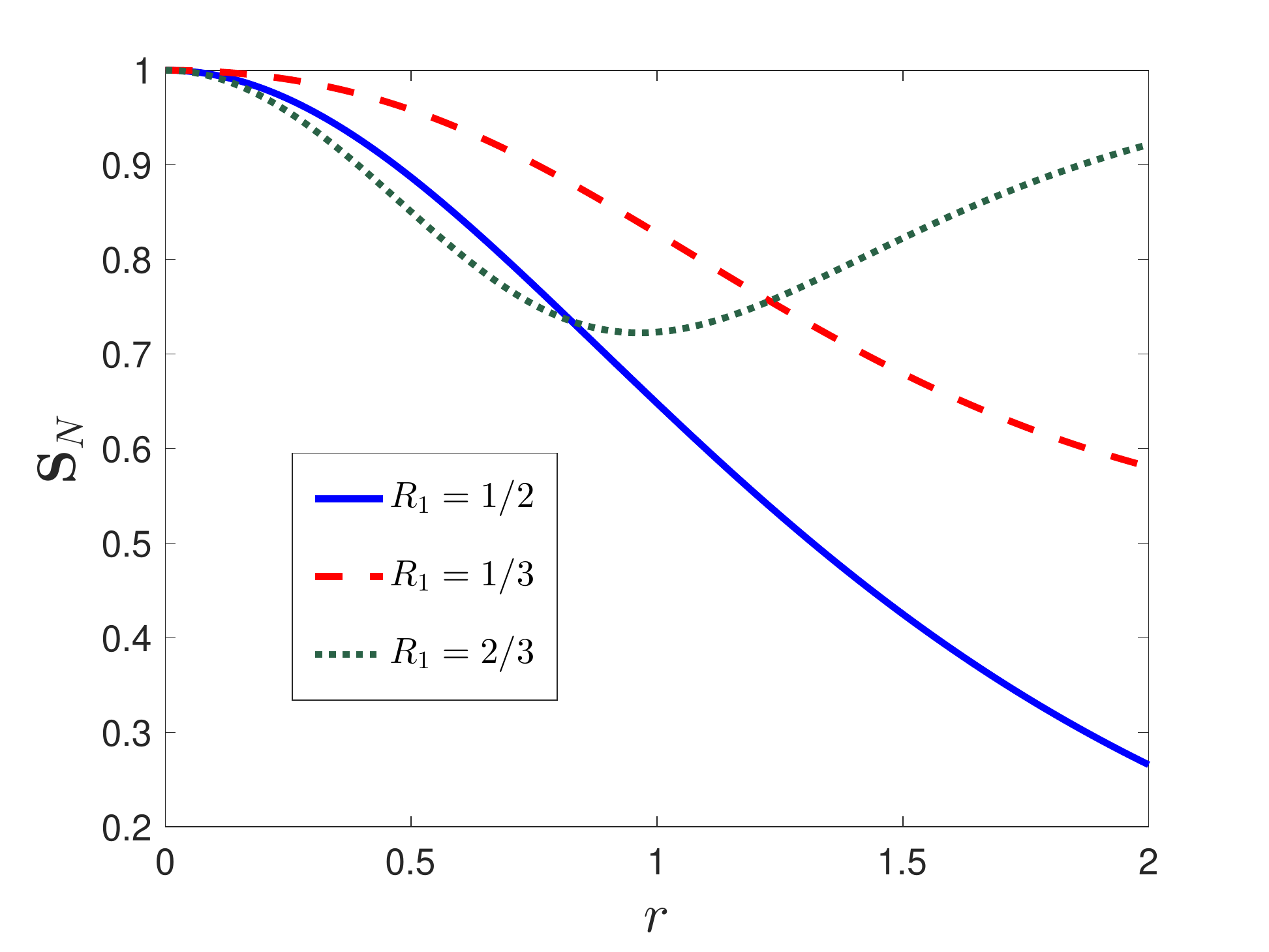}
\par\end{centering}
\caption{Genuine $N$-partite steering for the CV SS state: The plots shows
the optimal steering parameter $\mathbf{S}_{N}$ defined in Eqs. (\ref{eq:sr})
and (\ref{eq:steer-var-ss-1}) as a function of the squeezing parameter
$r$ and for different reflectivities $R_{1}$. Steering of system
$1$ is obtained when $S_{N}\equiv S_{1|\{2,..,N\}}<1$. It is evident
that genuine $N$-partite steering is obtained for sufficiently large
$r$, implying mutual steering between all subsystems, as given by
Criterion 1b. The blue solid line is the result given by Eq. (\ref{eq:steer-var-ss-1}).
\label{fig:splitss-steer}}
\end{figure}

\begin{figure}[H]
\begin{centering}
\includegraphics[width=0.93\columnwidth]{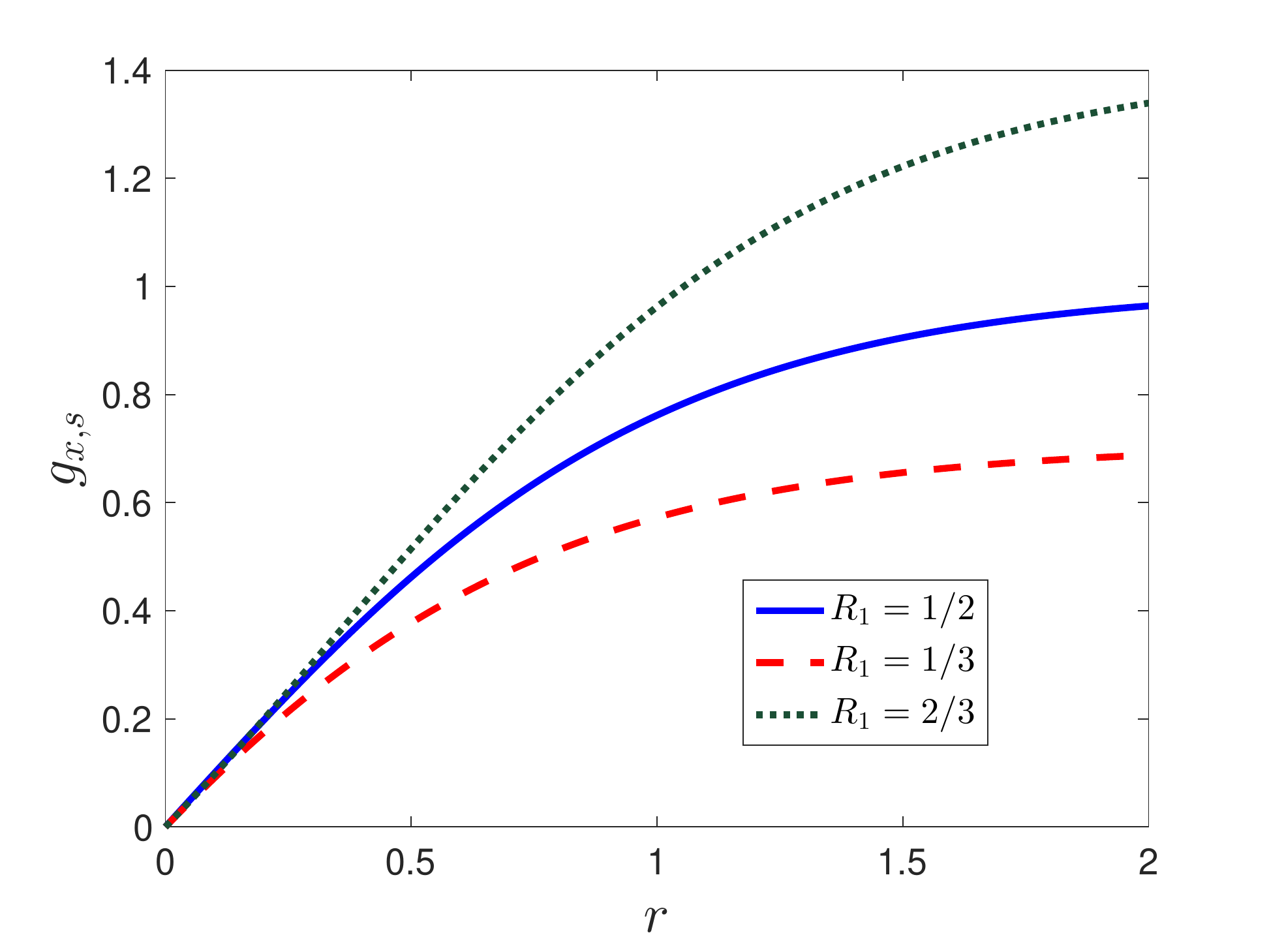}
\par\end{centering}
\caption{Optimal gains: The gains $g_{x,s}$ required for the steering in Figure
\ref{fig:splitss-steer} as given in Eq. (\ref{eq:g_xs}), as a function
of the squeezing parameter $r$, for different reflectivities $R_{1}$.
\label{fig:splitsssteer-gains}The plots for $g_{p,s}$ are obtained
on replacing $R_{1}$ with $T_{1}$.}
\end{figure}
We plot the bipartite steering product $\mathbf{S}_{2}$ in Figures
\ref{fig:splitss-steer} and \ref{fig:splitsssteer-gains}, versus
$r$. One may prove by differentiation for each fixed $r$ that the
optimal choice to minimize $\mathbf{S}_{2}$ is $R_{1}=1/2$, which
gives 
\begin{eqnarray}
g_{x,s} & =g_{p,s}= & \frac{1-e^{-2r}}{1+e^{-2r}}\label{eq:gsoln-1}
\end{eqnarray}
and 
\begin{eqnarray}
\Delta^{2}(x_{1}-g_{x,s}x'_{2}) & = & \frac{2}{1+e^{-2r}}\nonumber \\
\Delta^{2}(p_{1}+g_{p,s}p'_{2}) & = & \frac{2e^{-2r}}{1+e^{-2r}}\thinspace.\label{eq:varsolnss}
\end{eqnarray}
This leads to $\mathbf{S}_{2}=\frac{1}{\cosh r}$, indicating
steering of system $1$ for all $r$, as given by the steering condition
Eq. (\ref{eq:epr-cond-1}). We note that although $\mathbf{S}_{2}$
vanishes for significantly large $r$, the value of $\mathbf{S}_{2}$
remains close to $1$ for the experimental achievable set of squeezing
parameters, $r<2$. This is particularly true where the reflectivity
deviates from the ideal value of $R_{1}=1/2$.

To produce tripartite entangled fields, the field $a'_{2}$ is then
split using a beam splitter labelled $BS2$, with a vacuum input $a_{3}^{(in)}$,
to give two new outputs $a_{2}=(\sqrt{R_{2}}a_{2}'+\sqrt{T_{2}}a_{3}^{(in)})$
and $a_{3}=(\sqrt{T_{2}}a'_{2}-\sqrt{R_{2}}a_{3}^{(in)})$, where
$R_{2}+T_{2}=1$, as above. This gives  $a_{2}'=\sqrt{R_{2}}a_{2}+\sqrt{T_{2}}a_{3}$.
Taking $R_{2}=\frac{1}{2}$, we see that 
\begin{eqnarray}
\Delta^{2}(x_{1}-\frac{\sqrt{R_{1}}}{\sqrt{2T_{1}}}(x_{2}+x_{3})) & = & 1/T_{1}\nonumber \\
\Delta^{2}(p_{1}+\frac{\sqrt{T_{1}}}{\sqrt{2R_{1}}}(p_{2}+p_{3})) & = & e^{-2r}/R_{1}\thinspace.\label{eq:bS-ss-ent}
\end{eqnarray}
The product is $e^{-2r}/R_{1}T_{1}$ which for a given $r$ minimizes
for $R_{1}=T_{1}$ i.e. for $R_{1}=\frac{1}{2}$. The output fields
$a_{1}$, $a_{2}$ and $a_{3}$ satisfy the condition Eq. (\ref{eq:cond-ent3})
for genuine tripartite entanglement, with equal gains for the second
and third modes.

Continuing, it is possible to select the reflectivities of the string
of beam splitters so that we obtain 
\begin{eqnarray}
\Delta^{2}(x_{1}-\frac{1}{\sqrt{N-1}}(x_{2}+x_{3}+..x_{N})) & = & 2\nonumber \\
\Delta^{2}(p_{1}+\frac{1}{\sqrt{N-1}}(p_{2}+p_{3}+..p_{N})) & = & 2e^{-2r}\thinspace.\label{eq:u_fixedgain_cvss}
\end{eqnarray}
This is done by selecting for the beam splitters, $R_{N-1}=1/2$,
$R_{N-2}=\frac{1}{3}$, $R_{N-r}=\frac{1}{r+1}$ for $r<N-1$, as
explained in \citep{PhysRevA.90.062337,vanLoock_PRA2003}, with $R_{1}=1/2$.
The outputs satisfy the condition for genuine $N$-partite entanglement
given by Eq. (\ref{eq:cond-ent1}).

Similarly, for the same string of beam splitters, we obtain
\begin{align}
\Delta^{2}(x_{1}-\frac{g_{x,s}}{\sqrt{N-1}}(x_{2}+x_{3}+..x_{N})) & =\frac{2}{(1+e^{-2r})}\nonumber \\
\Delta^{2}(p_{1}+\frac{g_{p,s}}{\sqrt{N-1}}(p_{2}+p_{3}+..p_{N})) & =\frac{2e^{-2r}}{(1+e^{-2r})}\thinspace.\label{eq:steer-result-1}
\end{align}
The value of the steering product defined by Eq. (\ref{eq:eprS-3})
for the $N$ output modes becomes
\begin{equation}
\mathbf{S}_{N}\equiv S_{1|\{2,..,N\}}=\Delta u_{1}\Delta v_{1}=\frac{1}{\cosh r}\thinspace.\label{eq:steer-var-ss-1}
\end{equation}
There is steering of system $1$ for all values of $r$, as we see
by examining the steering condition Eq. (\ref{eq:epr-cond-1}). A
criterion to reveal genuine $N$-partite steering is given by Criterion
1b, as above. The product $\mathbf{S}_{N}$ is plotted for the choice
of gains given by Eqs. (\ref{eq:gsoln}) and (\ref{eq:soln-gp}) in
Figure \ref{fig:splitss-steer}. These gains are optimized to detect
the steering of system $1$ i.e. to minimize the value of $\mathbf{S}_{N}$
rather than to optimize violation of the inequality of Criterion 1b.
We see that genuine $N$-partite steering is possible for sufficiently
large $r$, despite that only one squeezed vacuum state has been used
to generate the output fields. In order to achieve the condition given
by the Criterion 1b, we require 
\begin{align*}
\Delta u\Delta v=\frac{2e^{-r}}{(1+e^{-2r})} & <\frac{\left|g_{x,s}g_{p,s}\right|}{2}\,.
\end{align*}
This leads to the inequality $\cosh^{2}r-\left(N-1\right)\cosh r-1<0$.
For $N=3$,  this reduces to $r>1.53$.

\subsection{CV GHZ states}

\begin{figure}[H]
\begin{centering}
\includegraphics[width=0.8\columnwidth]{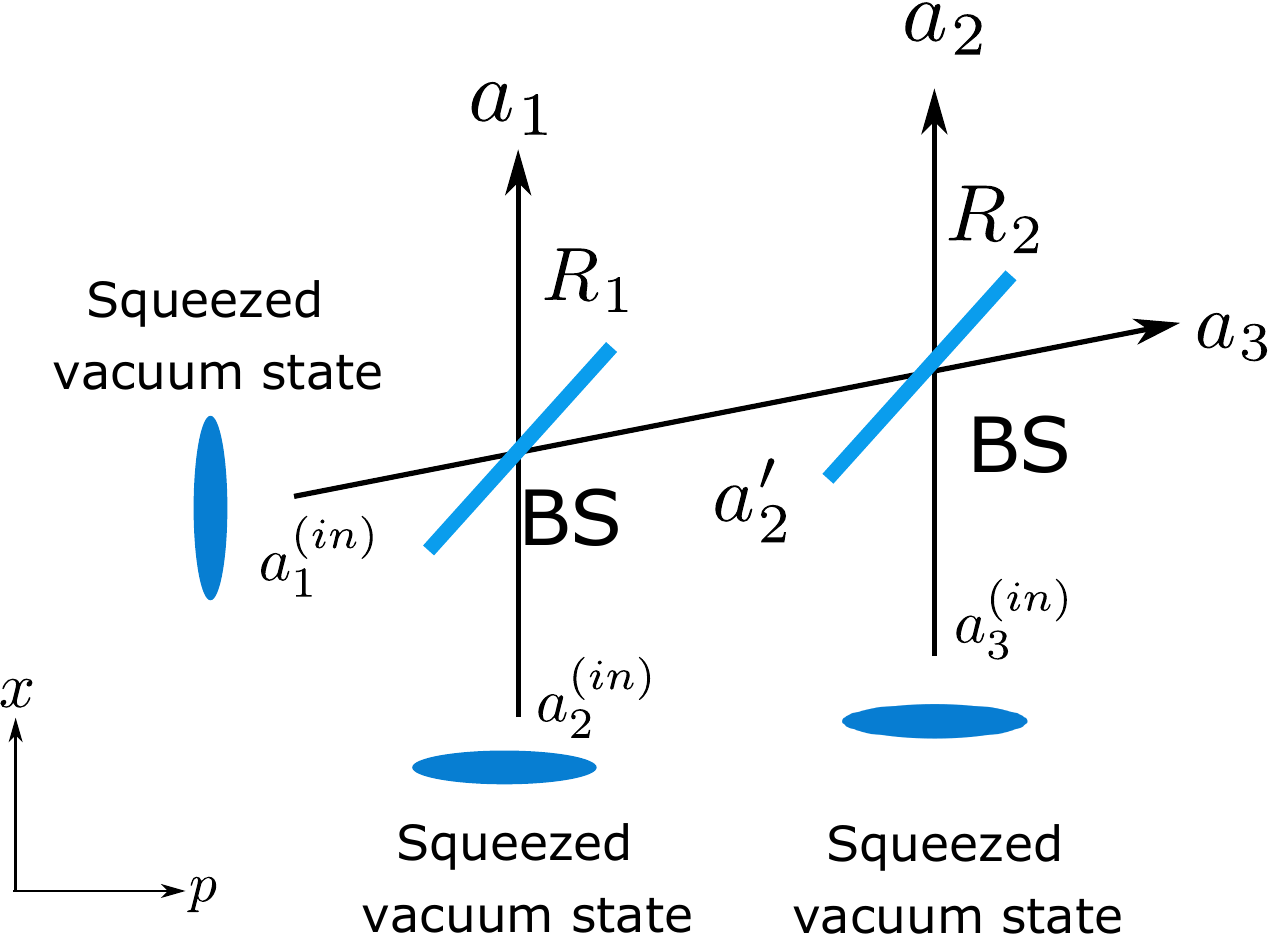}
\par\end{centering}
\caption{Generation of the tripartite-entangled CV GHZ state. The standard
configuration uses three squeezed-vacuum inputs and two beam splitters
(BS) with reflectivities $R_{1}=1/3$ and $R_{2}=1/2$. The $x_{i}$
and $p_{i}$ are the two orthogonal quadrature-phase amplitudes of
the spatially separated optical modes $i$ $(i=1,2,3)$. \label{fig:tripartite_ent_GHZ-1-1}}
\end{figure}

The CV GHZ state is generated by combining $N$ squeezed vacuum states
at the inputs of the $N-1$ beam splitters \citep{vanLoock_PRA2003,vanloock_PRA2001}.
The first mode is squeezed in the direction $p$ orthogonal to the
direction of squeezing $x$ of the remaining modes. The variances
are given as $\Delta^{2}X=e^{-2r}$, where $X$ is the squeezed quadrature
(Figure \ref{fig:tripartite_ent_GHZ-1-1}).

The final output variances are such that \citep{vanLoock_PRA2003}
\begin{equation}
\Delta^{2}(x_{i}-x_{j})=2e^{-2r}\label{eq:ghz-x}
\end{equation}
for any $i\neq j$ and
\begin{equation}
\Delta^{2}(p_{1}+p_{2}+p_{3}+..p_{N})=Ne^{-2r}\label{eq:ghz-p}
\end{equation}
which is the special case given by equation (35) of \citep{vanLoock_PRA2003}
with $g^{(N)}=1$. Steering of system $i$ is detected on selecting
$u=x_{i}-x_{j}$ and $v=p_{1}+p_{2}+p_{3}+..p_{N}$ and observing
$\Delta u\Delta v<1$, as we see by examining the steering condition
Eq. (\ref{eq:epr-cond-1}). Clearly, this is obtained for the expressions
of $u$ and $v$, for large $r$. For smaller $r$, a different choice
of gain is optimal, to allow steering for all $r$.

We now give a more detailed analysis. For just two modes, the correlations
are identical to those given in IV.A, for the CV EPR state. Explicitly,
from Eqs. (\ref{eq:var1}) and (\ref{eq:varp}), we get $\Delta^{2}(x_{1}-\frac{1}{\sqrt{2}}x_{2}^{'})=\frac{3}{2}e^{-2r}$
and $\Delta^{2}(p_{1}+\sqrt{2}p_{2}^{'})=3e^{-2r}$ for $R_{1}=1/3$
and $T_{1}=2/3$. Following the same procedure, using Eq. (\ref{eq:bs2x,p}),
the moments in the tripartite case are
\begin{align}
\Delta^{2}(x_{1}-\frac{1}{2}(x_{2}+x_{3})) & =\frac{3}{2}e^{-2r}\nonumber \\
\Delta^{2}(p_{1}+p_{2}+p_{3}) & =3e^{-2r}\,.\label{eq:cvGHZ_tripartite}
\end{align}

\begin{figure}[H]
\begin{centering}
\includegraphics[width=0.93\columnwidth]{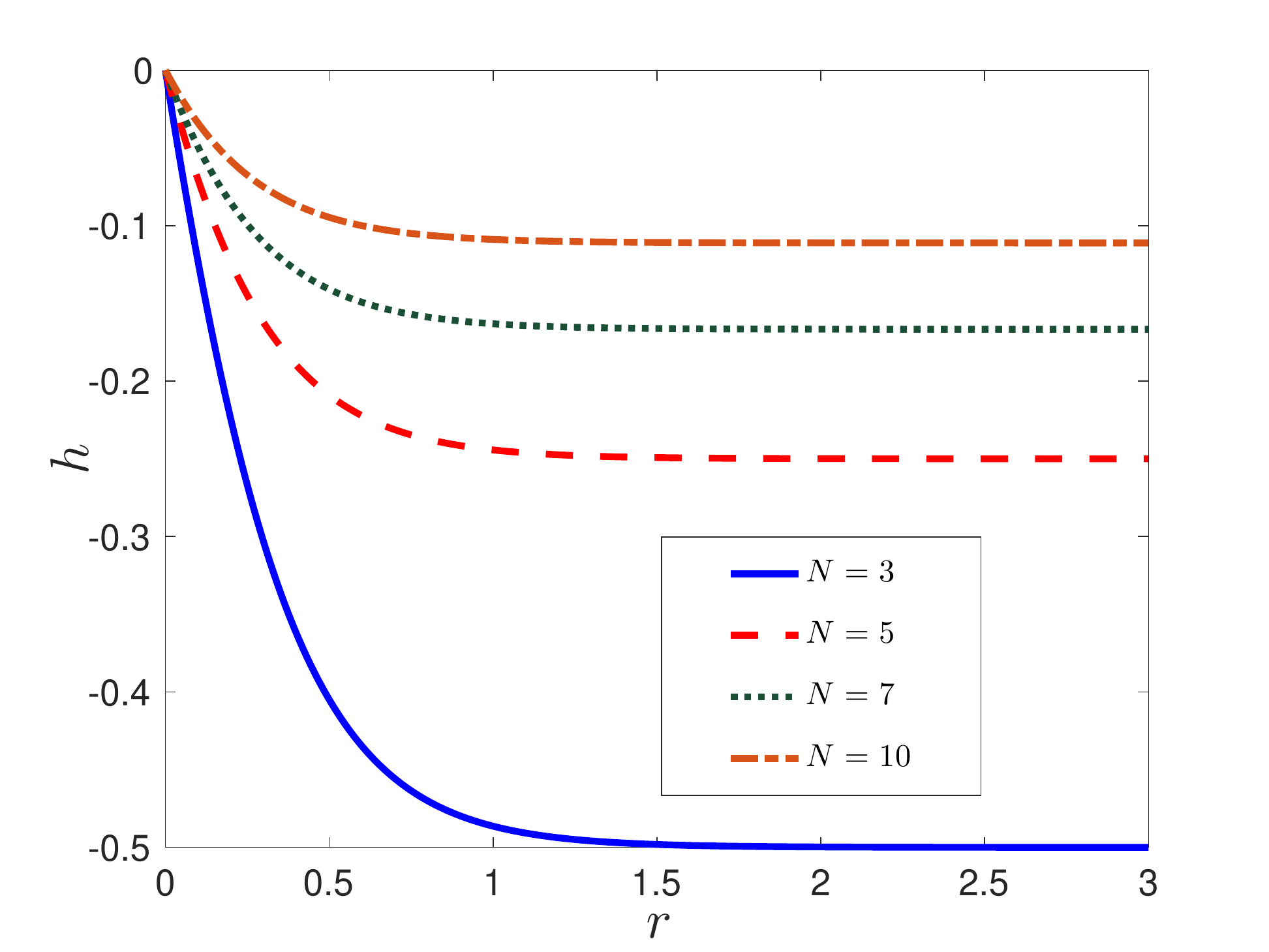}
\par\end{centering}
\caption{Optimal gains: The optimal gains $h$ that minimize $\Delta u\Delta v$
in Eq. (\ref{eq:var}) for CV GHZ states. The analytical expressions
for these gains are given by Eq. (\ref{eq:gains-N}). \label{fig:gains_CVGHZ-1}\textcolor{blue}{}}
\end{figure}

\begin{figure}[H]
\begin{centering}
\includegraphics[width=0.93\columnwidth]{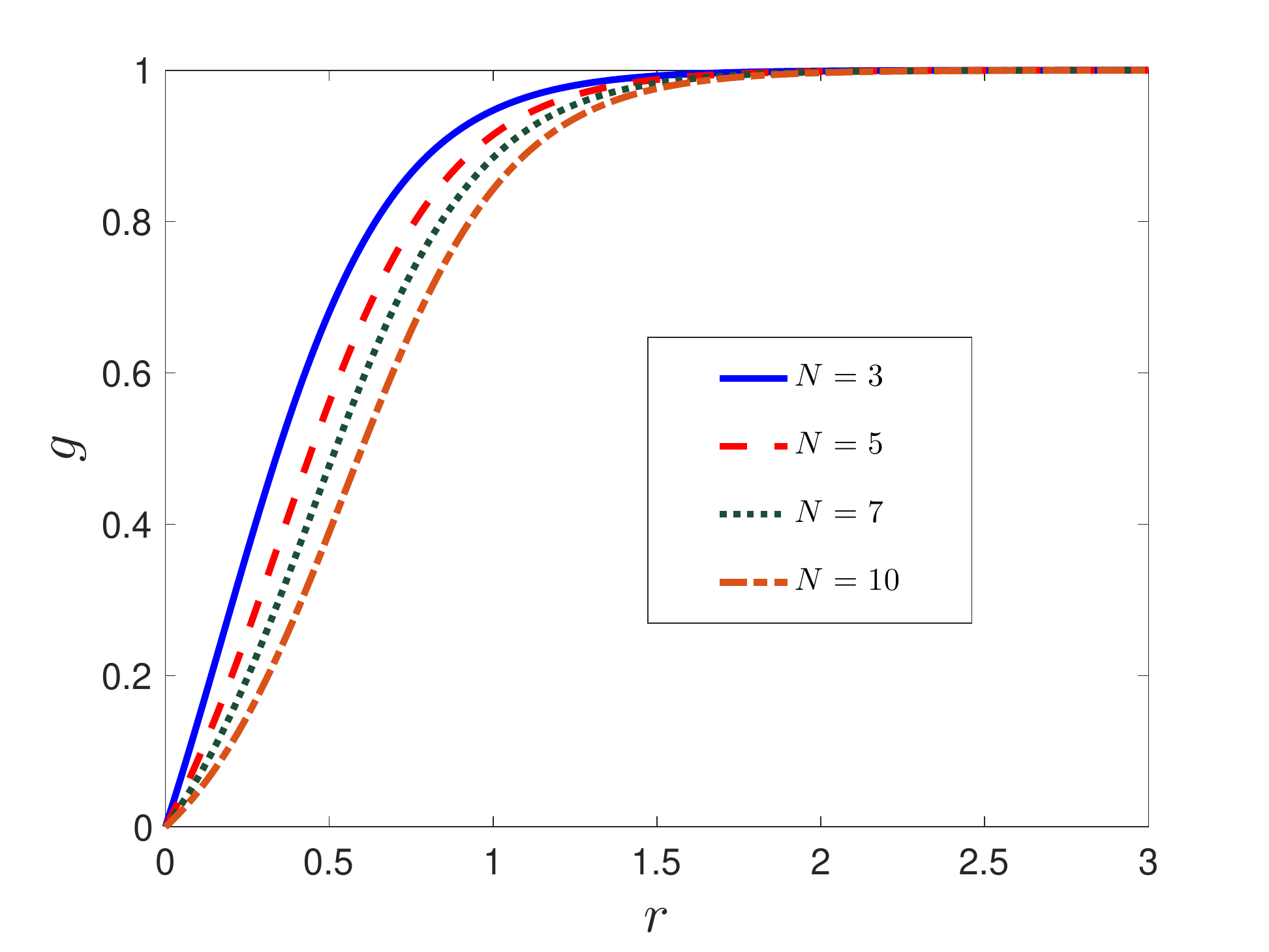}
\par\end{centering}
\caption{Optimal gains: The optimal gains $g$ that minimize $\Delta u\Delta v$
in (\ref{eq:var}) for CV GHZ states. The analytical expressions for
these gains are given by Eq. (\ref{eq:gains-N}). \label{fig:gains_CVGHZ}\textcolor{blue}{}}
\end{figure}

More generally, we allow different gains, and consider $\Delta^{2}\left[x_{1}+h\left(x_{2}+x_{3}\right)\right]$
and $\Delta^{2}\left[p_{1}+g\left(p_{2}+p_{3}\right)\right]$. Generalizing
to the $N$-partite CV GHZ state, there will be $N-1$ beam splitters
with reflectivities $R_{1}=1/N$, $R_{2}=1/\left(N-1\right)$, ...,
$R_{N-1}=1/2$.  The variances are
\begin{eqnarray}
\Delta^{2}(x_{1}+h(\sum_{j=2}^{N}x_{j})) & = & \frac{1}{N}\left[h\left(N-1\right)+1\right]^{2}\Delta^{2}x_{1}^{(in)}\nonumber \\
 &  & \ \ \ \ +\frac{\left(N-1\right)}{N}\left(h-1\right)^{2}\Delta^{2}x_{2}^{(in)}\nonumber \\
\Delta^{2}(p_{1}+g(\sum_{j=2}^{N}p_{j})) & = & \frac{1}{N}\left[g\left(N-1\right)+1\right]^{2}\Delta^{2}p_{1}^{(in)}\nonumber \\
 &  & \ \ \ \ +\frac{\left(N-1\right)}{N}\left(g-1\right)^{2}\Delta^{2}p_{2}^{(in)}\,\nonumber \\
\label{eq:var}
\end{eqnarray}
where $\Delta^{2}x_{1}^{(in)}=\Delta^{2}p_{2}^{(in)}=e^{2r}$ and
$\Delta^{2}x_{2}^{(in)}=\Delta^{2}p_{1}^{(in)}=e^{-2r}$, as provided
in equation (A5) of \citep{PhysRevA.90.062337}. The optimal gains
$g$ and $h$, on differentiation, are
\begin{align}
h & =-\frac{\Delta^{2}x_{1}^{(in)}-\Delta^{2}x_{2}^{(in)}}{\Delta^{2}x_{2}^{(in)}+\left(N-1\right)\Delta^{2}x_{1}^{(in)}}\nonumber \\
g & =-\frac{\Delta^{2}p_{1}^{(in)}-\Delta^{2}p_{2}^{(in)}}{\Delta^{2}p_{2}^{(in)}+\left(N-1\right)\Delta^{2}p_{1}^{(in)}}\,.\label{eq:gains-N}
\end{align}
For large $r$, the optimal values become $g\rightarrow1$ and $h\rightarrow-1/(N-1)$.
These optimal gains for different $N$ are plotted in Figures \ref{fig:gains_CVGHZ-1}
and \ref{fig:gains_CVGHZ} versus $r$.

The expressions in Eqs. (\ref{eq:var}) and (\ref{eq:gains-N}) give
\begin{align}
\Delta^{2}(x_{1}+h(x_{2}+...+x_{N})) & =\frac{N}{e^{-2r}+\left(N-1\right)e^{2r}}\nonumber \\
\Delta^{2}(p_{1}+g(p_{2}+...+p_{N})) & =\frac{N}{e^{2r}+\left(N-1\right)e^{-2r}}\thinspace.\label{eq:eqns-ghz}
\end{align}
Immediately, we see that the criterion $S_{N}=\Delta u\Delta v<\frac{1}{N-1}$
as given by Eq. (\ref{eq:cond-ent1}) and proved in \citep{PhysRevA.90.062337,vanLoock_PRA2003}
for $N$-partite entanglement is satisfied for large $r$.

\begin{figure}[H]
\begin{centering}
\includegraphics[width=0.93\columnwidth]{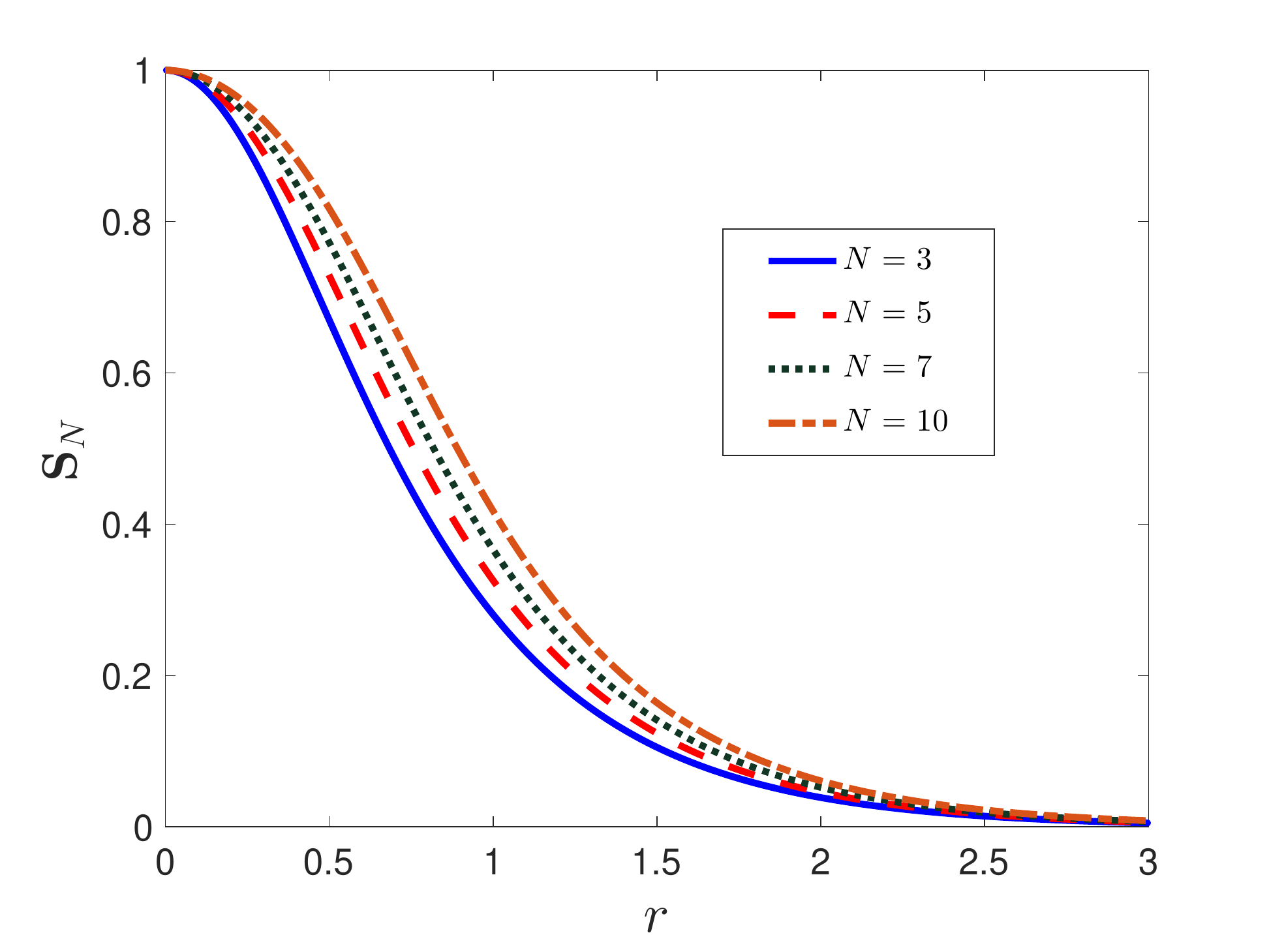}
\par\end{centering}
\caption{Genuine $N$-partite steering for the CV GHZ state: The value of $\mathbf{S}_{N}$
as a function of the squeezing parameter $r$, for CV GHZ states.
The analytical expression for $\mathbf{S}_{N}$ is given by Eq. (\ref{eq:s_n-equal-strenght}).
Steering of system $1$ is obtained when $S_{N}\equiv S_{1|\{2,..,N\}}<1$.
It is evident that genuine $N$-partite steering is obtained for sufficiently
large $r$, implying mutual steering between all subsystems, as given
by Criterion 1b. \label{fig:S_N_cvghz}}
\end{figure}

To examine the steering, we find that
\begin{align}
\mathbf{S}_{N} & \equiv S_{1|\{2,..N\}}\nonumber \\
 & =\frac{N}{\sqrt{N^{2}+4\left(N-1\right)\sinh^{2}2r}}\,.\label{eq:s_n-equal-strenght}
\end{align}
The analytical expression for $\mathbf{S}_{N}$ is plotted in Figure
\ref{fig:S_N_cvghz}. We see using the condition $S_{N}<1$ of Eq.
(\ref{eq:epr-cond-1}) that steering of system $1$ is possible for
all values of $r$. As $N$ becomes large, $\mathbf{S}_{N}\rightarrow\sqrt{N}/\left(2\sinh2r\right)$,
indicating genuine $N$-partite steering for sufficiently large $r$,
by Criterion 1b. The following condition is sufficient to reveal genuine
$N$-partite steering: 
\begin{eqnarray}
\mathbf{S}_{N}\equiv\Delta u_{1}\Delta v_{1} & < & \min\Bigl\{|gh|,|1-(N-2)|gh||\Bigl\}\thinspace.\nonumber \\
\label{eq:steering-cond-gen-2}
\end{eqnarray}
In the limit of large $r,$ this reduces to $\mathbf{S}_{N}<$$\frac{1}{N-1}$,
which is clearly satisfied for large $r$. 

\begin{figure}[H]
\begin{centering}
\includegraphics[width=0.93\columnwidth]{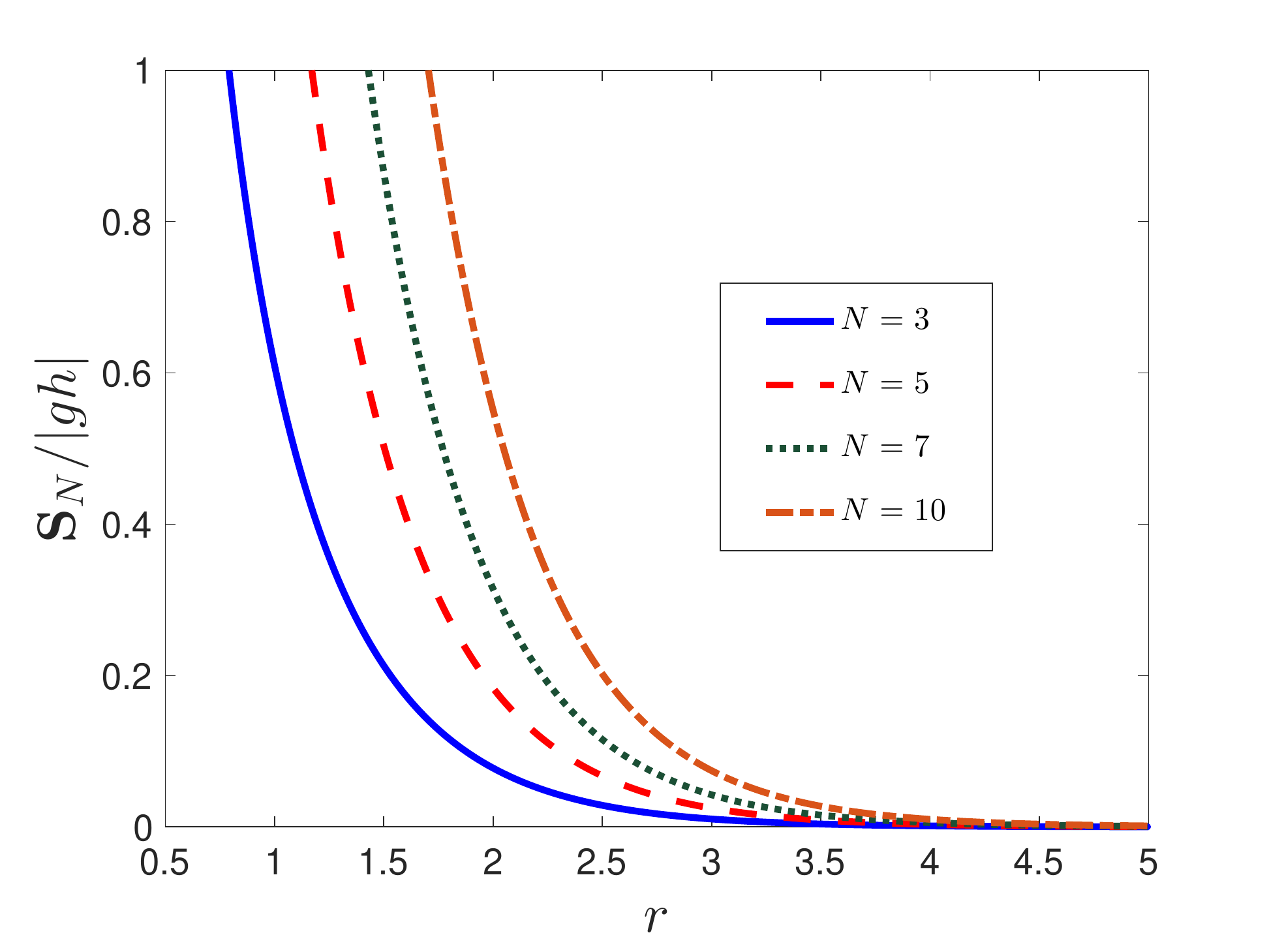}
\par\end{centering}
\caption{Genuine $N$-partite steering for the CV GHZ state: The value of $\mathbf{S}_{N}$
divided by the bound $|gh|$ as provided in Criterion \textit{\emph{1b}},
for CV GHZ states. The analytical expression for $\mathbf{S}_{N}$
and the optimal gains $g$ and $h$ are given by Eqs. (\ref{eq:s_n-equal-strenght})
and (\ref{eq:gains-N}) respectively. When the value is less than
$1$, there is genuine $N$-partite steering according to Criterion
\textit{\emph{1b}} given by Eq. (\ref{eq:steering-cond-gen-2}). \label{fig:S_N_cvghz-bound}}
\end{figure}
More generally, using the analytical expression for $\mathbf{S}_{N}$
and the optimal gains $g$ and $h$ given by Eqs. (\ref{eq:s_n-equal-strenght})
and (\ref{eq:gains-N}), we plot the value of $\mathbf{S}_{N}$ divided
by the bound provided in Criterion 1b, given by Eq. (\ref{eq:steering-cond-gen-2}).
As shown in Figure \ref{fig:S_N_cvghz-bound}, we see that genuine
$N$-partite steering is possible for large $N$. The minimum squeezing
parameter required to show this steering satisfies the inequality
$\sinh^{2}2r\geq(N^{2}\left(N-1\right)+N^{2}\sqrt{\left(N-1\right)^{2}+4)^{1/2}})/8$
in agreement with the results in Figure \ref{fig:S_N_cvghz-bound}.
Here, the CV GHZ state does not for a fixed $r$ give a smaller
value $\mathbf{S}_{N}$ than that obtained with the CV EPR state,
which has only two squeezed inputs. However, $\mathbf{S}_{N}$ is
an asymmetric parameter measuring steering of one mode only. The advantage
is that, unlike the CV EPR state, the GHZ state has symmetry with
respect to all modes.

\begin{figure}[H]
\begin{centering}
\includegraphics[width=0.93\columnwidth]{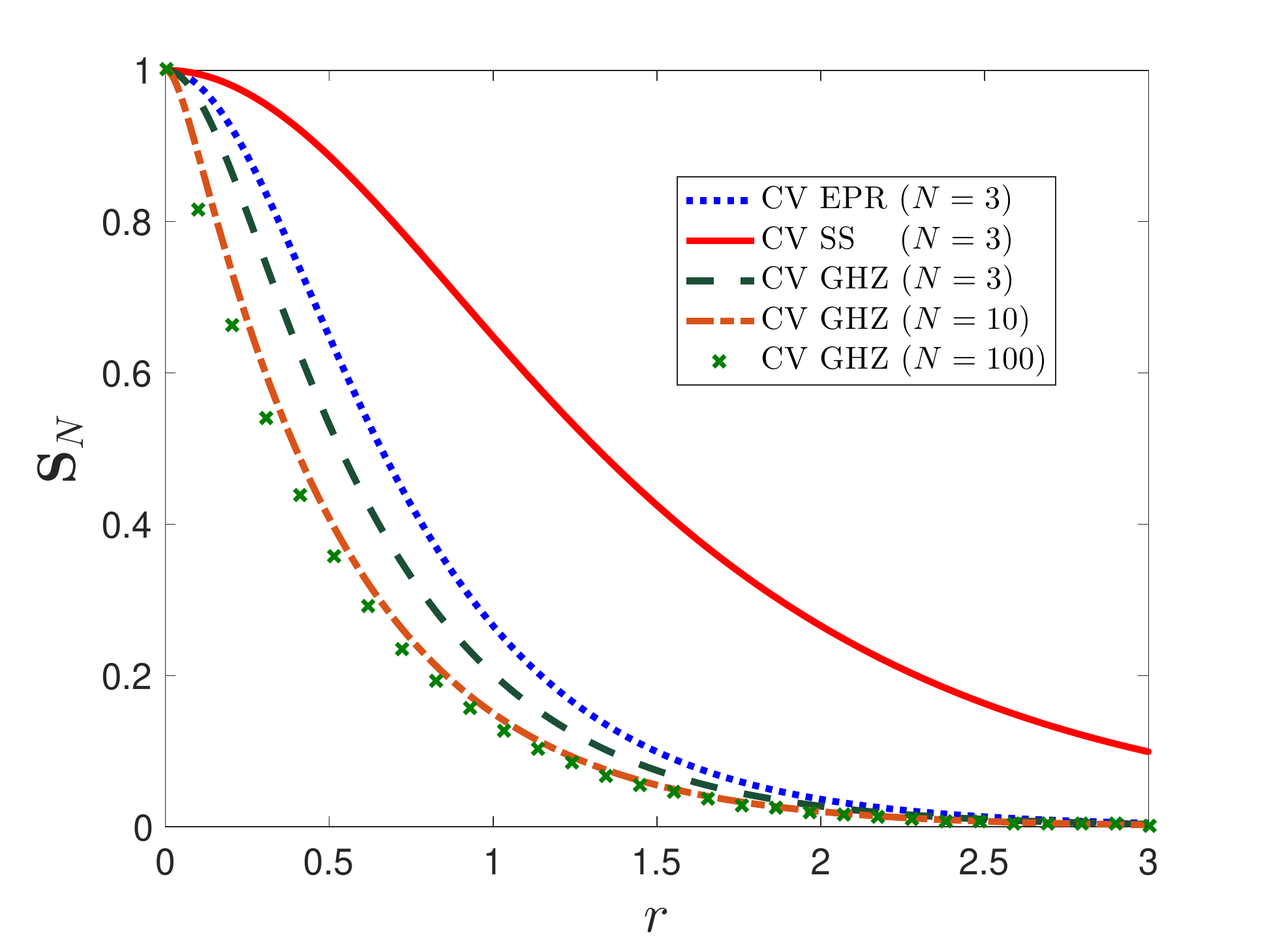}
\par\end{centering}
\caption{\textcolor{black}{Genuine $N$-partite steering for the CV GHZ state
with asymmetric squeezing strengths. The steering product $\mathbf{S}_{3}$
is also shown for CV EPR and CV SS states. The value of $\mathbf{S}_{N}$
is plotted for}\textcolor{red}{{} }the CV GHZ state with two squeezing
strengths, for $N=3$, $10$ and $100$. The two squeezing strengths
for the GHZ state are related by Eq. (\textcolor{black}{\ref{eq:squeezing_relation}).
In that case, $r$ as plotted corresponds to $r_{2}=r_{3}$. The values
of $\mathbf{S_{N}}\equiv S_{1|\{2,..,N\}}$ given by the green crosses
are very close to $e^{-2r}$, as predicted by the large $N$ limit
of $e^{-2r_{2}}$.} \textcolor{black}{Steering of system $1$ is obtained
when $\mathbf{S_{N}}<1$. }Genuine $N$-partite steering which implies
mutual steering of all subsystems is obtained for large $r$. \label{fig:S3_two_squeezing}\textcolor{blue}{}}
\end{figure}
On the other hand, the work of van Loock and Furusawa \citep{vanLoock_PRA2003}
reveals that asymmetric squeezing strengths can give stronger quantum
correlations for the CV GHZ state. The different squeezing strengths
are related by the expression (see equation (34) in \citep{vanLoock_PRA2003}):
\begin{align}
e^{\pm2r_{1}} & =\left(N-1\right)\sinh2r_{2}\Bigl(\sqrt{1+\frac{1}{\left(N-1\right)^{2}\sinh^{2}2r_{2}}}\pm1\Bigl)\label{eq:squeezing_relation}
\end{align}
where $r_{1}$ is the input squeeze parameter for mode $1$ and $r_{j}=r_{2}$
for $j\neq1$. We confirm this leads to a steering value $\mathbf{S}_{3}$
improved over the CV EPR case, for the same average squeezing value,
and relative to the squeezing parameter $r_{2}$ of the second and
third modes. For general $N$, following as above, we use
\begin{align}
\Delta^{2}\left[x_{1}+h\left(x_{2}+...+x_{N}\right)\right] & =\frac{N}{e^{-2r_{2}}+\left(N-1\right)e^{2r_{1}}}\nonumber \\
\Delta^{2}\left[p_{1}+g\left(p_{2}+...+p_{N}\right)\right] & =\frac{N}{e^{2r_{2}}+\left(N-1\right)e^{-2r_{1}}}
\end{align}
where the $g$ and $h$ are given by Eq. (116). This leads to 

\begin{align}
\mathbf{S}_{N} & =\frac{N}{\sqrt{\left(N-1\right)^{2}\sinh^{2}2r_{2}+1}+\left(N-1\right)\cosh2r_{2}}\,.\nonumber \\
\label{eq:S_n_two_squeezing_strengths}
\end{align}
For large $N$, the limit becomes $\mathbf{S}_{N}\rightarrow e^{-2r_{2}}$,
which is improved on the CV EPR and CV SS states. The CV GHZ states
require a single very strong squeezing at the first input $r_{1}$,
but for less squeezing in the $N-1$ inputs, a much higher steering
as measured by $\mathbf{S}_{N}$ is possible.

In Fig. 13, $\mathbf{S}_{N}$ is plotted for the CV GHZ state with
two squeezing strengths, for $N=3,10$ and $100$. Also plotted are
$\mathbf{S}_{3}$ for CV EPR and CV SS states.

\section{genuine tripartite steerable states}

We now consider in more detail the steering for the three types
of states considered in the last section, where $N=3$. Examples for
$N=4$ are given in the Supplemental Materials. It is useful to first
summarize criteria that detect steering across \emph{particular} bipartitions.
Here, we consider $u=h_{1}x_{1}+h_{2}x_{2}+h_{3}x_{3}$ and $v=g_{1}p_{1}+g_{2}p_{2}+g_{3}p_{3}$.
Using the Lemma 1, we detect steering of mode $k$ by $lm$ if ($k\neq l\neq m$)
$S_{k|lm}<1$ where 
\begin{align}
S_{k|lm} & =\frac{\Delta(h_{k}x_{k}+h_{l}x_{l}+h_{m}x_{m})\Delta(g_{k}p_{k}+g_{l}p_{l}+g_{m}p_{m})}{|g_{k}h_{k}|}\thinspace.\label{eq:S_klm}
\end{align}
Thus we consider minimizing $\Delta(x_{k}+\widetilde{h}_{l}x_{l}+\widetilde{h}_{m}x_{m})\Delta(p_{k}+\widetilde{g}_{l}p_{l}+\widetilde{g}_{m}p_{m})$
where $\tilde{g}_{i}=g_{i}/g_{k}$ and $\tilde{h}_{i}=h_{i}/h_{k}$.
This example is relevant for the application of secret sharing
where the collaborators $l$ and $m$ cannot be trusted. In order
to detect the steering across the bipartitions, one can select different
optimal choices of $\widetilde{g}_{i}$ and $\widetilde{h}_{i}$ for
each bipartition, $k-lm$. Similarly, from Lemma 1, we detect steering
of the combined systems $l$ and $m$, if $S_{lm|k}<1$ where 
\begin{align}
S_{lm|k} & =\frac{\Delta(h_{k}x_{k}+h_{l}x_{l}+h_{m}x_{m})\Delta(g_{k}p_{k}+g_{l}p_{l}+g_{m}p_{m})}{\left|g_{l}h_{l}+g_{m}h_{m}\right|}\,.\label{eq:S_lmk}
\end{align}
We note the generalization of the definition here of the steering
product given in Eq. (\ref{eq:eprS-3}) where there is the steering
of two systems. We also note that in this case, as the variance product
$S_{k|lm}$ goes to zero, so too will $S_{lm|k}$.

\subsection{CV GHZ and cluster states}

\subsubsection{Genuine tripartite steering using a single inequality}

The inequality of Criterion 1 is useful to detect the genuine tripartite
steering of the CV GHZ state. For a single choice of $g_{i}$ and
$h_{i}$ defining $u=h_{1}x_{1}+h_{2}x_{2}+h_{3}x_{3}$ and $v=g_{1}p_{1}+g_{2}p_{2}+g_{3}p_{3}$,
we wish to violate Eq. (\ref{eq:criterion-1-4-2}). Defining $\mathbf{S}_{3}=\Delta u\Delta v$,
we test each of the six inequalities $\mathbf{S}_{3}<\mathcal{B}$
provided by considering as the right side of the inequality, where
\begin{eqnarray}
\mathcal{B} & \in & \{\left|g_{1}h_{1}\right|,{\color{red}{\color{black}\left|g_{2}h_{2}+g_{3}h_{3}\right|,\left|g_{2}h_{2}\right|,}}\nonumber \\
 &  & \left|g_{1}h_{1}+g_{3}h_{3}\right|,\left|g_{3}h_{3}\right|,{\color{black}{\color{black}{\color{red}{\color{black}\left|g_{1}h_{1}+g_{2}h_{2}\right|\}}}}\thinspace.}\label{eq:bounds}
\end{eqnarray}
If each inequality $S_{3}<\mathcal{B}$ is satisfied, we indicate
steering across one of the bipartitions in a certain direction. Genuine
tripartite steering is confirmed if  \emph{all} six possibilities
are verified simultaneously, using a single set of gains e.g. from
the single inequality Eq. (\ref{eq:criterion-1-4-2}).
\begin{figure}[H]
\begin{centering}
\includegraphics[width=0.93\columnwidth]{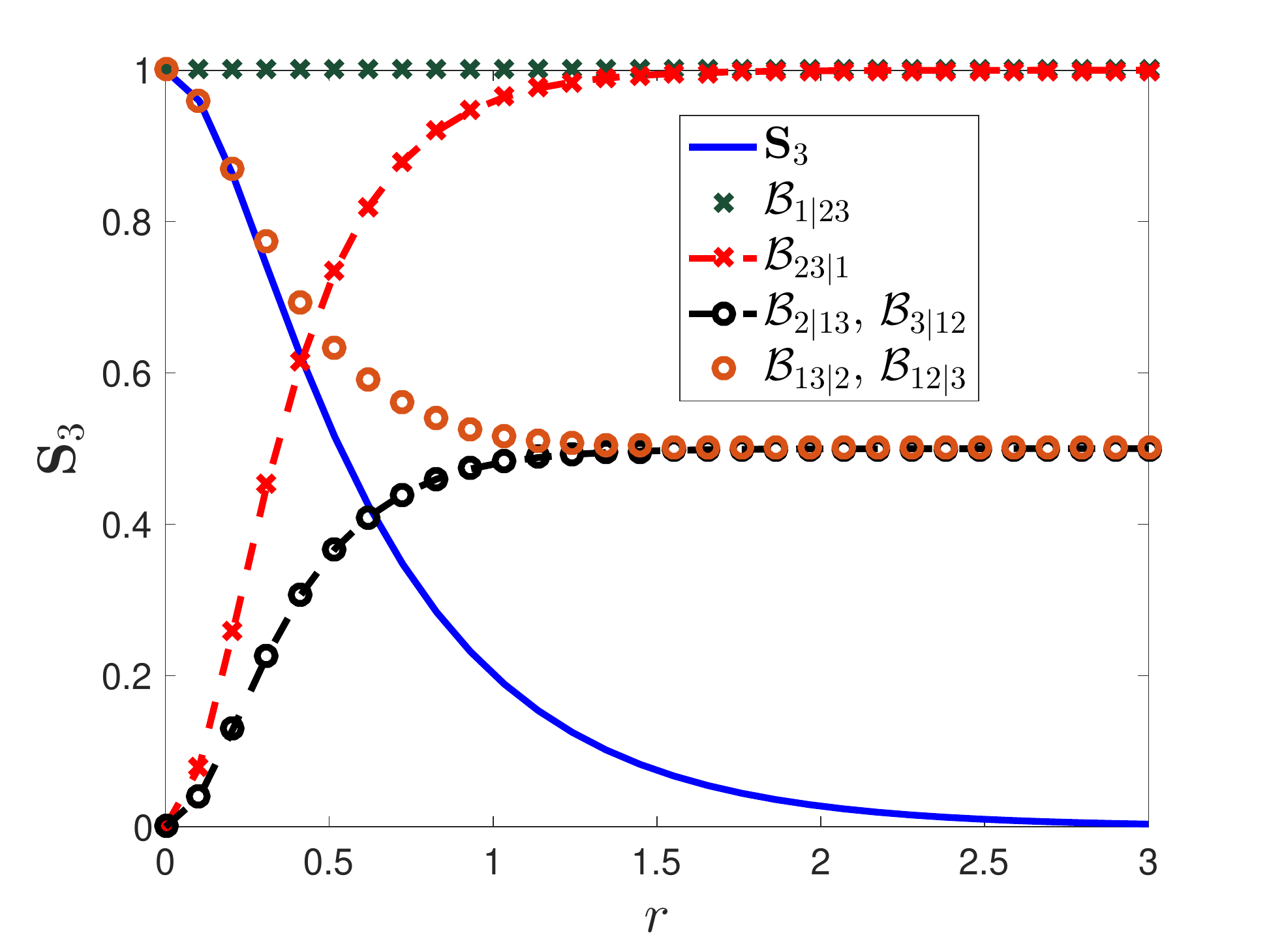}
\par\end{centering}
\caption{The value of $\mathbf{S}_{3}\equiv S_{1|2}=\Delta u\Delta v$\textcolor{red}{{}
}as a function of the squeezing parameter $r$, for the CV GHZ state,
for $g_{1}=h_{1}=1$, and $g_{2}=g_{3}=g$ and $h_{2}=h_{3}=h$. The
results here correspond to the case where two squeezing strengths
are used, with $r_{2}=r$ and $r_{1}$ is related to $r_{2}$ by the
relation Eq. (\ref{eq:squeezing_relation}). The gains $g$ and $h$
as a function of $r$ are given by the analytical expressions in Eq.
(\ref{eq:gains-N}). The blue solid line corresponds to the value
of $\mathbf{S}_{3}$, while other lines correspond to different bounds
on the right side of the inequality Eq. (\ref{eq:inequality_CVGHZ}).
The gains are optimized to minimize $\mathbf{S}_{3}$, and hence to
optimize the observation of the steering of system $1$. \label{fig:steering_CVGHZ-1}\textcolor{blue}{}}
\end{figure}

We select $g_{1}=h_{1}=1$, and $g_{2}=g_{3}=g$ and $h_{2}=h_{3}=h$
and optimize for $g$ and $h$ by minimizing $\mathbf{S}_{3}$ with
respect to these gains. The Criterion 1 given by Eq. (\ref{eq:criterion-1-4-2})
is then reduced to the violation of
\begin{align}
\mathbf{S}_{3} & \geq\text{min}\left\{ \mathcal{B}_{1|23},\mathcal{B}_{23|1},\mathcal{B}_{2|13},\mathcal{B}_{13|2},\mathcal{B}_{3|12},\mathcal{B}_{12|3}\right\} \,,\label{eq:inequality_CVGHZ}
\end{align}
where in this particular case $\mathcal{B}_{1|23}=1$, $\mathcal{B}_{23|1}=2\left|gh\right|$,
$\mathcal{B}_{2|13}=\mathcal{B}_{3|12}=\left|gh\right|$ and $\mathcal{B}_{12|3}=\mathcal{B}_{13|2}=\left|1+gh\right|$.
By differentiating $\mathbf{S}_{3}$ with respect to $g$ and $h$,
we select the optimal gains given by the analytical expressions Eq.
(\ref{eq:gains-N}). The numerical values of optimal gains as a function
of the squeezing parameter $r$ are tabulated in Table \ref{tab:fixed_gain_CVGHZ}
of the Appendix. The value of $\mathbf{S}_{3}$ as a function of the
squeezing parameter $r$ is given by Eq. (\ref{eq:s_n-equal-strenght}).
We calculate that genuine tripartite steering is detectable using
Criterion 1 when $r$ is sufficiently large ($r>0.8$). The choice
of gains used here may not be the optimal to observe genuine tripartite
steering for a fixed $r$, but nonetheless ensures steering of system
$1$ (or system $1$ combined with $2$ or $3$) for all $r$ values.

In Figure \ref{fig:steering_CVGHZ-1}, we plot $\mathbf{S}_{3}$ for
a CV GHZ state with two squeezing strengths, as given by Eq. (\ref{eq:S_n_two_squeezing_strengths}).
We also plot the different bounds on the right side of the inequality
Eq. (\ref{eq:inequality_CVGHZ}). The values of $g$ and $h$ are
given by Table \ref{tab:fixed_gain_CVGHZ-1-1}. We take $r_{2}=r_{3}=r$
and obtain a minimum $r_{2}=r_{3}$ of $0.633$ to observe steering
in all bipartitions. This corresponds to $r_{1}=0.95$ using the relation
Eq. (\ref{eq:squeezing_relation}).

\subsubsection{Full tripartite steering inseparability}

Next, we minimize the quantities $S_{k|lm}$ and $S_{lm|k}$ as defined
in Eqs. (\ref{eq:S_klm}) and (\ref{eq:S_lmk}), \emph{for each bipartition}.
For the steering of mode $1$ by $2$ and $3$, we minimize the quantity
$S_{1|23}$ by optimizing the gains $\tilde{h}_{2}$, $\tilde{h}_{3}$,
$\tilde{g}_{2}$ and $\tilde{g}_{3}$. On the other hand, for the
steering of modes $23$ by $1$, we minimize the quantity $S_{23|1}$,
with an independent choice of gains. For the steering of mode $2$
by $1$ and $3$, we minimize $S_{2|13}$ by optimizing the gains
$\tilde{h}_{1}$, $\tilde{h}_{3}$, $\tilde{g}_{1}$ and $\tilde{g}_{3}$.
The quantity $S_{13|2}$ is independently minimized for the steering
of modes $1$ and $3$, by $2$. We proceed similarly, for $S_{3|12}$
and $S_{12|3}$.

\begin{figure}[H]
\begin{centering}
\par\end{centering}
\begin{centering}
\includegraphics[width=0.93\columnwidth]{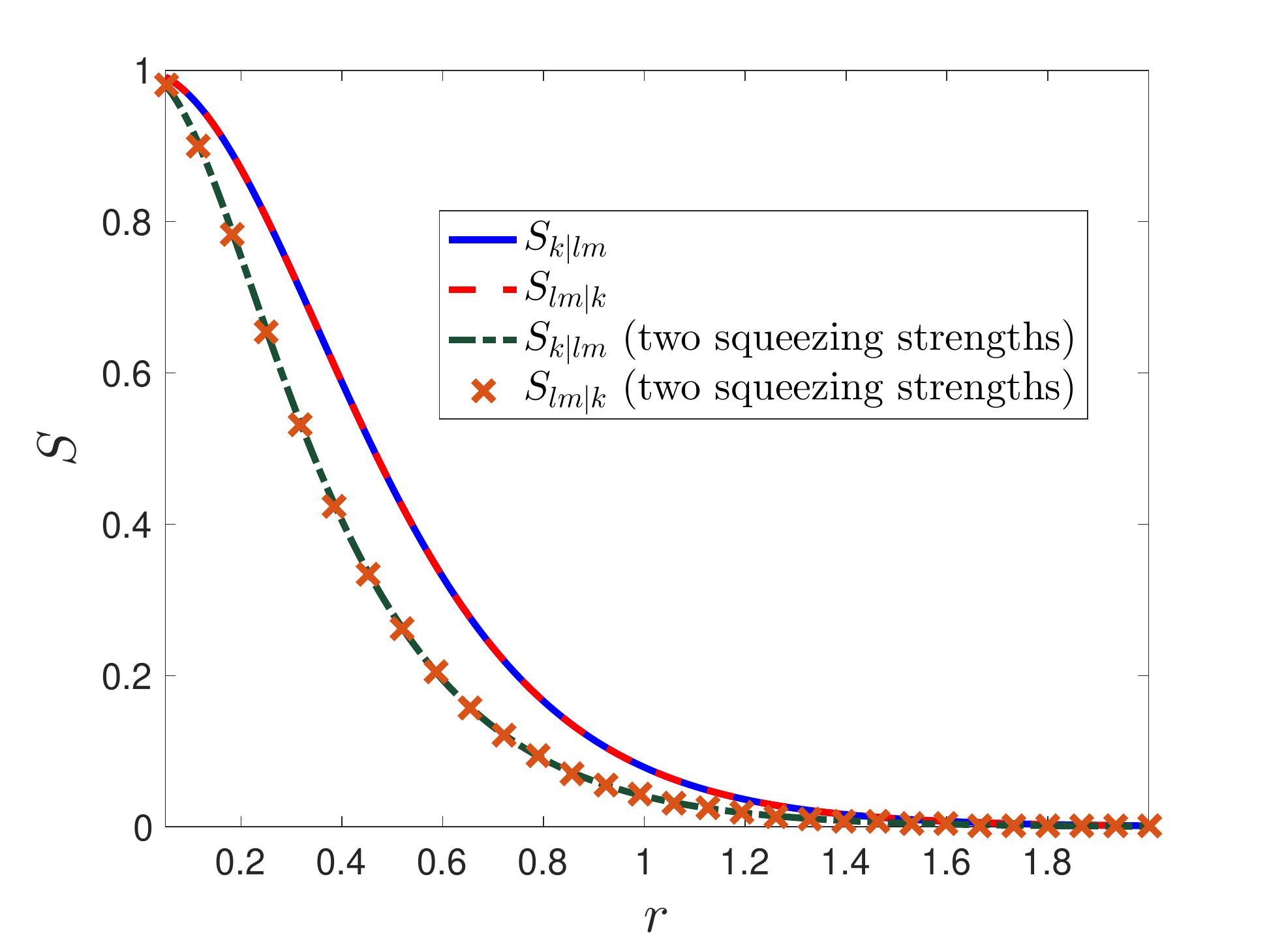}
\par\end{centering}
\caption{\textcolor{black}{The steering across each bipartition as a function
of squeezing parameter $r$, for the CV tripartite GHZ state}. Here
$k\protect\neq l\protect\neq m$. There is symmetry between all three
systems and the plots hold for each $k=1,2,$ and 3. The values of
$S=\Delta u\Delta v$ for $S_{k|lm}$ and $S_{lm|k}$ are determined
by the choice of gains in the Tables \ref{tab:gains_GHZ_1-23} and
\ref{tab:gains_GHZ_23-1}. Here, $S_{k|lm}$ and $S_{lm|k}$ coincide.
The gains for the case with two squeezing strengths are tabulated
in Tables \ref{tab:gains_GHZ_1-23-1} and \ref{tab:gains_GHZ_23-1-1}.\textcolor{red}{\label{fig:Steering_GHZ_numerical}}\textcolor{blue}{}}
\end{figure}
The gains are optimized independently in each case. All the optimal
gains are numerically computed using the \textit{fminsearch }function
in Matlab. The maximum number of iterations that calls the \textit{fminsearch}
function is chosen to be $10^{6}$. The tolerance of this Matlab function
is chosen to be $10^{-6}$, where no further iteration is taken if
the quantity to be minimized is smaller than $10^{-6}$ from one iteration
to the next. The gains are given in Tables \ref{tab:gains_GHZ_1-23}
and \ref{tab:gains_GHZ_23-1} of the Appendix.

Figure \ref{fig:Steering_GHZ_numerical} shows the results for the
steering across the bipartitions for the CV GHZ state with a single
squeezing strength $r$. For all $r$, two-way steering can be detected
across each bipartition, using the relevant inequalities with the
gains given in the Tables \ref{tab:gains_GHZ_1-23} and \ref{tab:gains_GHZ_23-1}.
For large $r$, the inferences improve with the correct choice of
gains, and all the relevant variances become zero. This gives a method
to confirm full tripartite two-way steering inseparability. The results
for asymmetric squeezing strengths $r_{1}$ and $r_{2}$ are also
given.

\subsubsection{Behavior in the highly squeezed limit}

Most interesting is the limit of large $r$ where the correlations
become ideal, implying zero variances, so that $S_{k|lm}\rightarrow0$
and $S_{lm|k}\rightarrow0$. For the standard CV GHZ state \citep{vanloock_PRA2001}
and for the CV GHZ state with asymmetric squeezing strengths \citep{vanLoock_PRA2003},
the optimal gain coefficients (refer Tables \ref{tab:gains_GHZ_1-23},
\ref{tab:gains_GHZ_23-1}, \ref{tab:gains_GHZ_1-23-1} and \ref{tab:gains_GHZ_23-1-1}
in the Appendix) are such that the steering criterion, for the steering
of system $k$, becomes
\begin{equation}
\Delta\left[x_{k}-\frac{\left(x_{l}+x_{m}\right)}{2}\right]\Delta\left[p_{k}+p_{l}+p_{m}\right]<1\label{eq:oldstineq-1}
\end{equation}
for each $k=1,2,3$ (recalling $k\neq l\ne m$, where $k,l,m\in\{1,2,3\}$).
For the standard CV GHZ state, we may also use $\Delta\left[x_{k}-x_{l}\right]\Delta\left[p_{k}+p_{l}+p_{m}\right]<1$,
or $\Delta\left[x_{k}-x_{m}\right]\Delta\left[p_{k}+p_{l}+p_{m}\right]<1$
\citep{vanloock_PRA2001}. For large $r$, the steering of system
$lm$ by $k$ is optimized by the same choice of gains. Noting that
in this limit, $\left|g_{l}h_{l}+g_{m}h_{m}\right|=1$, we see that
in fact the same inequality Eq. (\ref{eq:oldstineq-1}) detects steering
of $lm$ by $k$, and therefore detects two-way steering across the
bipartition $k-lm$.

\subsubsection{Genuine tripartite steering in a tripartite cluster state \label{subsec:steering_CV_cluster}}

We also compare with the results presented by Wang \emph{et al}. \citep{Wang_PRA2019}.
Here, the authors fixed $R_{1}=2/3$ and varied $R_{2}$ to optimize
the Gaussian steering parameter $\mathcal{G}^{A\rightarrow B}$ \citep{Kogias_PRL2015}.
In particular, their analysis for $R_{1}=2/3$ and $R_{2}=1/2$ corresponds
to that for a tripartite unweighted cluster state. This is of interest
here, as this corresponds to where no two modes can steer one another,
but where steering requires all three modes, in line with the notion
of genuine tripartite steering. They confirm full tripartite steering
inseparability (as we define it) for this state, based on the assumption
of a Gaussian state, using the Gaussian parameter $\mathcal{G}^{A\rightarrow B}$.
Here, we give a method sufficient to confirm tripartite steering inseparability
and genuine tripartite steering, \emph{without} the assumption of
Gaussian states. A summary of the correlations for this state is given
in the Supplemental Material. 

Before considering genuine tripartite steering for the cluster state,
we investigate the steering across all possible bipartitions separately.
For the bipartition $1-23$, in our approach, we consider
\begin{eqnarray}
u_{1}' & = & h'_{1,1}x_{1}-h'_{1,2}p_{2}+h'_{1,3}x_{3}\nonumber \\
v_{1}' & = & g'_{1,1}p_{1}+g'_{1,2}x_{2}+g'_{1,3}p_{3}\thinspace.\label{eq:uv-1-1-2-1}
\end{eqnarray}
As above, we analyse the steering for different bipartitions by
considering the quantity $S_{1P}'\equiv\Delta u_{1}'\Delta v_{1}'$.
Full details are given in the Supplemental Material. We obtain
two-way steering along this bipartition if $S_{1P}'<\mathcal{B}_{1|23}$
and $S_{1P}'<\mathcal{B}_{23|1}$ where the bounds are $\mathcal{B}_{1|23}=|g'_{1,1}h'_{1,1}|$
and $\mathcal{B}_{23|1}=\left|g'_{1,2}h'_{1,2}+g'_{1,3}h'_{1,3}\right|$.
These conditions become $S_{1|23}<1$ and $S_{23|1}<1$, on defining
the steering parameters as $S_{1|23}=S_{1P}'/\mathcal{B}_{1|23}$
and $S_{23|1}=S_{1P}'/\mathcal{B}_{23|1}$.
\begin{figure}[H]
\begin{centering}
\par\end{centering}
\begin{centering}
\includegraphics[width=0.93\columnwidth]{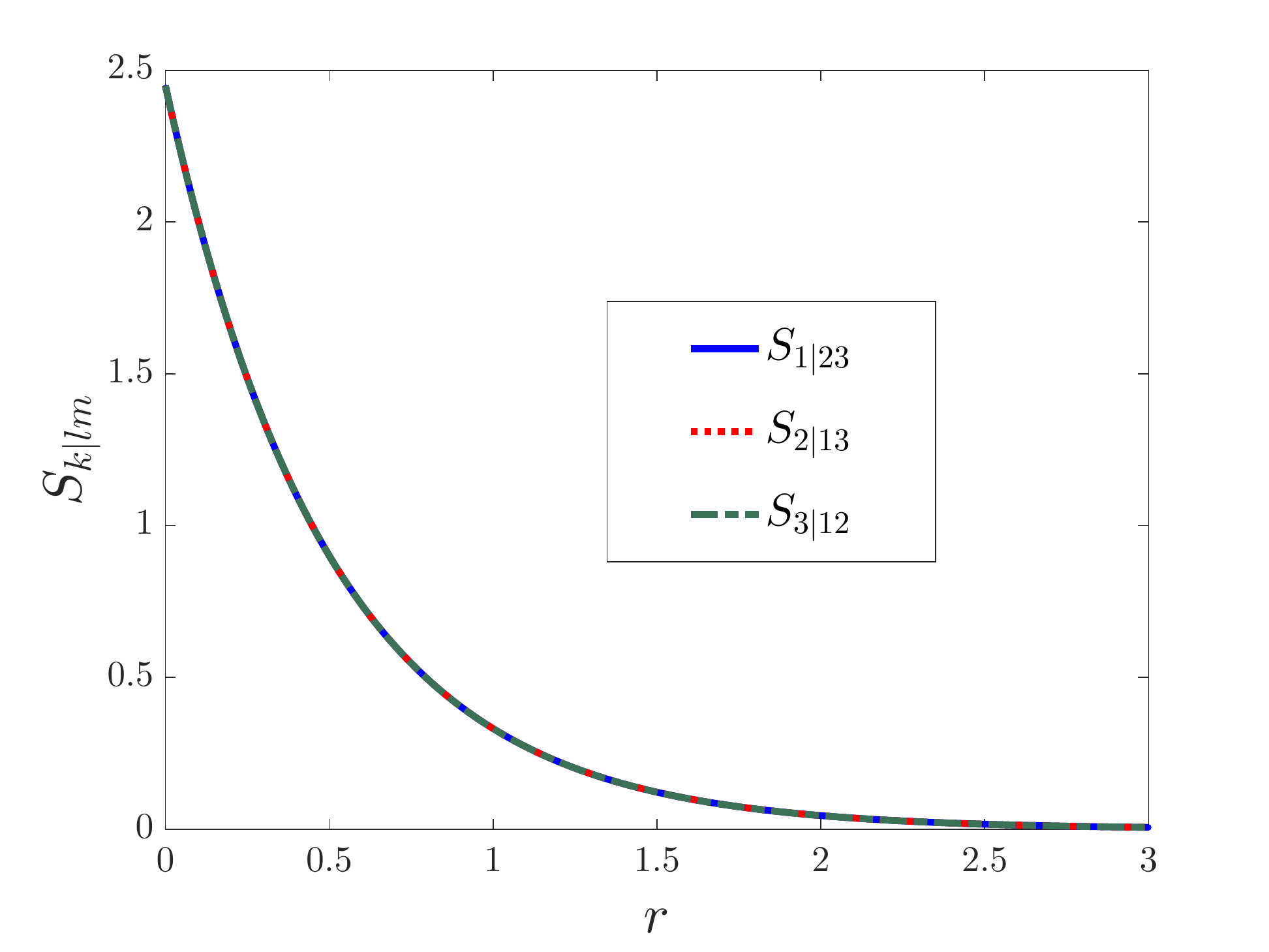}
\par\end{centering}
\caption{Steering parameter $S_{k|lm}$ as a function of the squeezing parameter
$r$, for the bipartitions of the cluster state considered by Wang
\emph{et al}. \citep{Wang_PRA2019}. Here, $R_{2}=1/2$ and $R_{1}=2/3$.
Steering occurs when $S_{k|lm}<1$. Here, all three lines conincide,
as expected from symmetry. There is two-way steering across all possible
bipartitions for $r\protect\geq0.44$. This occurs when $S_{k|lm}$
is below $1$. \textcolor{blue}{\label{fig:steering_bipartition_cluster-change-squeezing}}\textcolor{red}{}\textcolor{blue}{}}
\end{figure}

It has been shown in the work of Wang \emph{et al}. \citep{Wang_PRA2019}
that a choice of $h'_{1,1}=g'_{1,2}=-\sqrt{\left(1-R_{1}\right)}/\sqrt{R_{1}R_{2}}=-1$,
$h'_{1,3}=-\sqrt{\left(1-R_{2}\right)}/\sqrt{R_{2}}=-1$, $g'_{1,1}=-h'_{1,2}=1$
and $g'_{1,3}=0$ will lead to $S_{1P}'\rightarrow0$ for a large
squeezing parameter $r$. Here, we have taken $R_{2}=1/2$ and $R_{1}=2/3$.
These gains imply that $\text{min}\left\{ |g'_{1,1}h'_{1,1}|,\left|g'_{1,2}h'_{1,2}+g'_{1,3}h'_{1,3}\right|\right\} $
is $1$. The analytical expression for $S_{1P}'$ is $S_{1P}'=\sqrt{6}e^{-2r}\,$.
For large $r$, we see that the steering inequalities $S_{1P}'<\mathcal{B}_{1|23}$
and $S_{1P}'<\mathcal{B}_{23|1}$ are both satisfied for this choice
of gains. Both bounds $\mathcal{B}_{1|23}$ and $\mathcal{B}_{23|1}$
are $1$ for this choice of gains.

Similarly, there is two-way steering along the bipartition $2-13$
if $S_{2P}'<\mathcal{B}_{2|13}$ and $S_{2P}'<\mathcal{B}_{13|2}$.
Here, $S_{2P}'=\Delta u_{2}'\Delta v_{2}'$, where
\begin{eqnarray}
u_{2}' & = & h'_{2,1}x_{1}-h'_{2,2}p_{2}+h'_{2,3}x_{3}\nonumber \\
v_{2}' & = & g'_{2,1}p_{1}+g'_{2,2}x_{2}+g'_{2,3}p_{3}\thinspace.\label{eq:uv-1-1-2-1-1}
\end{eqnarray}
With the choice of $h'_{2,1}=g'_{2,2}=-\sqrt{\left(1-R_{1}\right)}/\sqrt{R_{1}R_{2}}=-1$,
$h'_{2,3}=-\sqrt{\left(1-R_{2}\right)}/\sqrt{R_{2}}=-1$, $g'_{2,1}=-h'_{2,2}=1$
and $g'_{2,3}=0$, $S_{2P}'=\sqrt{6}e^{-2r}\rightarrow0$ for a large
squeezing parameter $r$. These gains imply that $\text{min}\left\{ |g'_{2,2}h'_{2,2}|,\left|g'_{2,1}h'_{2,1}+g'_{2,3}h'_{2,3}\right|\right\} $
is $1$. We note that $S_{1P}'$ and $S_{2P}'$ have the same analytical
expression. The steering conditions for this bipartition become
$S_{2|13}<1$ and $S_{13|2}<1$, on defining $S_{2|13}=S'_{2P}/\mathcal{B}_{2|13}$
and $S_{13|2}=S_{2P}'/\mathcal{B}_{13|2}$ where $\mathcal{B}_{2|13}=|g'_{2,2}h'_{2,2}|$
and $\mathcal{B}_{13|2}=\left|g'_{2,1}h'_{2,1}+g'_{2,3}h'_{2,3}\right|$.
Both bounds are $1$ for this choice of gains.

Finally, in order to demonstrate steering $12\rightarrow3$ and $3\rightarrow12$,
the inequalities $S_{3P}'<\mathcal{B}_{3|12}$ and $S'_{3P}<\mathcal{B}_{12|3}$
are used, where $S_{3P}'=\Delta u_{3}'\Delta v_{3}'$ with
\begin{align}
u_{3}' & =h'_{3,1}x_{1}-h'_{3,2}p_{2}+h'_{3,3}x_{3}\nonumber \\
v_{3}' & =g'_{3,1}p_{1}+g'_{3,2}x_{2}+g'_{3,3}p_{3}\,,\label{eq:uv-1-1-2-2-2}
\end{align}
and $\mathcal{B}_{3|12}=|g'_{3,3}h'_{3,3}|$ and $\mathcal{B}_{12|3}=\left|g'_{3,1}h'_{3,1}+g'_{3,2}h'_{3,2}\right|$.
With the gains $h'_{3,1}=-\sqrt{\left(1-R_{1}\right)}/\sqrt{R_{1}R_{2}}=-1$,
$g'_{3,2}=h'_{3,3}=-\sqrt{\left(1-R_{2}\right)}/\sqrt{R_{2}}=-1$,
$-h'_{3,2}=g'_{3,3}=1$ and $g'_{3,1}=0$ can be used, the analytical
expression for $S_{3P}'$ is $S'_{3P}=\sqrt{6}e^{-2r}$. Also, we
find $\min\left\{ |g'_{3,3}h'_{3,3}|,\left|g'_{3,1}h'_{3,1}+g'_{3,2}h'_{3,2}\right|\right\} =1$.
The steering inequalities can be expressed as $S_{3|12}<1$ and $S_{12|3}<1$
where we define $S_{3|12}=S_{3P}'/\mathcal{B}_{3|12}$ and $S_{12|3}=S_{3P}'/\mathcal{B}_{12|3}$.
Both bounds are $1$ for this choice of gains.
\begin{figure}[H]
\begin{centering}
\par\end{centering}
\begin{centering}
\includegraphics[width=0.93\columnwidth]{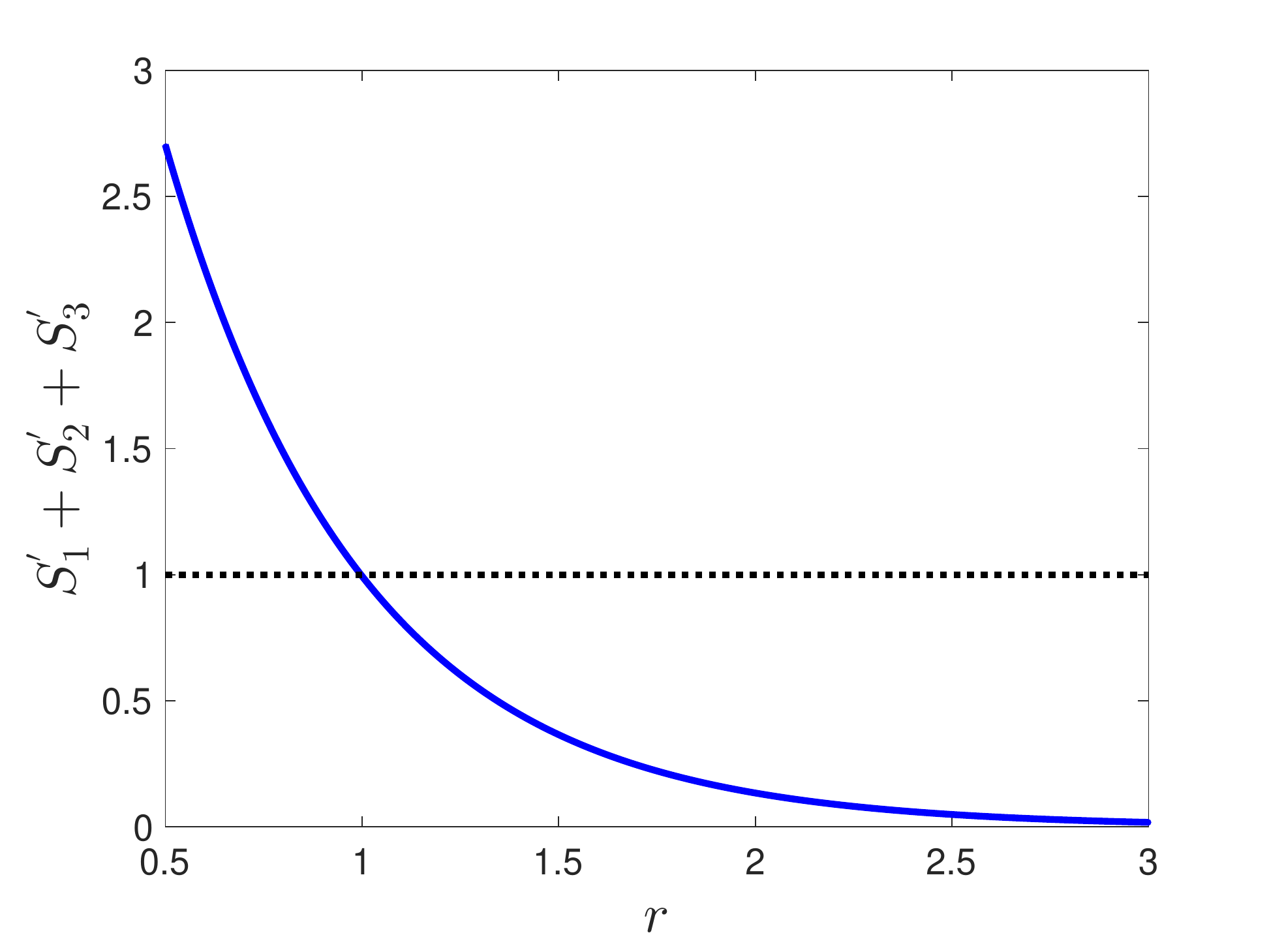}
\par\end{centering}
\caption{Genuine tripartite steering as a function of the squeezing parameter
$r$, using the criterion given by violation of the inequality Eq.
(\ref{eq:stinequ1-2}), for the cluster state studied by Wang \emph{et
al}. \citep{Wang_PRA2019}. The values of $S'_{1}+S'_{2}+S'_{3}$
are given by the blue solid line. The black dotted line corresponds
to the bound for the steering inequality. There is genuine tripartite
steering for $r>1$. \label{fig:Steering_squeezing_param-1}}
\end{figure}

The steering results for the cluster state are plotted in Figure \ref{fig:steering_bipartition_cluster-change-squeezing},
for a varying squeezing parameter $r$. This allows the determination
of the squeezing parameter $r$ required to detect two-way steering
across each bipartition. In fact, we see that there is two-way steering
across all possible bipartitions for $r\geq0.44$. This confirms that
full tripartite steering inseparability can be detected for large
$r$, using the inequalities with the given choice of gains. We note
that we have not optimized the gains, and therefore this may not be
the optimal criterion.

Having demonstrated steering across each bipartition, and hence full
tripartite steering inseparability, we are also able to demonstrate
genuine tripartite steering for this cluster state using Criterion
3.  On examining the gains $h'_{i,j}$ and $g'_{i,j}$ selected for
the $u'_{k}$ and $v'_{k}$ used in the definitions of $S_{k|lm}$
above, we see that $S_{k|lm}$ are identical to $S_{k|lm}$ defined
according to Eqs. (\ref{eq:defn-4}) and (\ref{eq:defn-4-1-1-1}),
provided we choose $h_{1,2}=1$, $h_{1,3}=1$, $g_{1,2}=-1$, $g_{1,3}=0$,
$h_{2,1}=1$, $h_{2,3}=1$, $g_{2,1}=-1$, $g_{2,3}=0$, $h_{3,1}=1$,
$h_{3,2}=1$, $g_{3,1}=0$ and $g_{3,2}=-1$ in the definitions of
$S_{k|lm}$. For those gains, all the relevant bounds on the right
side of the inequality of Criterion 3 are $1$. The inequality of
Criterion 3 reduces to
\begin{equation}
S_{1}^{'}+S_{2}^{'}+S_{3}^{'}\geq1\label{eq:stinequ1-2}
\end{equation}
where $S_{k}'=S{}_{k|lm}$ are given by Eqs. (\ref{eq:defn-4}) and
(\ref{eq:defn-4-1-1-1}). The analytical expressions for $S_{1}^{'}$,
$S_{2}^{'}$ and $S_{3}^{'}$ are solved above, as $S_{1}^{'}=S_{2}^{'}=S_{3}^{'}=\sqrt{6}e^{-2r}\,,$
respectively. Using these expressions, we investigate genuine tripartite
steering based on Eq. (\ref{eq:stinequ1-2}), as a function of squeezing
parameter. The result is plotted in Figure \ref{fig:Steering_squeezing_param-1}
and genuine tripartite steering is possible for $r>1$.

\subsection{CV EPR state}

\subsubsection{Genuine tripartite steering with a single inequality}

\begin{figure}[H]
\begin{centering}
\includegraphics[width=0.93\columnwidth]{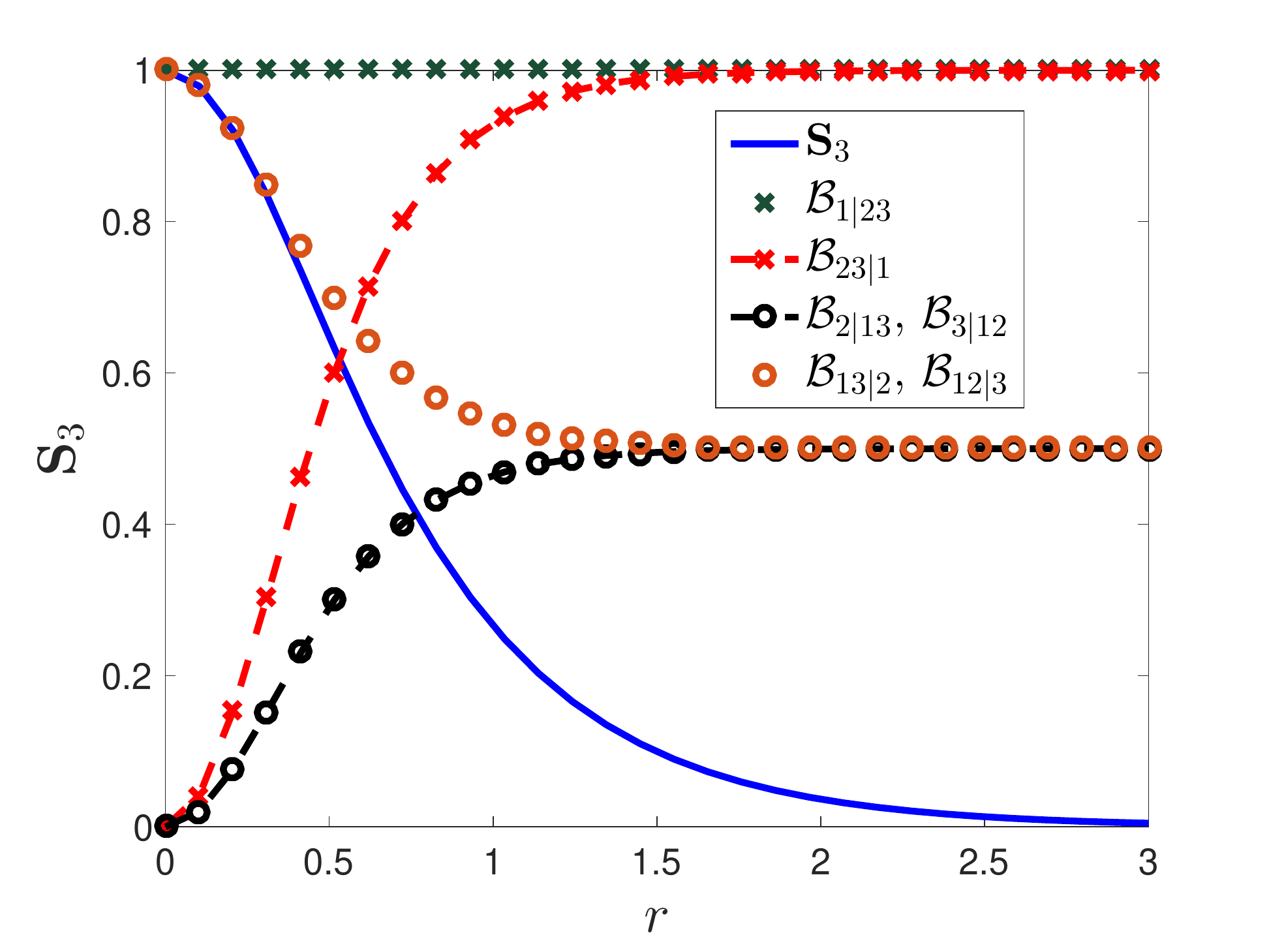}
\par\end{centering}
\caption{The value of $\mathbf{S}_{3}=\Delta u\Delta v$ as a function of the
squeezing parameter $r$, for the CV EPR state given by Figure 1.
The gains as a function of $r$ are given by the analytical expressions
in Eqs. (\ref{eq:g_xs}) and (\ref{eq:g_ps}), which here are optimized
to enhance observation of steering of system $1$. The blue solid
line corresponds to the value of $\mathbf{S}_{3}$. The remaining
lines correspond to different bounds on the right side of the inequality
Eq. (\ref{eq:inequality_CVGHZ}). When $\mathbf{S}$ is smaller than
all bounds, there is genuine tripartite steering. This is obtained
for $r>0.76$.\label{fig:steering_CVEPR-1}}
\end{figure}

To investigate genuine tripartite steering for the CV EPR state,
the Criterion 1 given by Eq. (\ref{eq:inequality_CVGHZ}) can be used.
We first select $g_{1}=h_{1}=1$, and $g_{1}=g_{2}=g$ and $h_{1}=h_{2}=h$
and optimize for $g$ and $h$ by minimizing $S$ with respect to
these gains. This optimization has been carried out in Section IV.A.
The optimal values are $g=g_{x,s}/\sqrt{2}$ and $h=g_{p,s}/\sqrt{2}$
where $g_{x,s}$ and $g_{p,s}$ are given by Eqs. (\ref{eq:g_xs})
and (\ref{eq:g_ps}). The value of $\mathbf{S}_{3}$ as a function
of the squeezing parameter $r$ is plotted in Figure \ref{fig:steering-product}
and in Figure \ref{fig:steering_CVEPR-1}, relative to the bounds
of Eq. (\ref{eq:bounds}). We see that for this choice of gains, there
is steering of system $1$ for all $r$ values.

Genuine tripartite steering is detectable with Criterion 1, for sufficiently
large $r>0.76$. The experiment of Walk \emph{et al}. \citep{Walk_optica2016}
reports a maximum squeezing of $-6.5\text{dB}$, corresponding to
a squeeze parameter of $r=0.75$. This  suggests that detection of
genuine tripartite steering may be feasible using this approach in
the near future.

\begin{figure}[H]
\begin{centering}
\includegraphics[width=0.93\columnwidth]{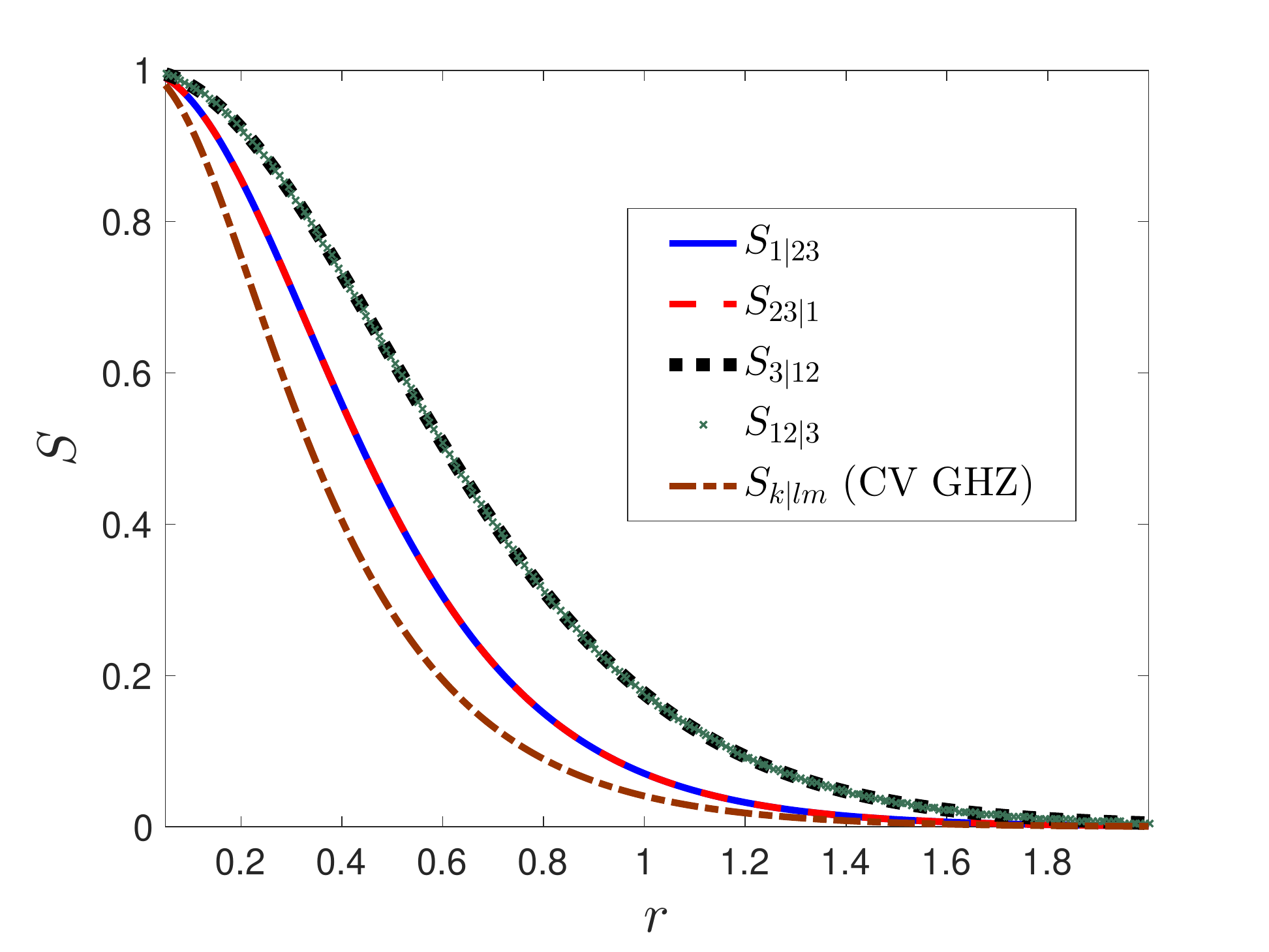}
\par\end{centering}
\caption{The steering for each bipartition as a function of squeezing parameter
$r$, for the CV EPR state given by Figure 1. Here, $S=\Delta u\Delta v$
with the choice of gains $g$ and $h$ given in Tables \ref{tab:gains_EPR_1-23}
and \ref{tab:gains_EPR_23-1}, for the various quantities $S=S_{k|lm}$
or $S_{km|l}$ indicated. A value of $S$ less than $1$ implies steering.
We see that there is two-way steering across each bipartition. There
is symmetry with respect to subsystems $2$ and $3$ (refer Figure
\ref{fig:tripartite_ent_EPR-2}) and hence the plots for $S_{2|13}$
and $S_{13|2}$ are identical to those of $S_{3|12}$ and $S_{12|3}$.
For comparison, we also plot (brown dotted line) the value of $S_{k|lm}$,
which is identical for all values of $k$ (and $l\protect\neq m\protect\neq k$),
for the CV GHZ state with two squeezing parameters. \label{fig:steering_CVEPR}\textcolor{blue}{}}
\end{figure}

\subsubsection{Full tripartite steering inseparability}

Next, we examine full tripartite steering inseparability. We numerically
compute gains that minimize $S_{k|lm}$ and $S_{lm|k}$ (Eqs. (\ref{eq:S_klm})
and (\ref{eq:S_lmk}) respectively) for \emph{each} bipartition. The
values of these gains are given in Tables \ref{tab:gains_EPR_1-23}
and \ref{tab:gains_EPR_23-1} of the Appendix. In particular, the
gains for $S_{1|23}$ are as above. As for the CV GHZ state, there
is two-way steering for all the bipartitions, and hence full tripartite
steering inseparability, for all values of $r$. This is evident in
Figure \ref{fig:steering_CVEPR}. We note that in the limit of large
$r$, the correlations become ideal, implying zero variances, so that
$S_{k|lm}\rightarrow0$ and $S_{lm|k}\rightarrow0$.

\subsubsection{Behavior in the highly squeezed limit}

In the limit of large $r$ where the correlations become ideal and
the variances small, the steering of system $lm$ by $k$ is optimized
by the same choice of gains as the steering of $k$ by $lm$. The
optimal gain coefficients are such that the best steering criterion
for the steering of system $k=1$ is, in the limit of large $r$,
\begin{equation}
\Delta\left[x_{k}-\frac{\left(x_{l}+x_{m}\right)}{\sqrt{2}}\right]\Delta\left[p_{k}+\frac{\left(p_{l}+p_{m}\right)}{\sqrt{2}}\right]<1\thinspace,\label{eq:oldstineq-1-2}
\end{equation}
as expected from the analysis of Section IV. Using that $\left|g_{l}h_{l}+g_{m}h_{m}\right|=1$,
we see that the same inequality Eq. (\ref{eq:oldstineq-1-2}) confirms
steering of $lm$ by $k$, and therefore detects two-way steering
across the bipartition $k-lm$.

\subsubsection{\textup{Experimental observation}}

The observation of steering of a mode $k$ using the criterion Eq.
(\ref{eq:oldstineq-1-2}) has been reported in the experimental system
of Armstrong et al. \citep{Armstrong_Nature2015}, using the set-up
of Figure \ref{fig:tripartite_ent_EPR-2} for EPR states. They reported
steering of each mode in a tripartite system using the violation of
$S_{k|lm}\geq1$, where $g_{k}=h_{k}=1$ for the steered mode. This
corresponds to a realization of full tripartite steering inseparability,
according to the definition given in Section II.C. They found agreement
with the theoretical predictions, with $S_{1|23}=0.78$, $S_{2|13}=S_{3|12}=0.87$
(refer to Table I in the supplementary information of \citep{Armstrong_Nature2015}).
From the gains provided in the same table of their paper, we obtain
\begin{equation}
S_{23|1}=\frac{\Delta(h_{2}x_{2}+h_{3}x_{3}+x_{1})\Delta(g_{2}p_{2}+g_{3}p_{3}+p_{1})}{\left|g_{2}h_{2}+g_{3}h_{3}\right|}=0.95\thinspace,\label{eq:armstrong}
\end{equation}
which indicates two-way steering for the bipartition $1-23$. However,
we estimate $S_{13|2}=S_{12|3}=1.73>1$, which does not satisfy the
requirement for full tripartite two-way steering inseparability.The
estimated maximum squeeze parameter for the experiment was $r\sim0.47$
based on a noise suppression of $-4.1\text{dB}$. The experiment was
also constructed for up to $8$ modes, with the steering of each mode
of the $8$ mode system observed.

\subsection{CV Split squeezed state}

The CV split squeezed state (CV SS) arises from a single squeezed
input, as in Figure \ref{fig:tripartite_ent_EPR-1-1}. Full tripartite
inseparability can be detected using criteria for entanglement, as
explained in \citep{Teh_PRA2019,Teh_erratum_PRA2020}. Here, however
we optimize the correlations further, by selecting a \emph{different}
choice of beam splitter reflectivity, as in Figure \ref{fig:tripartite_ent_EPR-1-1}.

\subsubsection{Genuine tripartite steering using a single inequality}

\begin{figure}[H]
\begin{centering}
\par\end{centering}
\begin{centering}
\includegraphics[width=0.93\columnwidth]{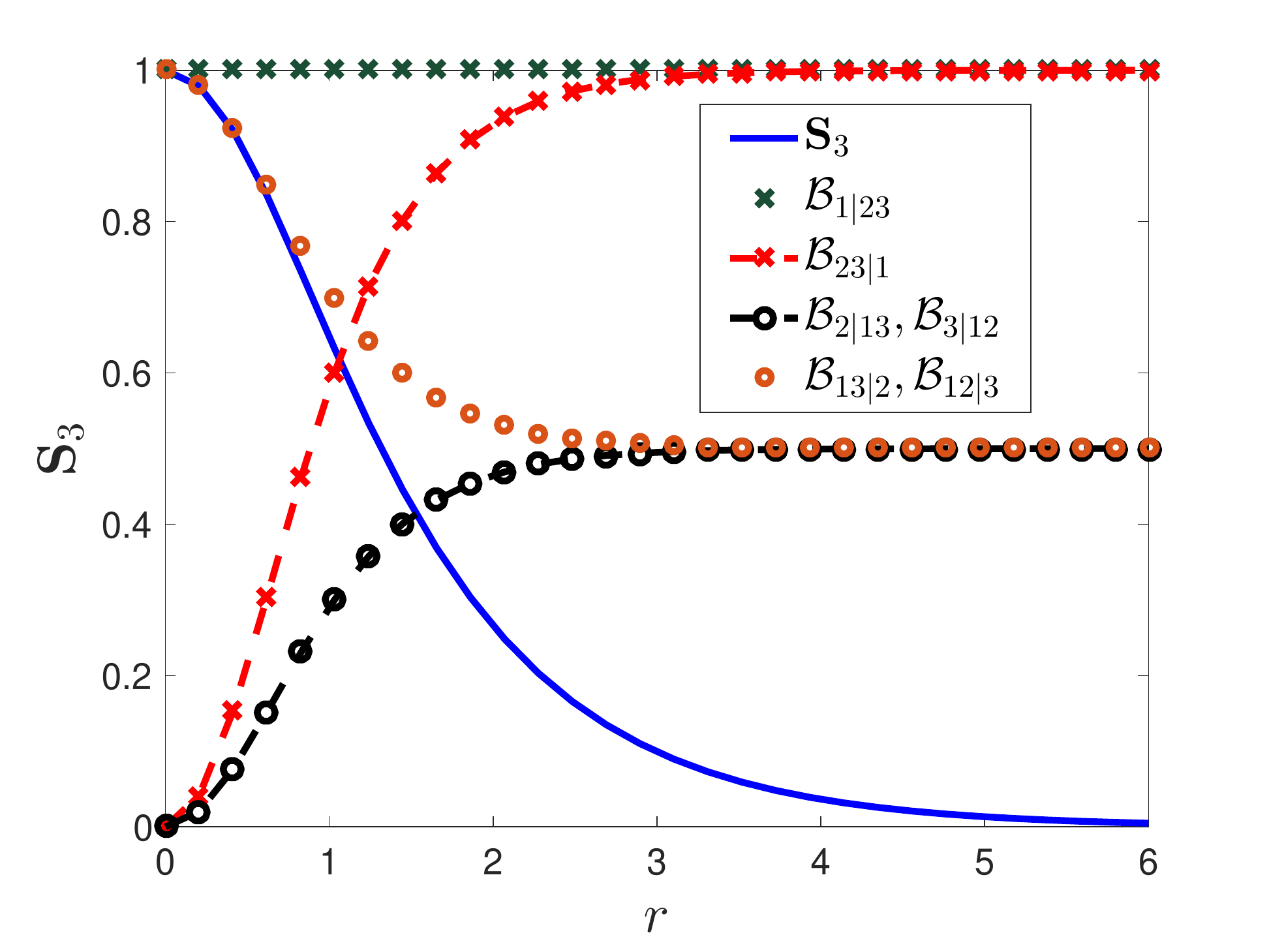}
\par\end{centering}
\caption{The value of $\mathbf{S}_{3}=\Delta u\Delta v$ as a function of the
squeezing parameter $r$, for the CV split squeezed state. The gains
as a function of $r$ are given by the analytical expressions in Eqs.
(\ref{eq:gsoln}) and (\ref{eq:soln-gp}), which are optimized for
the observation of steering of system $1$ (refer to Figure \ref{fig:tripartite_ent_EPR-1-1}).
The blue solid line corresponds to the value of $\mathbf{S}_{3}$.
The remaining lines correspond to different bounds on the right side
of the inequality Eq. (\ref{eq:inequality_CVGHZ}). When $\mathbf{S}_{3}$
is smaller than all the bounds, there is genuine tripartite steering.
This is obtained for $r>1.53$. \label{fig:steering_CVsplit}\textcolor{blue}{}}
\end{figure}

From Figure \ref{fig:steering_CVsplit}, we see that genuine tripartite
steering is detectable using the Criterion 1 for $r>1.56$. This is
expected from the results of Section IV. Here, we use Criterion 1
as given by the inequality Eq. (\ref{eq:inequality_CVGHZ}), where
we select $g_{1}=h_{1}=1$, and $g_{1}=g_{2}=g$ and $h_{1}=h_{2}=h$
and optimize for $g$ and $h$. The optimal gains are derived in Section
IV. With only one squeezed input, the amount of steering for a given
$r$ is reduced, but nonetheless perfect steering is possible for
large $r$.

\begin{figure}[H]
\begin{centering}
\par\end{centering}
\begin{centering}
\includegraphics[width=0.93\columnwidth]{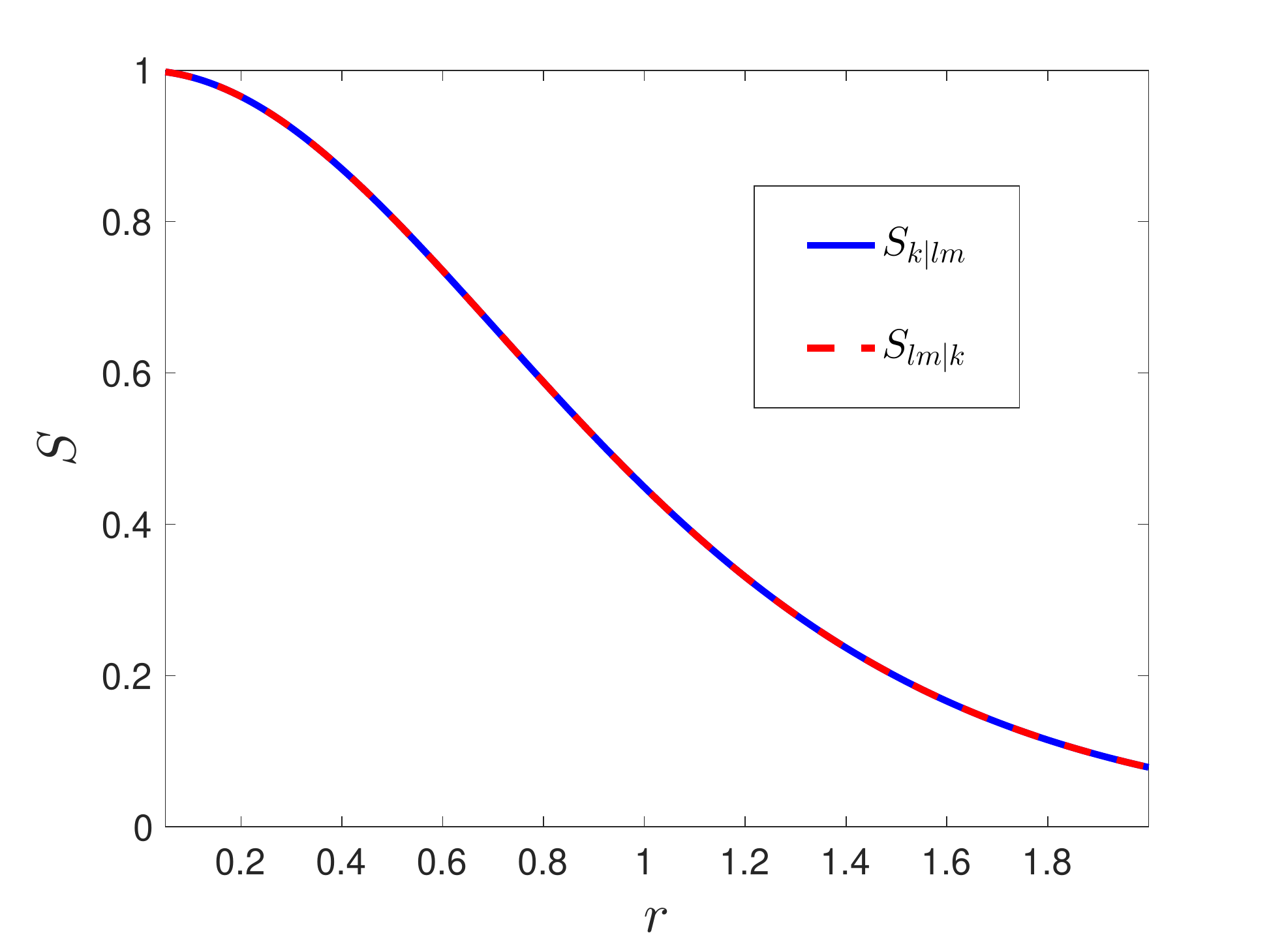}
\par\end{centering}
\caption{\textcolor{black}{The steering for each bipartition as a function
of squeezing parameter $r$, for the CV tripartite SS stat}e. Here,
$S=S_{k|lm}$ or $S_{lm|k}$ as indicated. Here, the lines coincide.
We numerically compute $S_{k|lm}$ and $S_{lm|k}$, as defined in
Eqs. (\ref{eq:S_klm}) and (\ref{eq:S_lmk}), with the choice of gains
provided in Tables \ref{tab:gains_SS_1-23} and \ref{tab:gains_SS_23-1}.
 \textcolor{black}{Here $k\protect\neq l\protect\neq m$ and $k=1,2,$
or $3$. }\textcolor{blue}{}\textcolor{black}{$S<1$ implies steering.
Steering is observed in both directions across each bipartition.}\textcolor{red}{{}
\label{fig:steering_CVSS}}}
\end{figure}

\subsubsection{Full tripartite steering inseparability}

Full tripartite steering inseparability can be detected, for all
$r$. This is seen from Figure \ref{fig:steering_CVSS}. We numerically
obtain the gains that minimize $S_{k|lm}$ and $S_{lm|k}$, as given
in Eqs. (\ref{eq:S_klm}) and (\ref{eq:S_lmk}) respectively, for
each bipartition. When $S_{k|lm},S_{lm|k}<1$, there is steering for
the corresponding bipartition. We see from the figure that there is
two-way steering across all bipartitions. Hence, it is possible to
detect full tripartite two-way steering inseparability. 

\subsection{Experimental genuine tripartite steering using the van Loock-Furuswa-type
inequalities}

We are able to apply the criteria derived in this paper to confirm
from experimental data the realization of genuine tripartite steering
for cluster states. The criteria that we use are based on the van
Loock-Furusawa varainces, which have been measured experimentally.

First, we note from the form of the inequality of Criterion 1 and
from the symmetry of the CV GHZ state that in the limit of large $r$,
the CV GHZ state will give a zero value for the van Loock-Furusawa
quantities $B_{i}$ and $S_{i}$. These quantities were defined by
the van Loock-Furusawa-type Criteria 4 and 5 in Section III.C. This
is also seen directly, from the predictions Eqs. (\ref{eq:ghz-x})
and (\ref{eq:ghz-p}) for the GHZ variances. The CV EPR state will
also give, in the limit of large $r$, a zero for the product $S_{i}$,
as has been shown in \citep{PhysRevA.90.062337,vanLoock_PRA2003}
where similar inequalities were used to detect genuine tripartite
entanglement and steering. Hence, the van Loock-Furusawa-type steering
inequalities of Criteria 4 and 5 can also be used to detect full tripartite
steering inseparability and genuine tripartite steering.

This is useful for interpreting the level of steering generated in
previous CV experiments, which measure the van Loock-Furusawa inequalities.
Previous experiments report full multipartite inseparability \citep{Armstrong_Nature2012}.
In the experiment of Armstrong \emph{et al}. \citep{Armstrong_Nature2012},
the measured variances in the tripartite case are $B_{I}$ and $B_{II}$
of Eq. (\ref{eq:threeineq}). They reported $B_{I}=B_{II}=0.14<2$
(note a different scaling of quadrature amplitudes) for the CV EPR
state. This implies an experimental confirmation of full tripartite
two-way steering inseparability according to Criterion 4b.

In order to demonstrate genuine tripartite steering, Criterion 5b
can be used. However, $B_{III}$ was not directly measured in the
experiment \citep{Armstrong_Nature2012}. Here, the inequality of
Criterion 6c, involving just $B_{I}$ and $B_{II}$ is useful. 
Armstrong \emph{et al}. \citep{Armstrong_Nature2012} measured the
van Loock-Furusawa entanglement inequalities for a CV cluster state.
In the following, we apply van Loock-Furusawa-type inequalities of
Section III.C to show genuine tripartite steering for the cluster
state. The variances measured in the experiment are $B_{I}^{'}\equiv\Delta^{2}\left(p_{1}-x_{2}\right)+\Delta^{2}\left(p_{2}-x_{1}-x_{3}\right)$
and $B_{II}^{'}\equiv\Delta^{2}\left(p_{3}-x_{2}\right)+\Delta^{2}\left(p_{2}-x_{1}-x_{3}\right)$.

Following the proofs given for Criteria 5 and 6c, and the result
Eq. (\ref{eq:prod-sum}), we see that the violation of the inequality
\begin{align}
B_{I}^{'}+B_{II}^{'} & \geq2P_{1}+2P_{2}+2P_{3}\geq2\label{eq:crit-two-vlf-cluster}
\end{align}
implies genuine tripartite steering. Following the proof for Criterion
7, we see that the same violation also implies genuine tripartite
steering by the stricter Definition 3.  In the experiment, Armstrong
\emph{et al}. \citep{Armstrong_Nature2012} obtained $\Delta^{2}\left(p_{1}-x_{2}\right)+\Delta^{2}\left(p_{2}-x_{1}-x_{3}\right)=0.12$
and $\Delta^{2}\left(p_{3}-x_{2}\right)+\Delta^{2}\left(p_{2}-x_{1}-x_{3}\right)=0.18$,
which violates the above inequality and hence demonstrates \emph{experimentally}
genuine tripartite steering, by Definition 1 and Definition 3.

We note that the CV GHZ will satisfy the Criterion 7 of Section III.D
for large $r$. This follows from the predictions Eq. (\ref{eq:ghz-x})
and Eq. (\ref{eq:ghz-p}) for the variances. This gives an avenue
to generate and detect the strict form of genuine tripartite steering
(Definition 3) for these states, which applies to networks of only
one trusted site. The extension to $N$ systems would seem straightforward.

\section{Monogamy relations}

In this section, we summarize how the bipartite entanglement and
steering is distributed among the subsystems of the tripartite steerable
states. It is known that for three qubit systems, the bipartite entanglement
between any two of them is limited by monogamy relations \citep{Coffman_PRA2000}.
This result can be extended to $N$-qubit systems \citep{Osborne_PRL2006}
and to nonlocality and steering \citep{Koashi_monogamy_PRA2004,Toner_monogamy_2009,Streltsov_monogamy_PRL2012,Reid_monogamy_PRA2013,Bai_monogamy_PRL2014,Ji_monogamy_JPA2015,Lami_Winter_PRL2016,Xiang_monogamy_PRA2017}.
The monogamy for higher dimensional systems is more complex, and has
been investigated for CV systems \citep{Adesso_tangle_NJP2006,Hiroshima_monogamy_PRL2007,Reid_monogamy_PRA2013}.

\begin{figure}[H]
\begin{centering}
\includegraphics[width=0.93\columnwidth]{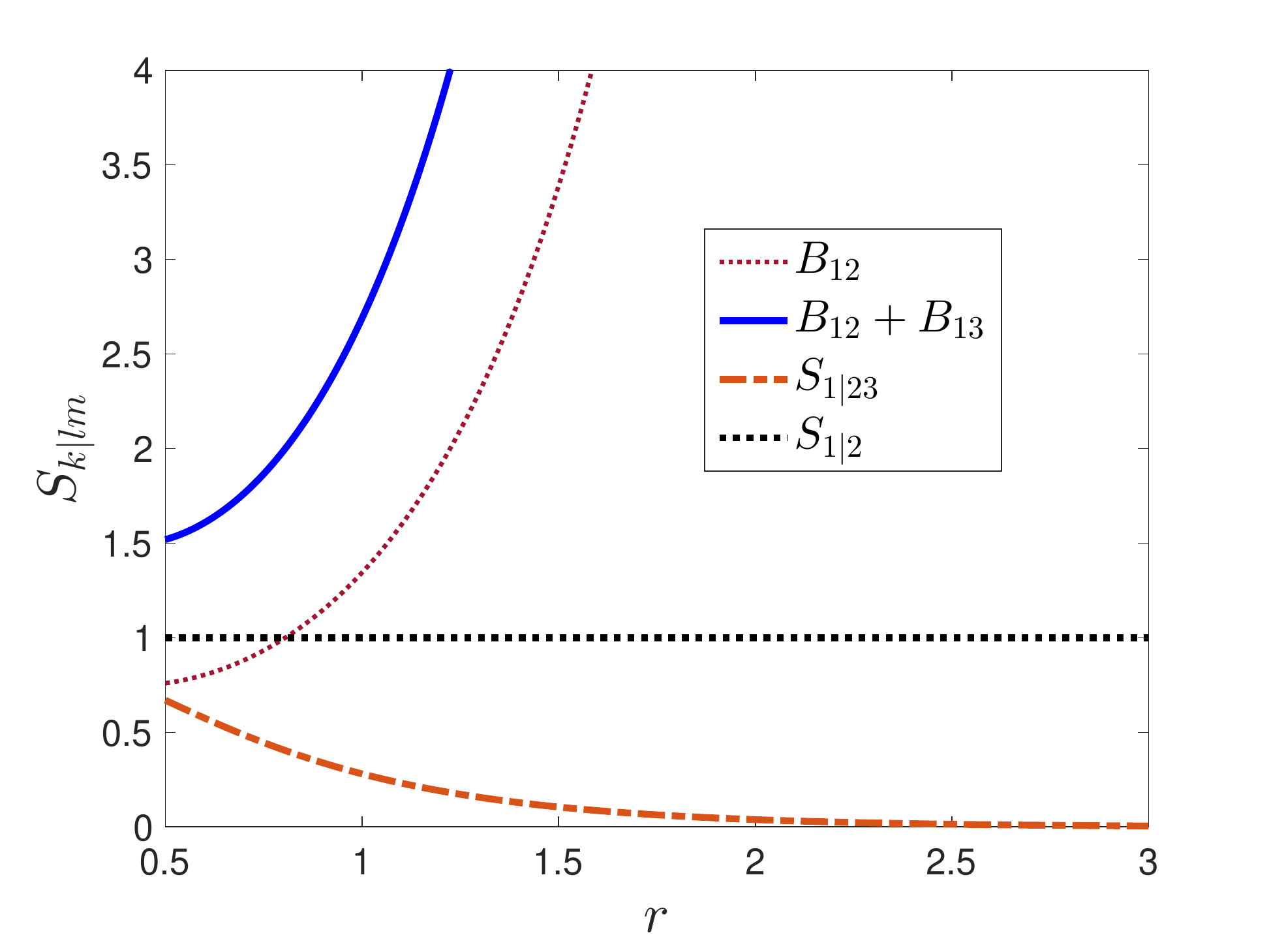}
\par\end{centering}
\caption{Steering and entanglement monogamy for the CV GHZ state depicted in
Figure \ref{fig:tripartite_ent_GHZ-1-1}. Here, all modes are symmetric
e.g. $S_{1|23}=S_{2|13}=S_{3|12}$ and $S_{1|2}=S_{1|3}$. Steering
of system $k$ by both steering parties $l$ and $m$ is optimal for
large $r$, as $S_{k|lm}\rightarrow0$, given by the dashed-dotted
brown line. However, no bipartite steering as measurable by the criterion
$S_{i|j}<1$ is possible for any mode $i$.\textcolor{red}{{} }\textcolor{blue}{}Similarly,
there is no DGCZ bipartite entanglement measurable for large $r$,
as $B_{ij}>1$ in this limit. We observe that the steering monogamy
relation Eq. (\ref{eq:monogamy-full}) is saturated for all $k$,
with $S_{k|l}S_{k|m}=1$ (black dotted line). The entanglement monogamy
inequalities Eq. (\ref{eq:mong-ent}) and Eq. (\ref{eq:mono1}) are
satisfied. \label{fig:monogamy_DAB_CVGHZ}\textcolor{blue}{}}
\end{figure}

We first examine the distribution of bipartite steering among the
three systems created in CV GHZ, CV EPR and CV SS states. For steering,
it is known that \citep{Reid_monogamy_PRA2013}
\begin{equation}
\mathcal{S}_{A|B}\mathcal{S}_{A|C}\geq1\label{eq:monogamy-1}
\end{equation}
where $\mathcal{S}_{i|j}$ is defined in Section II.F for an arbitrary
observable of the steering parties. In this paper, the observables
taken for the steering parties are linear combinations of quadrature
phase amplitudes, in which case we write $S\equiv\mathcal{S}$. The
steering parameter is then $S_{k|l}=\Delta\left(x_{k}-h_{kl}x_{l}\right)\Delta\left(p_{k}+g_{kl}p_{l}\right)$
where the gains $h_{kl}$ and $g_{kl}$ are optimized to minimize
the value of $S_{k|l}$. Regardless of the choice of gains however,
steering is obtained when $S_{k|l}<1$ \citep{Cavalcanti_PRA2009,Reid_PRA1989}.
For the choice of optimal gains, $S_{k|l}<1$ becomes a necessary
and sufficient condition to demonstrate steering for Gaussian states
and measurements \citep{Jones_PRA2007}. The specific monogamy inequality
\begin{equation}
S_{k|l}S_{k|m}\geq\max\{1,S_{k|lm}^{2}\},\label{eq:monogamy-full}
\end{equation}
where $S_{k|lm}$ is defined with the optimal linear gains $h_{k(lm)}$
and $g_{k(lm)}$ that minimize the value of $S_{k|lm}$, follows from
the definitions and Eq. (\ref{eq:monogamy-1}) (without the assumption
of Gaussian states \citep{laura_monogamy,Reid_monogamy_PRA2013,Armstrong_Nature2015}).
The relation has been verified experimentally for the CV EPR state
\citep{Bowen_PRL_monogamy,Armstrong_Nature2015}. Where there is
collective steering such that $S_{1|23}<1$, the monogamy relation
gives $S_{1|2}S_{1|3}\geq1$.
\begin{figure}[H]
\begin{centering}
\includegraphics[width=0.93\columnwidth]{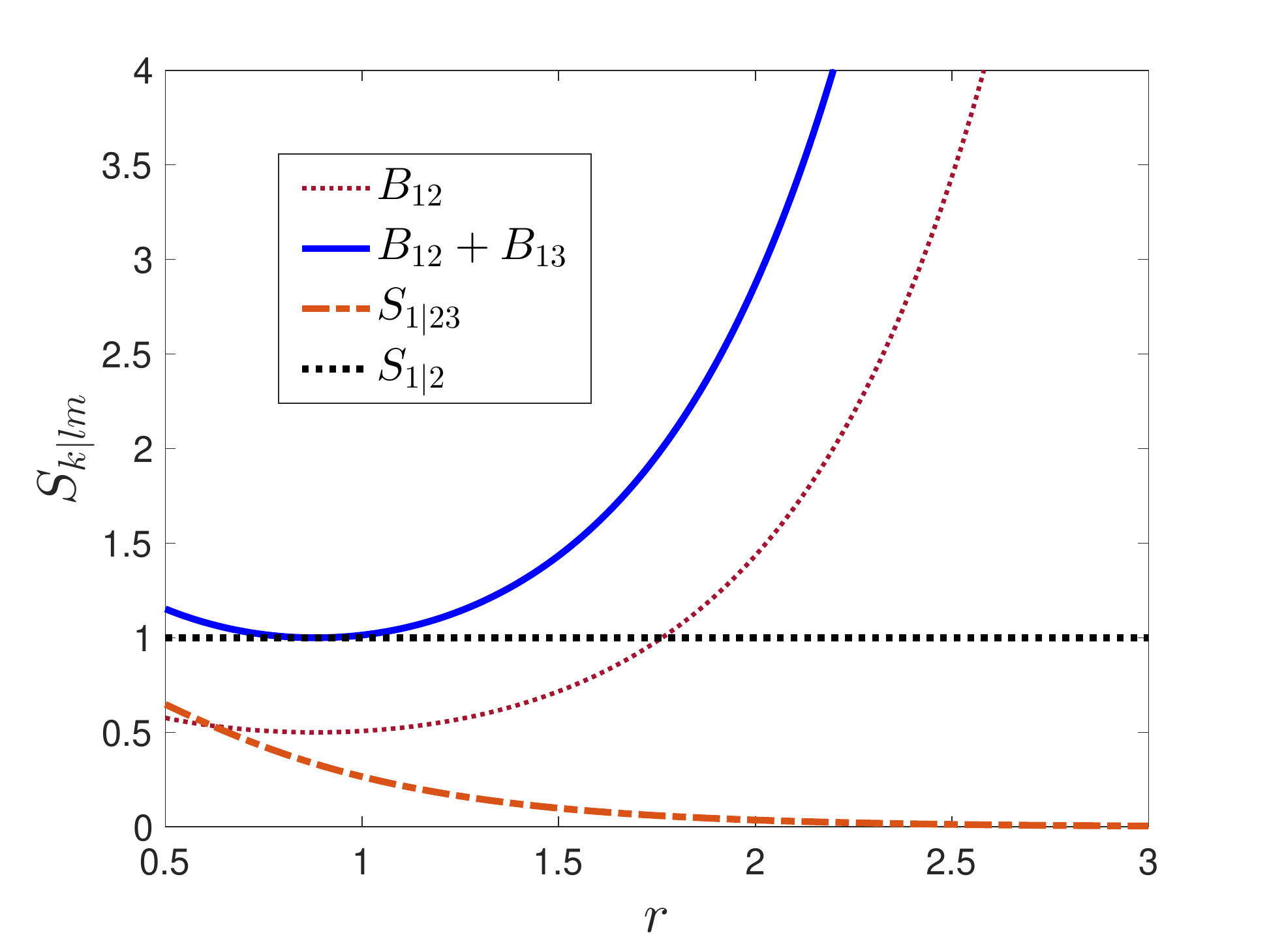}
\par\end{centering}
\caption{Steering and entanglement monogamy for the CV EPR state depicted in
Figure \ref{fig:tripartite_ent_EPR-2}. Modes $2$ and $3$ are symmetric
e.g. $S_{1|2}=S_{1|3}$. Steering of system $1$ by both $2$ and
$3$ is optimal for large $r$, but no bipartite steering measured
by $S_{1|2}<1$ is possible.\textcolor{blue}{{} }The steering monogamy
relation Eq. (\ref{eq:monogamy-full}) is saturated for $k=1$, with
$S_{1|2}S_{1|3}=1$. A similar result is obtained for $k=3$.\textcolor{blue}{{}
}\label{fig:monogamy_DAB_CVEPR}\textcolor{blue}{}\textcolor{teal}{}}
\end{figure}

We see from the monogamy relation Eq. (\ref{eq:monogamy-full}) that
where there is symmetry with respect to fields $l$ and $m$, so that
$S_{k|l}=S_{k|m}$,  then $S_{k|l}\geq1$ and $S_{k|m}\geq1$. Thus
for the CV GHZ state, which has three symmetrical modes, there can
be no bipartite steering\emph{ }as witnessed by $S_{k|l}<1$, for
any system mode $k=1,2,3$ (Figure \ref{fig:monogamy_DAB_CVGHZ}).
This implies there is no such steering of $k$ by just one of the
other single subsystems $l$ or $m$. There is however maximal steering
of any mode $k$ if one considers collectively both of the other modes,
since $S_{k|lm}\rightarrow0$ for large $r$. The steering witness
$S_{k|lm}$ refers to the error in the estimate of $x$ and $p$ of
system $k$ by systems $l$ and $m$, and hence this property is of
value for CV secret sharing.

For the CV EPR system, there is symmetry between the modes $2$ and
$3$, as depicted in Figure 1. Therefore, $S_{1|2}=S_{1|3}\geq1$,
as evident in Figure \ref{fig:monogamy_DAB_CVEPR}. We see however
that $S_{1|23}\rightarrow0$ for large $r$, implying that mode $1$
steered collectively by $2$ and $3$ is useful for secret sharing.

\begin{figure}[H]
\begin{centering}
\includegraphics[width=0.93\columnwidth]{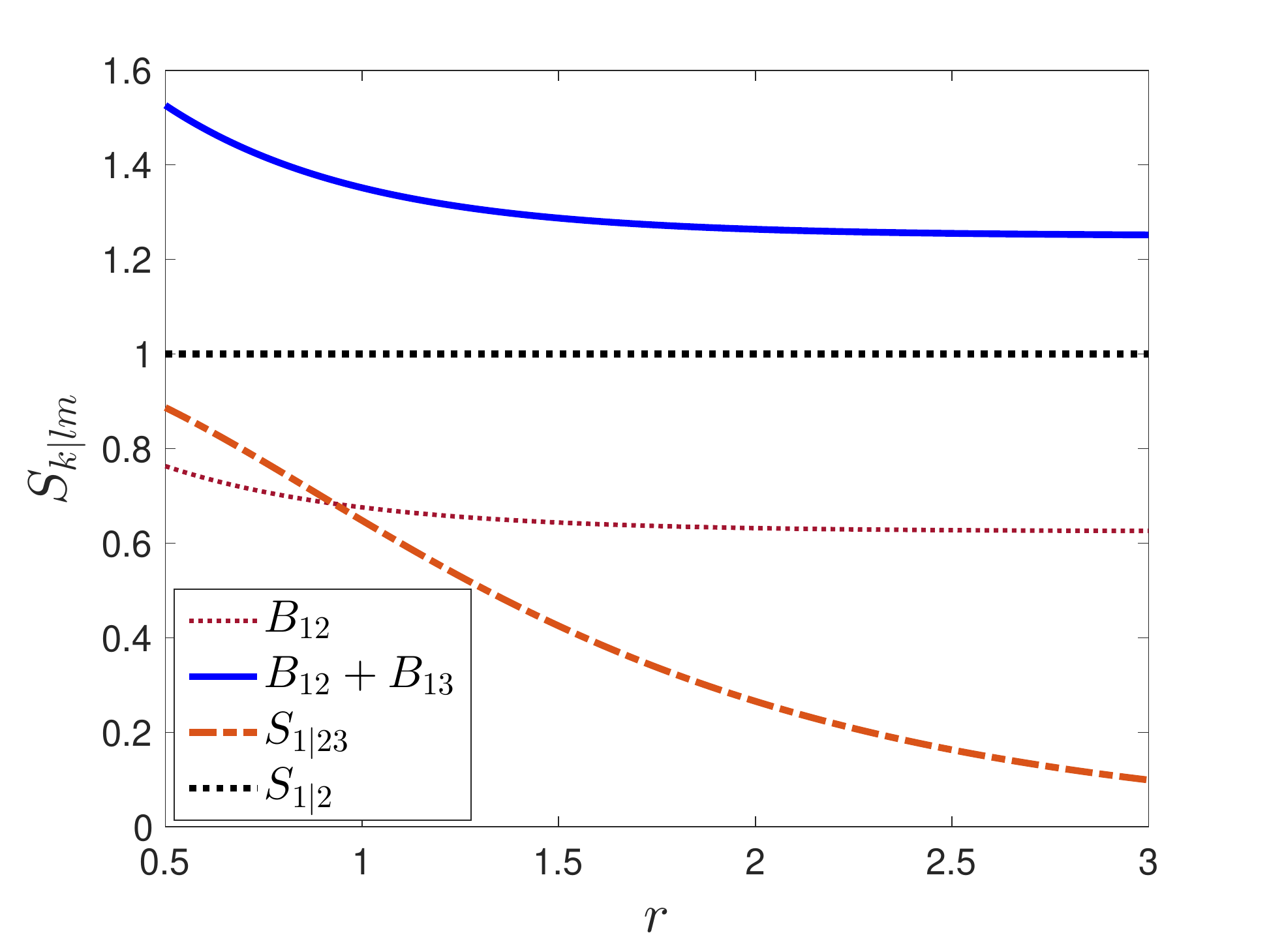}
\par\end{centering}
\caption{Steering and entanglement monogamy for the CV SS state depicted in
Figure \ref{fig:tripartite_ent_EPR-1-1}. Notation as for Figure \ref{fig:monogamy_DAB_CVGHZ}.
Although bipartite steering according to $S_{1|j}$ is not possible,
we observe bipartite entanglement $B_{12}=B_{13}\sim0.63$ for large
$r$.\textcolor{blue}{{} }The steering monogamy relation Eq. (\ref{eq:monogamy-full})
is saturated for $k=1$, with $S_{1|2}S_{1|3}=1$.\textcolor{blue}{{}
}A similar result is obtained for $k=3$. \label{fig:steering-monogamy_CVSS}\textcolor{blue}{}\textcolor{teal}{}}
\end{figure}

The CV split squeezed state also has symmetry between the modes $2$
and $3$, as depicted in Figure \ref{fig:tripartite_ent_EPR-1-1}.
Again, we obtain $S_{1|23}\rightarrow0$ for large $r$, and see that
$S_{1|2}=S_{1|3}\geq1$. We show this relation in Figure \ref{fig:steering-monogamy_CVSS}.

Further monogamy relations have been derived for Gaussian systems
\citep{Ji_monogamy_JPA2015,Lami_Winter_PRL2016,Xiang_monogamy_PRA2017},
using the Gaussian steering quantifier $\mathcal{G^{B\rightarrow A}}$
\citep{Kogias_PRL2015} that quantifies the steerability of mode $A$
by $B$. For two-mode Gaussian steering, the measure of Gaussian
steering $\mathcal{G^{B\rightarrow A}}$ may be mapped from $S_{A|B}$,
since $S_{A|B}<1$ is a necessary and sufficient condition for steering
\citep{Jones_PRA2007}. Explicitly, the steerability of mode $A$
by $B$ by the steering parameter $S_{A|B}$ and the Gaussian steering
quantifier $\mathcal{G}^{\mathcal{B}\rightarrow\mathcal{A}}$ are
related by the expression
\begin{align}
S_{A|B} & =e^{-2\mathcal{G}^{\mathcal{B}\rightarrow\mathcal{A}}}\,.\label{eq:Gaussian_steering_relation}
\end{align}
We note that $S_{A|B}=0$ corresponds to maximum steering. In the
present paper, we restrict study to the condition $S_{A|B}<1$ which
confirms steering without the assumption of Gaussian states, giving
an advantage for applications relating to secure quantum communication.
However, the Gaussian monogamy relations derived in \citep{Kogias_PRL2015}
and verified experimentally \citep{Deng_monogamy_PRL2017} will hold
for the CV systems we examine in this paper, which are examples of
Gaussian states.

\begin{figure}[H]
\begin{centering}
\includegraphics[width=0.95\columnwidth]{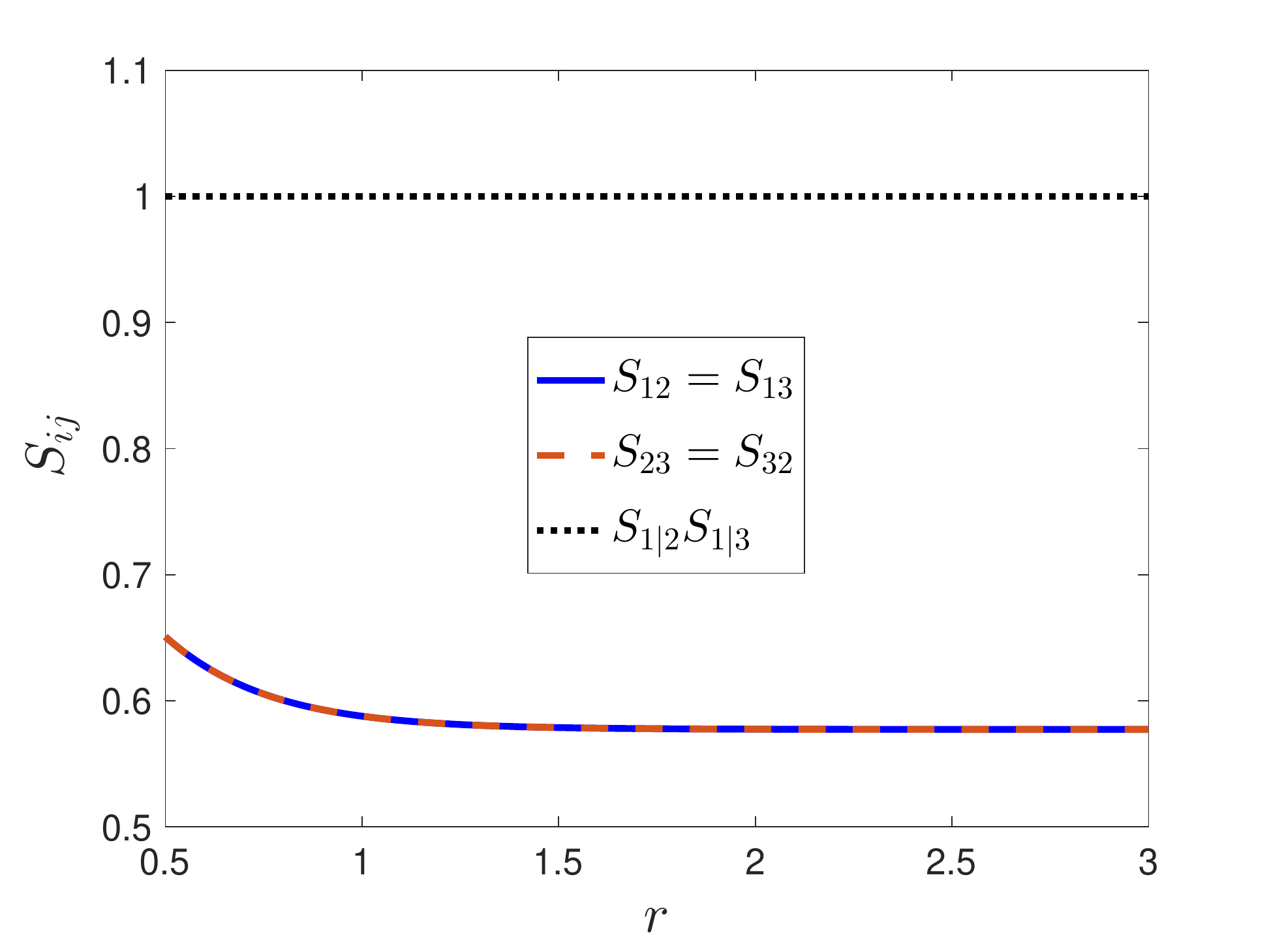}
\par\end{centering}
\caption{Bipartite entanglement and entanglement monogamy for the CV GHZ state
depicted in Figure \ref{fig:tripartite_ent_GHZ-1-1}. The system is
symmetric with respect to all modes. Entanglement between $i$ and
$j$ is confirmed when $S_{ij}<1$. Bipartite entanglement is observed
between all pairs of parties. The values of $S_{1|2}S_{1|3}$ are
given by the upper black dashed line. Saturation of the generalised
monogamy entanglement inequality Eq. (\ref{eq:mono2}) is observed
for all $k$.\textcolor{blue}{} \label{fig:Bipartite-ent-monogamy-GHZ}\textcolor{red}{}}
\end{figure}
We now turn to examine the distribution of bipartite entanglement
among the tripartite steerable systems. Rosales-Zarate \emph{et al}.
derived four monogamy inequalities for entanglement \citep{laura_monogamy}.
A monogamy relation exists for the bipartite CV variance
\begin{equation}
B_{ij}\equiv\frac{1}{4}\left[(\Delta\left(X_{i}-X_{j}\right))^{2}+(\Delta\left(P_{i}+P_{j}\right))^{2}\right]\label{eq:B_ij}
\end{equation}
of Duan, Giedke, Cirac and Zoller (DGCZ) \citep{Duan_PRL2000}. The
criterion $B_{ij}<1$ is sufficient to confirm entanglement between
states $i$ and $j$ \citep{Duan_PRL2000}. We note $B_{12}$ is
given by $B_{I}$ of Eq. (\ref{eq:threeineq}) with $g_{3}=0$, and
$B_{13}$ is $B_{III}$ in Eq. (\ref{eq:threeineq}) with $g_{2}=0$
(apart from a renormalisation factor). For any tripartite state of
systems labelled $1$, $2$ and $3$, the monogamy inequalities
\begin{equation}
B_{12}+B_{13}\geq1\,\label{eq:mong-ent}
\end{equation}
and 
\begin{align}
B_{12}+B_{13} & \geq\text{max}\left\{ 1,S_{1|23}\right\} \,\label{eq:mono1}
\end{align}
will always hold \citep{laura_monogamy}.

As pointed out by Rosales-Zarate \emph{et al}. \citep{laura_monogamy},
the monogamy relation Eq. (\ref{eq:mong-ent}) implies that where
there is symmetry between modes $2$ and $3$ (so that $B_{12}=B_{13}$),
it follows that $B_{12}\geq0.5$ and $B_{13}\geq0.5$. Maximum bipartite
entanglement ($B_{12}=0$, $B_{13}=0$) is not possible. Since the
CV GHZ is fully symmetric for all three modes, this limit applies
to all DGCZ bipartite entanglement for the CV GHZ state, as observed
in Figure \ref{fig:monogamy_DAB_CVGHZ}. In fact, for large $r$ there
is no DGCZ bipartite entanglement for the CV GHZ state. The constraint
$B_{12},B_{13}\geq0.5$ also applies to the two symmetric modes of
the CV EPR and CV SS states (Figures \ref{fig:monogamy_DAB_CVEPR}
and \ref{fig:steering-monogamy_CVSS}). The optimal limit $B_{12}=B_{13}=0.5$
is not obtained for any of the three CV states, because of the enhanced
fluctuations arising from the anti-squeezed quadrature of the highly
squeezed inputs. The CV SS state however has fewer squeezed inputs,
and Figure \ref{fig:steering-monogamy_CVSS} shows that for this case,
$B_{12}=B_{13}\sim0.625$ for large $r$. The monogamy relations Eq.
(\ref{eq:mong-ent}) and Eq. (\ref{eq:mono1}) plotted in Figure \ref{fig:monogamy_DAB_CVEPR}
agree with the results presented by Rosales-Zarate \emph{et al}. \citep{laura_monogamy}
(Figure 3 with the parameter $\eta_{0}=0.5$ corresponds to the CV
EPR state). Rosales-Zarate \emph{et al}. \citep{laura_monogamy} showed
how the value $B_{12}=B_{13}=0.5$ can be obtained for the CV EPR
state, if mode $1$ is attenuated to reduce the increased vacuum fluctuations
entering from the squeezed input. The full calculations are given
in the Supplemental Material.

A more general version of the inequality Eq. (\ref{eq:mong-ent})
can be given in terms of the entanglement parameter 
\begin{equation}
S_{kl}=\Delta\left(x_{k}-h_{kl}x_{l}\right)\Delta\left(p_{k}+g_{kl}p_{l}\right)/\left(1+h_{kl}g_{kl}\right)\thinspace.\label{eq:S_ij}
\end{equation}
Here, $g_{ij}$ and $h_{ij}$ are gain factors, selected to minimise
$S_{kl}$. The condition $S_{ij}<1$ (for any $g_{ij}$, $h_{ij}$)
confirms entanglement between modes $i$ and $j$ without the assumption
of Gaussian states, as proved by Giovannetti, Mancini, Vitali and
Tombesi \citep{Giovannetti_PRA2003}. The monogamy relation 
\begin{align}
S_{kl}S_{km} & \geq\frac{\text{max}\left\{ 1,S_{k|lm}^{2}\right\} }{\left(1+h_{kl}g_{kl}\right)\left(1+h_{km}g_{km}\right)}\,\label{eq:mono2}
\end{align}
holds for all states. The proof of the inequality Eq. (\ref{eq:mono2})
extends the work of \citep{laura_monogamy} and is given in the Supplementary
Material. This inequality is \emph{equivalent} to the steering inequality,
Eq. (\ref{eq:monogamy-full}). The criterion $S_{ij}<1$ with optimal
gains was shown equivalent to the Simon-Peres condition for entanglement,
provided there is symmetry between $X$ and $P$ such that the correlation
matrix elements satisfy $\langle X_{1},X_{2}\rangle=-\langle P_{1},P_{2}\rangle$
in which case $h_{ij}=g_{ij}$ \citep{He_steeringdirection_PRL2015}.
For two-mode  Gaussian systems $i$ and $j$, the Simon-Peres condition
is necessary and sufficient to confirm entanglement \citep{Simon_PRL2000,Marian_JOPA2018}).
The monogamy relation for this special case was derived in \citep{laura_monogamy}.
By considering the variance expressions for observables $\hat{X}=\alpha_{1}\hat{x}_{1}-\alpha_{2}\hat{x}_{2}$
and $\hat{P}=\beta_{1}\hat{p}_{1}+\beta_{2}\hat{p}_{2}\,$, which
involve an extra parameter, Marian and Marian generalised the results
to show the how EPR variance criteria reduce to the Simon-Peres condition
for the two-mode Gaussian states \citep{Marian_JOPA2018}.

\begin{figure}[H]
\begin{centering}
\includegraphics[width=0.95\columnwidth]{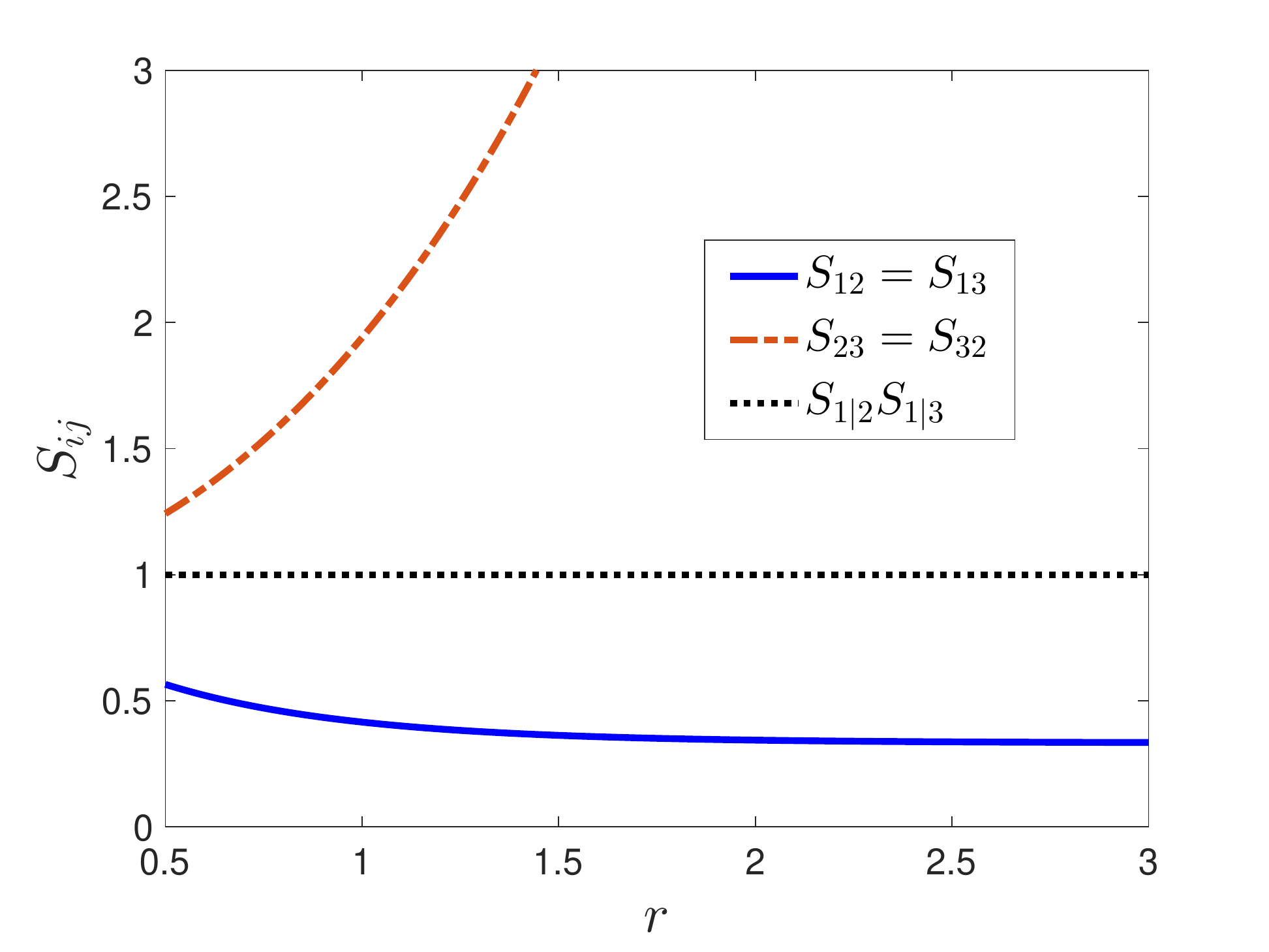}
\par\end{centering}
\caption{Bipartite entanglement and entanglement monogamy for the CV EPR state
depicted in Figure \ref{fig:tripartite_ent_EPR-2}. Entanglement between
$i$ and $j$ is observed when $S_{ij}<1$. The values of $S_{1|2}S_{1|3}$
are given by the upper black dashed line, and correspond to saturation
of the generalised monogamy entanglement inequality Eq. (\ref{eq:mono2})
for $k=1$. Here, there is bipartite entanglement detectable between
systems $1$ and 2, and between systems 1 and $3$.\label{fig:Bipartite-entanglement-and-monogamy-epr}}
\end{figure}
\begin{figure}[H]
\begin{centering}
\includegraphics[width=0.95\columnwidth]{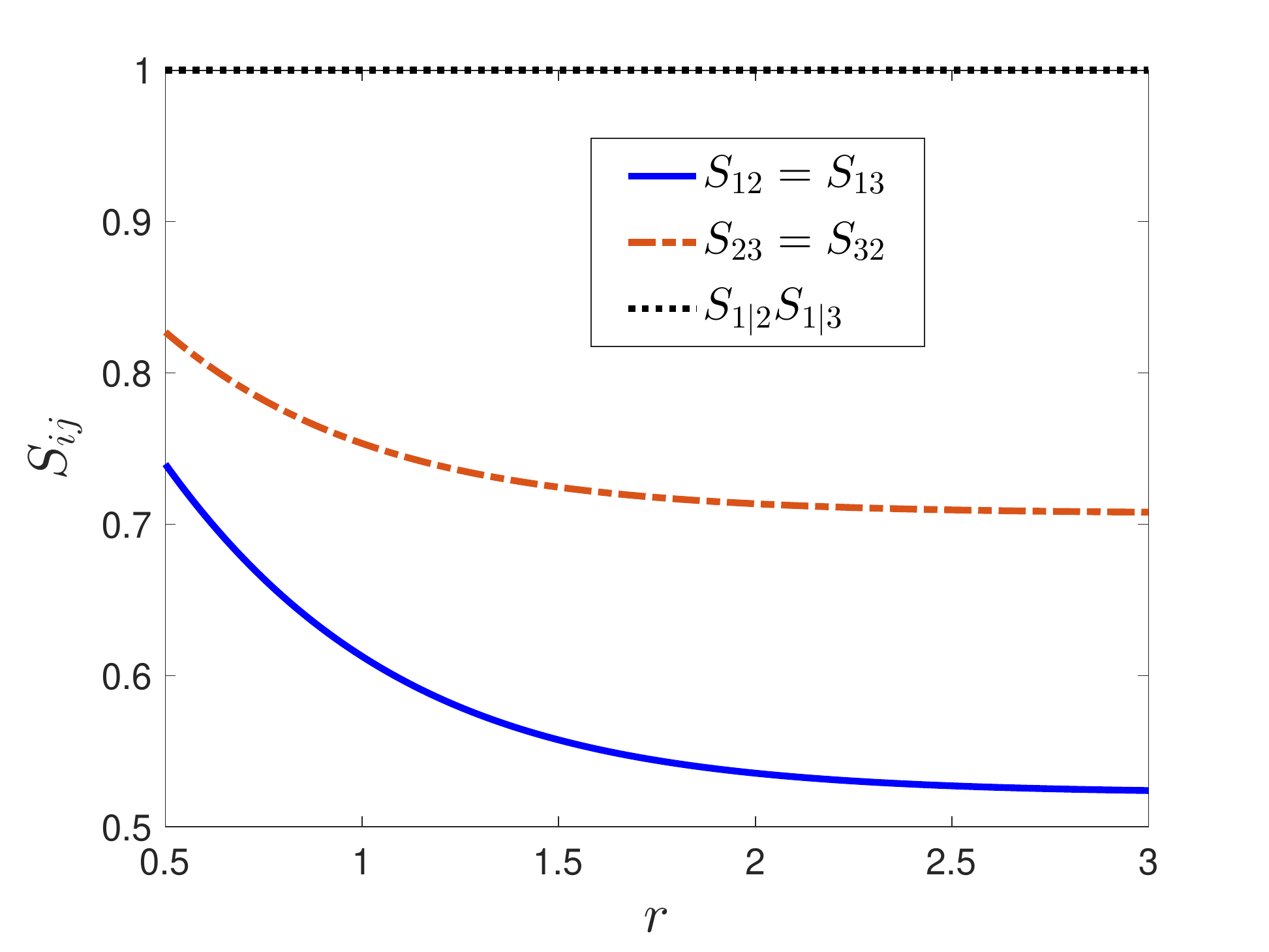}
\par\end{centering}
\caption{Bipartite entanglement and entanglement monogamy for the CV SS state
depicted in Figure \ref{fig:tripartite_ent_EPR-1-1}. Entanglement
between $i$ and $j$ is observed when $S_{ij}<1$. Here, bipartite
entanglement is detectable between all pairs of systems. The values
of $S_{1|2}S_{1|3}$ are given by the upper black dashed line, and
correspond to saturation of the generalised monogamy entanglement
inequality Eq. (\ref{eq:mono2}) for $k=1$.\label{fig:Bipartite-entanglement-and-monogamy-CVSS}}
\end{figure}

In Figures \ref{fig:Bipartite-ent-monogamy-GHZ}-\ref{fig:Bipartite-entanglement-and-monogamy-CVSS},
we see bipartite entanglement $S_{kl}<1$ to be possible for all three
states (the CV GHZ, CV EPR and CV SS states). This extends the results
of \citep{laura_monogamy}. The bipartite entanglement in these figures
is numerically computed where $S_{kl}$  is minimized with the optimal
gains $h_{kl}$ and $g_{kl}$, using the \textit{fminsearch} function
in Matlab. We found that $S_{kl}=S_{lk}$ with the gains satisfying
the relation $h_{kl}=1/h_{lk}$ and $g_{kl}=1/g_{lk}$, as proved
in \citet{He_PRL2015}. The relation Eq. (\ref{eq:mono2}) confirms
it is possible to obtain $S_{ij}<1$, even in the presence of symmetric
modes. However, the optimal EPR-correlated bipartite states have $h_{1j}=g_{1j}=1$,
in which case for symmetrical fields where $S_{1i}=S_{1j}$, we will
find $S_{1j}\geq0.5$. In fact, we observe the approximate relation
$S_{ij}\apprge0.5$ for each of the three types of CV states. 

For the states in this work, $S_{1|23}$ is equal or smaller than
$1$, and the generalized monogamy inequality becomes $S_{12}S_{13}\geq1/\left[\left(1+h_{12}g_{12}\right)\left(1+h_{13}g_{13}\right)\right]$.
A saturation of the generalized entanglement monogamy inequality is
then observed when $S_{1|2}S_{1|3}=1$, this being equivalent to the
saturation of the steering monogamy relation Eq. (\ref{eq:monogamy-full}).
The authors of \citep{laura_monogamy} observed that while there is
no saturation of the monogamy relations Eq. (\ref{eq:mong-ent}) for
the CV EPR and CV GHZ states, saturation of the generalized monogamy
relation Eq. (\ref{eq:mono2}) can be observed for CV EPR states.
However, that result was limited by the restriction $h_{ij}=g_{ij}$.
Here, we extend this result, by examining the CV GHZ and CV SS state,
as given in Figures \ref{fig:Bipartite-ent-monogamy-GHZ} and \ref{fig:Bipartite-entanglement-and-monogamy-CVSS}.
These figures clearly show saturation of the generalized monogamy
relation Eq. (\ref{fig:Bipartite-entanglement-and-monogamy-CVSS})
to be possible for each of these states, for all $r$.

\section{Conclusion}

In this paper we have examined the concept of multipartite steering
for multi-mode CV systems, deriving inequalities that if violated
signify the presence of genuine $N$-partite steering. In contrast
with much previous work, the inequalities are derived without the
assumption of Gaussian states. This gives an advantage for protocols
which rely on the rigorous confirmation of multipartite entanglement
or steering (e.g. quantum key distribution).

Our classification of $N$-partite steering carefully distinguishes
between full tripartite steering inseparability and genuine tripartite
steering. In fact, because of the asymmetry of the steering correlation
with respect to parties which may have different levels of trust,
a multitude of definitions are possible. Here, we use a conceptual
definition based on the requirement to have steering both ways along
each of the different bipartitions of the $N$ systems. However, alternative
definitions are likely to be useful, especially where the $N$ systems
are nodes of a network with fixed levels of trust. In this paper,
we also consider one such alternative definition (Definition 3) of
genuine multipartite steering, motivated by the example of networks
with only one trusted subsystem. We derive inequalities to detect
this strict type of steering.

In Section IV, strategies are identified to generate genuine multipartite
steering based on the coherent beam-splitter mixing of one, two, and
N squeezed beams with vacuum modes. We examine the CV GHZ, CV EPR
and CV split-squeezed states that use $N$, two and one squeezed beams
as inputs for a network of $N$ subsystems, respectively. It is shown
that the genuine $N$-partite steering for each of these states can
be detected by the inequalities derived in this paper, provided the
squeezing of the inputs is sufficiently large. Specific examples
for $N=3$ are given in Section V, and the genuine $N$-partite steering
of a tripartite CV cluster state is also analyzed. We also show that
these systems predict the strict form of genuine multipartite steering
given by Definition 3, and that this type of steering can be detected
by a van Loock-Furusawa-type inequality.

Using the inequalities derived in this paper, we confirm that full
tripartite steering inseparability has been experimentally generated
for three optical modes. Armstrong et al. \citep{Armstrong_Nature2012}
produced CV EPR states and tested the van Loock-Furusawa type inequalities,
measuring $B_{I}$ and $B_{II}$ as given by Eq. (\ref{eq:threeineq})
of our paper. Using Criterion 4b, the experimental values show violation
of both $B_{I}\geq2$ and $B_{II}\geq2$, implying full tripartite
two-way steering inseparability. In a second experiment, Armstrong
et al. \citep{Armstrong_Nature2015} measured the steering along each
bipartition of the three modes $A$, $B$, and\textbf{ $C$ }by violating
the steering inequalities $S_{A|BC}\geq1$, $S_{B|AC}\geq1$ and $S_{C|AB}\geq1$
(Eqs. (\ref{eq:fulloneway}) and (\ref{eq:S_klm})). This confirms
full tripartite steering inseparability, by our definition.  The
experimental results of \citep{Armstrong_Nature2012} also indicate
that genuine tripartite steering has been generated for a CV cluster
state. This is based on the data reported for the van Loock-Furusawa
inequalities. In Section V.D, we show that Criteria 6c and 7 (with
an adjusted phase choice) certify both forms of genuine tripartite
steering for this data, by Definitions 1 and 3. Deng et al. \citep{Deng_monogamy_PRL2017}
generated four-mode cluster states and examined full steering inseparability
with the focus on monogamy relations between these modes, but within
the Gaussian-state assumption.

For tripartite states, the distributions of bipartite steering and
of bipartite entanglement among the three subsystems will be constrained.
In Section VI, we present new results for the monogamy of steering
and for the monogamy of entanglement, as certified with respect to
a particular witness. We study the monogamy properties for the tripartite
CV GHZ, CV EPR and CV SS states. We find that a limited amount of
bipartite entanglement may exist between the nodes of the tripartite
network, for each of the three types of states, but that bipartite
steering as measured by the CV EPR-steering criterion can not. The
monogamy relations we derive do not require the assumption of Gaussian
systems and may therefore be useful for CV secret sharing protocols.
\begin{acknowledgments}
This research has been supported by the Australian Research Council
Discovery Project Grants schemes under Grant DP180102470. This work
was funded by LabEx ENS-ICFP Grants No. ANR-10-LABX-0010 and No. ANR-10-IDEX-0001-02
PSL{*}. This work received funding from MCIN and AEI under Project
No. PID2020-115761RJ-I00, from a fellowship from the la Caixa Foundation
(Grant No. 100010434), and from the European Union\textquoteright s
Horizon 2020 research and innovation program under Marie Sk\l odowska-Curie
Grant No. 847648 (Fellowship Code No. LCF/BQ/PI21/11830025). The authors
wish to thank Nippon Telegraph and Telephone (NTT) Research for their
financial and technical support
\end{acknowledgments}

\section*{Appendix}

\subsection{Gains for CV GHZ states\label{subsec:gains_GHZ}}

\begin{table}[H]
\begin{centering}
\begin{tabular}{|c|c|c|}
\hline 
\multirow{2}{*}{\textbf{r}} & \multicolumn{2}{c|}{\textbf{CV GHZ}}\tabularnewline
 & \multicolumn{1}{c}{$h$} & $g$\tabularnewline
\hline 
\hline 
0 & 0 & 0\tabularnewline
\hline 
0.25 & -0.27 & 0.36\tabularnewline
\hline 
0.50 & -0.40 & 0.68\tabularnewline
\hline 
0.75 & -0.46 & 0.86\tabularnewline
\hline 
1.00 & -0.49 & 0.95\tabularnewline
\hline 
1.50 & -0.50 & 0.99\tabularnewline
\hline 
2.00 & -0.50 & 1.00\tabularnewline
\hline 
\end{tabular}
\par\end{centering}
\caption{Values of the gains $h$ and $g$ that minimize the variance product
in Criterion Eq. (\ref{eq:inequality_CVGHZ}).  \label{tab:fixed_gain_CVGHZ}}
\end{table}

\begin{table}[H]
\begin{centering}
\begin{tabular}{|c|c|c|c|c|c|c|c|c|}
\hline 
\multirow{2}{*}{\textbf{r}} & \multicolumn{4}{c|}{Gains for $S_{1|23}$} & \multicolumn{4}{c|}{Gains for $S_{2|13}$}\tabularnewline
\cline{2-9} 
 & \multicolumn{1}{c}{$\tilde{h}_{2}$} & \multicolumn{1}{c}{$\tilde{g}_{2}$} & \multicolumn{1}{c}{$\tilde{h}_{3}$} & $\tilde{g}_{3}$ & \multicolumn{1}{c}{$\tilde{h}_{1}$} & \multicolumn{1}{c}{$\tilde{g}_{1}$} & \multicolumn{1}{c}{$\tilde{h}_{3}$} & $\tilde{g}_{3}$\tabularnewline
\hline 
\hline 
0.25 & $-0.27$ & $0.36$ & $-0.27$ & $0.36$ & $-0.27$ & $0.36$ & $-0.27$ & $0.36$\tabularnewline
\hline 
0.50 & $-0.40$ & $0.68$ & $-0.40$ & $0.68$ & $-0.40$ & $0.68$ & $-0.40$ & $0.68$\tabularnewline
\hline 
0.75 & $-0.46$ & $0.86$ & $-0.46$ & $0.86$ & $-0.46$ & $0.86$ & $-0.46$ & $0.86$\tabularnewline
\hline 
1.00 & $-0.49$ & $0.95$ & $-0.49$ & $0.95$ & $-0.49$ & $0.95$ & $-0.49$ & $0.95$\tabularnewline
\hline 
1.50 & $-0.50$ & $0.99$ & $-0.50$ & $0.99$ & $-0.50$ & $0.99$ & $-0.50$ & $0.99$\tabularnewline
\hline 
2.00 & $-0.50$ & $1.00$ & $-0.50$ & $1.00$ & $-0.50$ & $1.00$ & $-0.50$ & $1.00$\tabularnewline
\hline 
\end{tabular}
\par\end{centering}
\caption{The optimal gains for CV GHZ state. The optimal gains for $S_{3|12}$
are identical to the gains for $S_{2|13}$ with $\tilde{h}_{3}=\tilde{h}_{2}$
and $\tilde{g}_{3}=\tilde{g}_{2}$. \label{tab:gains_GHZ_1-23}}
\end{table}

\begin{table}[H]
\begin{centering}
\begin{tabular}{|c|c|c|c|c|c|c|c|c|}
\hline 
\multirow{2}{*}{\textbf{r}} & \multicolumn{4}{c|}{Gains for $S_{23|1}$} & \multicolumn{4}{c|}{Gains for $S_{13|2}$}\tabularnewline
\cline{2-9} 
 & \multicolumn{1}{c}{$\tilde{h}_{2}$} & \multicolumn{1}{c}{$\tilde{g}_{2}$} & \multicolumn{1}{c}{$\tilde{h}_{3}$} & $\tilde{g}_{3}$ & \multicolumn{1}{c}{$\tilde{h}_{1}$} & \multicolumn{1}{c}{$\tilde{g}_{1}$} & \multicolumn{1}{c}{$\tilde{h}_{3}$} & $\tilde{g}_{3}$\tabularnewline
\hline 
\hline 
0.25 & $-1.37$ & $1.87$ & $-1.37$ & $1.87$ & $-1.37$ & $1.87$ & $-1.37$ & $1.87$\tabularnewline
\hline 
0.50 & $-0.73$ & $1.23$ & $-0.73$ & $1.23$ & $-0.73$ & $1.23$ & $-0.73$ & $1.23$\tabularnewline
\hline 
0.75 & $-0.58$ & $1.08$ & $-0.58$ & $1.08$ & $-0.58$ & $1.08$ & $-0.58$ & $1.08$\tabularnewline
\hline 
1.00 & $-0.53$ & $1.03$ & $-0.53$ & $1.03$ & $-0.53$ & $1.03$ & $-0.53$ & $1.03$\tabularnewline
\hline 
1.50 & $-0.50$ & $1.00$ & $-0.50$ & $1.00$ & $-0.50$ & $1.00$ & $-0.50$ & $1.00$\tabularnewline
\hline 
2.00 & $-0.50$ & $1.00$ & $-0.50$ & $1.00$ & $-0.50$ & $1.00$ & $-0.50$ & $1.00$\tabularnewline
\hline 
\end{tabular}
\par\end{centering}
\caption{The optimal gains for CV GHZ state. The optimal gains for $S_{12|3}$
are identical to the gains for $S_{13|2}$ with $\tilde{h}_{3}=\tilde{h}_{2}$
and $\tilde{g}_{3}=\tilde{g}_{2}$. \label{tab:gains_GHZ_23-1}}
\end{table}
\begin{table}[H]
\begin{centering}
\begin{tabular}{|c|c|c|}
\hline 
\multirow{2}{*}{\textbf{$r_{2}=r$}} & \multicolumn{2}{c|}{\textbf{CV GHZ}}\tabularnewline
 & \multicolumn{1}{c}{$h$} & $g$\tabularnewline
\hline 
\hline 
0 & 0 & 0\tabularnewline
\hline 
0.25 & -0.34 & 0.51\tabularnewline
\hline 
0.50 & -0.45 & 0.80\tabularnewline
\hline 
0.75 & -0.48 & 0.97\tabularnewline
\hline 
1.00 & -0.49 & 0.99\tabularnewline
\hline 
1.50 & -0.50 & 1.00\tabularnewline
\hline 
2.00 & -0.50 & 1.00\tabularnewline
\hline 
\end{tabular}
\par\end{centering}
\caption{Values of the gains $h$ and $g$, as given by the analytical expressions
in Eq. (\ref{eq:gains-N}), that minimize the variance product $\mathbf{S}_{3}$.
Here, the squeezing strengths $r_{2}=r$ and $r_{1}$ is related to
$r_{2}$ by the relation Eq. (\ref{eq:squeezing_relation}). For large
$r$, the gains are identical to those in the case where there is
only one squeezing strength.  \label{tab:fixed_gain_CVGHZ-1-1}}
\end{table}
\begin{table}[H]
\begin{centering}
\begin{tabular}{|c|c|c|c|c|c|c|c|c|}
\hline 
\multirow{2}{*}{\textbf{r}} & \multicolumn{4}{c|}{Gains for $S_{1|23}$} & \multicolumn{4}{c|}{Gains for $S_{2|13}$}\tabularnewline
\cline{2-9} 
 & \multicolumn{1}{c}{$\tilde{h}_{2}$} & \multicolumn{1}{c}{$\tilde{g}_{2}$} & \multicolumn{1}{c}{$\tilde{h}_{3}$} & $\tilde{g}_{3}$ & \multicolumn{1}{c}{$\tilde{h}_{1}$} & \multicolumn{1}{c}{$\tilde{g}_{1}$} & \multicolumn{1}{c}{$\tilde{h}_{3}$} & $\tilde{g}_{3}$\tabularnewline
\hline 
\hline 
0.25 & $-0.34$ & $0.51$ & $-0.34$ & $0.51$ & $-0.34$ & $0.51$ & $-0.34$ & $0.51$\tabularnewline
\hline 
0.50 & $-0.45$ & $0.80$ & $-0.45$ & $0.80$ & $-0.45$ & $0.80$ & $-0.45$ & $0.80$\tabularnewline
\hline 
0.75 & $-0.48$ & $0.93$ & $-0.48$ & $0.93$ & $-0.48$ & $0.93$ & $-0.48$ & $0.93$\tabularnewline
\hline 
1.00 & $-0.49$ & $0.97$ & $-0.49$ & $0.97$ & $-0.49$ & $0.97$ & $-0.49$ & $0.97$\tabularnewline
\hline 
1.50 & $-0.49$ & $0.97$ & $-0.49$ & $0.97$ & $-0.49$ & $0.97$ & $-0.49$ & $0.97$\tabularnewline
\hline 
2.00 & $-0.50$ & $1.00$ & $-0.50$ & $1.00$ & $-0.50$ & $1.00$ & $-0.50$ & $1.00$\tabularnewline
\hline 
\end{tabular}
\par\end{centering}
\caption{The optimal gains for CV GHZ state with two squeezing strengths. Here,
$r=r_{2}$ while $r_{1}$ is related to $r_{2}$ by Eq. (\ref{eq:squeezing_relation}).
The optimal gains for $S_{3|12}$ are identical to the gains for $S_{2|13}$
with $\tilde{h}_{3}=\tilde{h}_{2}$ and $\tilde{g}_{3}=\tilde{g}_{2}$.
\label{tab:gains_GHZ_1-23-1}\textcolor{blue}{}}
\end{table}

\begin{table}[H]
\begin{centering}
\begin{tabular}{|c|c|c|c|c|c|c|c|c|}
\hline 
\multirow{2}{*}{\textbf{r}} & \multicolumn{4}{c|}{Gains for $S_{23|1}$} & \multicolumn{4}{c|}{Gains for $S_{13|2}$}\tabularnewline
\cline{2-9} 
 & \multicolumn{1}{c}{$\tilde{h}_{2}$} & \multicolumn{1}{c}{$\tilde{g}_{2}$} & \multicolumn{1}{c}{$\tilde{h}_{3}$} & $\tilde{g}_{3}$ & \multicolumn{1}{c}{$\tilde{h}_{1}$} & \multicolumn{1}{c}{$\tilde{g}_{1}$} & \multicolumn{1}{c}{$\tilde{h}_{3}$} & $\tilde{g}_{3}$\tabularnewline
\hline 
\hline 
0.25 & $-0.98$ & $1.48$ & $-0.98$ & $1.48$ & $-0.98$ & $1.48$ & $-0.98$ & $1.48$\tabularnewline
\hline 
0.50 & $-0.62$ & $1.12$ & $-0.62$ & $1.12$ & $-0.62$ & $1.12$ & $-0.62$ & $1.12$\tabularnewline
\hline 
0.75 & $-0.54$ & $1.04$ & $-0.54$ & $1.04$ & $-0.54$ & $1.04$ & $-0.54$ & $1.04$\tabularnewline
\hline 
1.00 & $-0.51$ & $1.01$ & $-0.51$ & $1.01$ & $-0.51$ & $1.01$ & $-0.51$ & $1.01$\tabularnewline
\hline 
1.50 & $-0.50$ & $1.00$ & $-0.50$ & $1.00$ & $-0.50$ & $1.00$ & $-0.50$ & $1.00$\tabularnewline
\hline 
2.00 & $-0.50$ & $1.00$ & $-0.50$ & $1.00$ & $-0.50$ & $1.00$ & $-0.50$ & $1.00$\tabularnewline
\hline 
\end{tabular}
\par\end{centering}
\caption{The optimal gains for CV GHZ state with two squeezing strengths. Here,
$r=r_{2}$ while $r_{1}$ is related to $r_{2}$ by Eq. (\ref{eq:squeezing_relation}).
The optimal gains for $S_{12|3}$ are identical to the gains for $S_{13|2}$
with $\tilde{h}_{3}=\tilde{h}_{2}$ and $\tilde{g}_{3}=\tilde{g}_{2}$.
\label{tab:gains_GHZ_23-1-1}\textcolor{blue}{}}
\end{table}

\subsection{Gains for CV EPR states\label{subsec:gains_EPR}}

\begin{table}[H]
\begin{centering}
\begin{tabular}{|c|c|c|c|c|c|c|c|c|}
\hline 
\multirow{2}{*}{\textbf{r}} & \multicolumn{4}{c|}{$S_{1|23}$} & \multicolumn{4}{c|}{$S_{2|13}$}\tabularnewline
\cline{2-9} 
 & \multicolumn{1}{c}{$\tilde{h}_{2}$} & \multicolumn{1}{c}{$\tilde{g}_{2}$} & \multicolumn{1}{c}{$\tilde{h}_{3}$} & $\tilde{g}_{3}$ & \multicolumn{1}{c}{$\tilde{h}_{1}$} & \multicolumn{1}{c}{$\tilde{g}_{1}$} & \multicolumn{1}{c}{$\tilde{h}_{3}$} & $\tilde{g}_{3}$\tabularnewline
\hline 
\hline 
0.25 & $-0.33$ & $0.33$ & $-0.33$ & $0.33$ & $-0.35$ & $0.35$ & $0.06$ & $0.06$\tabularnewline
\hline 
0.50 & $-0.54$ & $0.54$ & $-0.54$ & $0.54$ & $-0.65$ & $0.65$ & $0.21$ & $0.21$\tabularnewline
\hline 
0.75 & $-0.64$ & $0.64$ & $-0.64$ & $0.64$ & $-0.90$ & $0.90$ & $0.40$ & $0.40$\tabularnewline
\hline 
1.00 & $-0.68$ & $0.68$ & $-0.68$ & $0.68$ & $-1.08$ & $1.08$ & $0.58$ & $0.58$\tabularnewline
\hline 
1.50 & $-0.70$ & $0.70$ & $-0.70$ & $0.70$ & $-1.28$ & $1.28$ & $0.82$ & $0.82$\tabularnewline
\hline 
2.00 & $-0.71$ & $0.71$ & $-0.71$ & $0.71$ & $-1.36$ & $1.36$ & $0.93$ & $0.93$\tabularnewline
\hline 
\end{tabular}
\par\end{centering}
\caption{The optimal gains for CV EPR state. The optimal gains for $S_{3|12}$
are identical to the gains for $S_{2|13}$ with $\tilde{h}_{3}=\tilde{h}_{2}$
and $\tilde{g}_{3}=\tilde{g}_{2}$. \label{tab:gains_EPR_1-23}}
\end{table}

\begin{table}[H]
\begin{centering}
\begin{tabular}{|c|c|c|c|c|c|c|c|c|}
\hline 
\multirow{2}{*}{\textbf{r}} & \multicolumn{4}{c|}{$S_{23|1}$} & \multicolumn{4}{c|}{$S_{13|2}$}\tabularnewline
\cline{2-9} 
 & \multicolumn{1}{c}{$\tilde{h}_{2}$} & \multicolumn{1}{c}{$\tilde{g}_{2}$} & \multicolumn{1}{c}{$\tilde{h}_{3}$} & $\tilde{g}_{3}$ & \multicolumn{1}{c}{$\tilde{h}_{1}$} & \multicolumn{1}{c}{$\tilde{g}_{1}$} & \multicolumn{1}{c}{$\tilde{h}_{3}$} & $\tilde{g}_{3}$\tabularnewline
\hline 
\hline 
0.25 & $-1.53$ & $1.53$ & $-1.53$ & $1.53$ & $-2.98$ & $2.98$ & $0.52$ & $0.52$\tabularnewline
\hline 
0.50 & $-0.93$ & $0.93$ & $-0.93$ & $0.93$ & $-1.71$ & $1.71$ & $0.56$ & $0.56$\tabularnewline
\hline 
0.75 & $-0.78$ & $0.78$ & $-0.78$ & $0.78$ & $-1.40$ & $1.40$ & $0.63$ & $0.63$\tabularnewline
\hline 
1.00 & $-0.73$ & $0.73$ & $-0.73$ & $0.73$ & $-1.31$ & $1.31$ & $0.70$ & $0.70$\tabularnewline
\hline 
1.50 & $-0.71$ & $0.71$ & $-0.71$ & $0.71$ & $-1.32$ & $1.32$ & $0.85$ & $0.85$\tabularnewline
\hline 
2.00 & $-0.71$ & $0.71$ & $-0.71$ & $0.71$ & $-1.37$ & $1.37$ & $0.93$ & $0.93$\tabularnewline
\hline 
\end{tabular}
\par\end{centering}
\caption{The optimal gains for CV EPR state. The optimal gains for $S_{13|2}$
are identical to the gains for $S_{12|3}$ with $\tilde{h}_{3}=\tilde{h}_{2}$
and $\tilde{g}_{3}=\tilde{g}_{2}$. \label{tab:gains_EPR_23-1}}
\end{table}

\subsection{Gains for CV SS state\label{subsec:gains_CVSS}}

\begin{table}[H]
\begin{centering}
\begin{tabular}{|c|c|c|c|c|c|c|c|c|}
\hline 
\multirow{2}{*}{\textbf{r}} & \multicolumn{4}{c|}{\textbf{$S_{1|23}$}} & \multicolumn{4}{c|}{$S_{2|13}$}\tabularnewline
\cline{2-9} 
 & \multicolumn{1}{c}{\textbf{$\tilde{h}_{2}$}} & \multicolumn{1}{c}{$\tilde{g}_{2}$} & \multicolumn{1}{c}{$\tilde{h}_{3}$} & $\tilde{g}_{3}$ & \multicolumn{1}{c}{$\tilde{h}_{1}$} & \multicolumn{1}{c}{$\tilde{g}_{1}$} & \multicolumn{1}{c}{$\tilde{h}_{3}$} & $\tilde{g}_{3}$\tabularnewline
\hline 
\hline 
\noun{0.25} & \textbf{$-0.15$} & $0.18$ & $-0.15$ & $0.18$ & $-0.15$ & $0.18$ & $-0.15$ & $0.18$\tabularnewline
\hline 
\noun{0.50} & \textbf{$-0.27$} & $0.36$ & $-0.27$ & $0.36$ & $-0.27$ & $0.36$ & $-0.27$ & $0.36$\tabularnewline
\hline 
\noun{0.75} & \textbf{$-0.35$} & $0.54$ & $-0.35$ & $0.54$ & $-0.35$ & $0.54$ & $-0.35$ & $0.54$\tabularnewline
\hline 
\noun{1.00} & \textbf{$-0.40$} & $0.68$ & $-0.40$ & $0.68$ & $-0.40$ & $0.68$ & $-0.40$ & $0.68$\tabularnewline
\hline 
\noun{1.50} & \textbf{$-0.46$} & $0.86$ & $-0.46$ & $0.86$ & $-0.46$ & $0.86$ & $-0.46$ & $0.86$\tabularnewline
\hline 
\noun{2.00} & \textbf{$-0.49$} & $0.95$ & $-0.49$ & $0.95$ & $-0.49$ & $0.95$ & $-0.49$ & $0.95$\tabularnewline
\hline 
\end{tabular}
\par\end{centering}
\caption{The optimal gains for CV SS state. The optimal gains for $S_{3|12}$
are identical to the gains for $S_{2|13}$ with $\tilde{h}_{3}=\tilde{h}_{2}$
and $\tilde{g}_{3}=\tilde{g}_{2}$. \label{tab:gains_SS_1-23}}
\end{table}

\begin{table}[H]
\begin{centering}
\begin{tabular}{|c|c|c|c|c|c|c|c|c|}
\hline 
\multirow{2}{*}{\textbf{r}} & \multicolumn{4}{c|}{\textbf{$S_{23|1}$}} & \multicolumn{4}{c|}{$S_{13|2}$}\tabularnewline
\cline{2-9} 
 & \multicolumn{1}{c}{$\tilde{h}_{2}$} & \multicolumn{1}{c}{$\tilde{g}_{2}$} & \multicolumn{1}{c}{$\tilde{h}_{3}$} & $\tilde{g}_{3}$ & \multicolumn{1}{c}{$\tilde{h}_{1}$} & \multicolumn{1}{c}{$\tilde{g}_{1}$} & \multicolumn{1}{c}{$\tilde{h}_{3}$} & $\tilde{g}_{3}$\tabularnewline
\hline 
\hline 
0.25 & $-2.81$ & $3.31$ & $-2.81$ & $3.31$ & $-2.81$ & $3.31$ & $-2.81$ & $3.31$\tabularnewline
\hline 
0.50 & $-1.37$ & $1.87$ & $-1.37$ & $1.87$ & $-1.37$ & $1.87$ & $-1.37$ & $1.87$\tabularnewline
\hline 
0.75 & $-0.93$ & $1.43$ & $-0.93$ & $1.43$ & $-0.93$ & $1.43$ & $-0.93$ & $1.43$\tabularnewline
\hline 
1.00 & $-0.73$ & $1.23$ & $-0.73$ & $1.23$ & $-0.73$ & $1.23$ & $-0.73$ & $1.23$\tabularnewline
\hline 
1.50 & $-0.58$ & $1.08$ & $-0.58$ & $1.08$ & $-0.58$ & $1.08$ & $-0.58$ & $1.08$\tabularnewline
\hline 
2.00 & $-0.53$ & $1.03$ & $-0.53$ & $1.03$ & $-0.53$ & $1.03$ & $-0.53$ & $1.03$\tabularnewline
\hline 
\end{tabular}
\par\end{centering}
\caption{The optimal gains for CV SS state. The optimal gains for $S_{13|2}$
are identical to the gains for $S_{12|3}$ with $\tilde{h}_{3}=\tilde{h}_{2}$
and $\tilde{g}_{3}=\tilde{g}_{2}$. \label{tab:gains_SS_23-1}}
\end{table}

\bibliographystyle{apsrev4-1}
\bibliography{multipartite_steering-final}

\end{document}